\documentclass[twocolumn]{IEEEtran}

\usepackage[T1]{fontenc}

\ifCLASSINFOpdf
\else
\fi

\usepackage{url} 

\usepackage{dblfloatfix}    
\usepackage{amsfonts}
\usepackage{graphicx}
\usepackage{epstopdf}
\usepackage{amsthm}
\usepackage{amssymb}
\usepackage{mathtools}
\usepackage{amsmath} 

\theoremstyle{definition}

\usepackage{mathtools, cuted} 
\usepackage{cuted}

\usepackage{url} 
\usepackage{blindtext, graphicx}

\usepackage{float} 
\usepackage{subcaption} 

\usepackage{multirow}
\usepackage{array}

\usepackage[numbers,sort&compress]{natbib}

\usepackage{booktabs}
\usepackage{tabularx}

\usepackage{array,multirow}

\usepackage{pbox}

\usepackage{longtable}

\usepackage{dblfloatfix}

\usepackage{enumitem}

\usepackage[linesnumbered,ruled,vlined]{algorithm2e}

\usepackage[table]{xcolor} 

\usepackage{amssymb} 
\usepackage{amsfonts}
\usepackage{bbm}

\usepackage{bm}

\DeclareMathAlphabet{\mathcal}{OMS}{cmsy}{m}{n}

\usepackage[normalem]{ulem} 

\newcounter{probNum}

\usepackage{pbox}

\usepackage{pifont}

\usepackage{changepage}

\usepackage{totcount}
\newtotcounter{citenum}
\def\oldcite{}
\let\oldcite=\bibcite
\def\bibcite{\stepcounter{citenum}\oldcite}

\definecolor{anti-flashwhite}{rgb}{0.95, 0.95, 0.96}

\begin{document}
	%
	\title{User-centric Cell-free Massive MIMO Networks: A Survey of Opportunities, Challenges and Solutions}
	
	\author{Hussein~A.~Ammar\IEEEauthorrefmark{1},~\IEEEmembership{Student Member,~IEEE}, 
		Raviraj~Adve\IEEEauthorrefmark{1},~\IEEEmembership{Fellow,~IEEE},
		Shahram~Shahbazpanahi\IEEEauthorrefmark{2}\IEEEauthorrefmark{1},~\IEEEmembership{Senior Member,~IEEE},
		Gary~Boudreau\IEEEauthorrefmark{3},~\IEEEmembership{Senior Member,~IEEE},
		and~Kothapalli~Venkata~Srinivas\IEEEauthorrefmark{3},~\IEEEmembership{Member,~IEEE}
		\thanks{This work was supported in part by the Natural Sciences and Engineering Research Council (NSERC) of Canada and in part by Ericsson Canada.}
		\thanks{
			\IEEEauthorrefmark{1}H. A. Ammar and R. Adve are with the Edward S. Rogers Sr. Department of Electrical and Computer Engineering, University of Toronto, Toronto, ON M5S 3G4, Canada (e-mail: ammarhus@ece.utoronto.ca; rsadve@comm.utoronto.ca).
		}
		\thanks{
			\IEEEauthorrefmark{2}S. Shahbazpanahi is with the Department of Electrical, Computer, and Software Engineering, University of Ontario Institute of Technology, Oshawa, ON L1H 7K4, Canada. He also holds a Status-Only position with the Edward S. Rogers Sr. Department of Electrical and Computer Engineering, University of Toronto.
		}
		\thanks{
			\IEEEauthorrefmark{3}G. Boudreau and K. Srinivas are with Ericsson Canada, Ottawa, ON L4W 5K4, Canada.
		}
	}


	\maketitle 

	\makeatletter
	\def\tagform@#1{\maketag@@@{\normalsize(#1)\@@italiccorr}}
	\makeatother
	
	\begin{abstract}
		Densification of network base stations is indispensable to achieve the stringent Quality of Service (QoS) requirements of future mobile networks. However, with a dense deployment of transmitters, interference management becomes an arduous task. To solve this issue, exploring radically new network architectures with intelligent coordination and cooperation capabilities is crucial. This survey paper investigates the emerging user-centric cell-free massive Multiple-input multiple-output (MIMO) network architecture that sets a foundation for future mobile networks. Such networks use a dense deployment of distributed units (DUs) to serve users; the crucial difference from the traditional cellular paradigm is that a specific serving cluster of DUs is defined for each user. This framework provides macro diversity, power efficiency, interference management, and robust connectivity. Most importantly, the user-centric approach eliminates cell edges, thus contributing to uniform coverage and performance for users across the network area. We present here a guide to the key challenges facing the deployment of this network scheme and contemplate the solutions being proposed for the main bottlenecks facing cell-free communications. Specifically, we survey the literature targeting the fronthaul, then we scan the details of the channel estimation required, resource allocation, delay, and scalability issues. Furthermore, we highlight some technologies that can provide a management platform for this scheme such as distributed software-defined network (SDN). Our article serves as a check point that delineates the current status and indicates future directions for this area in a comprehensive manner.
	\end{abstract}

	\IEEEpeerreviewmaketitle
	
\section{Introduction} 
A user-centric cell-free network architecture defines a cooperative serving cluster of spatially distributed transmitters specifically for each user, hence the name user-centric; this in turn, eliminates conventional cell-edges, and is, hence, termed cell-free. Based on this definition, each user in the network is served by its neighboring set of transmitters, so each user is found at the effective center of its serving cluster~\cite{cellFreeUserCentricPower8901451}. The user-centric cell-free scheme can be seen as a practical method to deploy cell-free communications, and the concept is often coupled with Multiple-input multiple-output (MIMO) network concepts, and recently, massive MIMO concepts thus branding the architecture as cell-free massive MIMO~\cite{cellFreeVersusSmallCells7827017, differentCooperationLevels8845768, 8385475, han2019sparse, 8901196, powerControlCellFree7917284, HardwareImpairements8891922, LocaPartialZFBF9069486}. The growing interest in this technology relates to seeing it as one of the drivers for future mobile networks~\cite{9113273}; this interest motivates us to review the developments, current trends and challenges in this topic.

\begin{figure}[t]
	\centering
	\includegraphics[width=0.8\columnwidth]{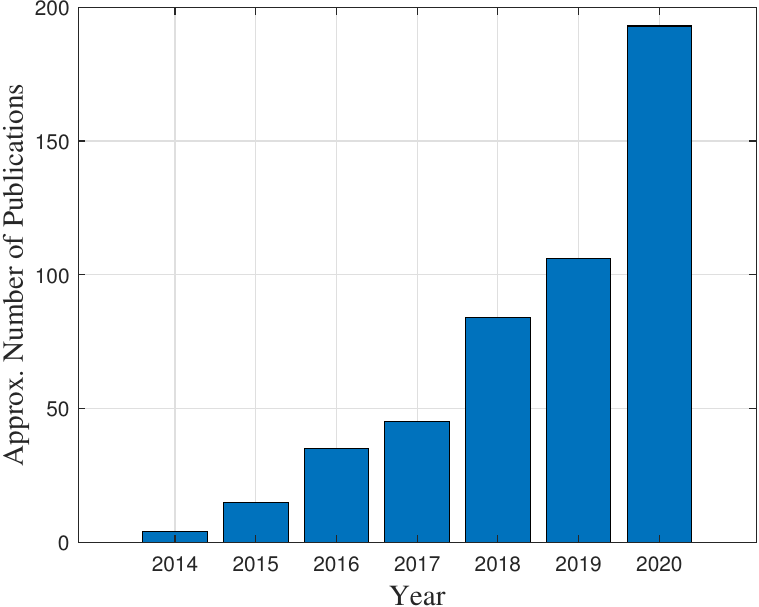}
	\caption{Number of publications targeting the cell-free network architecture. The growing number shows a wide interest in this topic that is a candidate architecture for future networks. Note that the term "cell-free" started to appear in the publications in 2015.}
	\label{fig:NofPublications}
	\vspace{-1em}
\end{figure}

MIMO technology has been one of the most important advancements in wireless communications. It provides the capability for a transmitter to send concurrent signals using the installed multiple antennas, thus allowing for the vectorization of the transmission~\cite{heath2018foundations}. MIMO provides beamforming capabilities where the power of the transmit signals can be focused toward specific users. It also provides antenna diversity where in a nutshell, the same transmitted signal passes through different channels, allowing the receiver to select the best version of the signal, hence enhancing the quality of the communication. Furthermore, it allows for spatial multiplexing, where different data streams can be sent at the same time/frequency resource and recovered successfully. Massive MIMO in its turn greatly increases the number of antennas (more specifically at a base station) which enables aggressive multiplexing and boosting of data rate alongside other benefits~\cite{marzetta2016fundamentals}. One important benefit is channel hardening, where the randomness in wireless channels is eliminated.

\begin{table}[t]
	\centering
	\scriptsize
	\begin{tabular}
		{|p{0.08\textwidth}|p{0.35\textwidth}|}
		\hline
		\hline
		\multicolumn{1}{|l|}{ \textit{\textbf{Acronym}}}& \multicolumn{1}{l|}{ \textit{\textbf{Definition}}} \\
		\hline
		\hline
		5G & Fifth-generation of mobile communication networks
		\\
		\hline
		BBU & Baseband processing unit
		\\
		\hline
		BS & Base station
		\\
		\hline
		CFO & Carrier frequency offset
		\\
		\hline
		CoMP & Coordinated multi-point
		\\
		\hline
		C-RAN & Cloud Radio Access Network
		\\
		\hline
		CSI & Channel state information
		\\
		\hline
		CU & Central unit
		\\
		\hline
		D2D & Device-to-device
		\\
		\hline
		DNN & Deep neural network
		\\
		\hline
		DU & Distributed unit
		\\
		\hline
		EE & Energy efficiency
		\\
		\hline
		eMBB & Enhanced Mobile Broadband
		\\
		\hline
		FSO & Free space optical
		\\
		\hline
		gNB & next-generation NodeB
		\\
		\hline
		GP & Geometric programming
		\\
		\hline
		IAB & Integrated access and backhaul
		\\
		\hline
		IoT & Internet of things
		\\
		\hline
		LoS & Line of Sight
		\\
		\hline
		LSFD & Large-scale fading decoding
		\\
		\hline
		LTE & Long-term evolution
		\\
		\hline
		MEC & Mobile edge computing
		\\
		\hline
		MIMO & Multiple-input multiple-output
		\\
		\hline
		mMTC & Massive machine-type communications
		\\
		\hline
		MRC & Maximum ratio combining
		\\
		\hline
		MSE & Mean squared error
		\\
		\hline
		NOMA & Non-orthogonal multiple-access
		\\
		\hline
		OFDM & Orthogonal frequency division multiplexing
		\\
		\hline
		PA & Pilot assignment
		\\
		\hline
		PDP & Power delay profile
		\\
		\hline
		PTP & IEEE 1588v2 precision time protocol
		\\
		\hline
		QoS & Quality of Service
		\\
		\hline
		QSI & Queue state information
		\\
		\hline
		IRS & Intelligent reflecting surfaces
		\\
		\hline
		RRH & Remote radio head
		\\
		\hline
		SDN & software-defined network
		\\
		\hline
		SE & Spectral efficiency
		\\
		\hline
		SIC & Successive interference cancellation
		\\
		\hline
		SDNR & signal to distortion noise ratio
		\\
		\hline
		SINR & Signal to interference and noise ratio
		\\
		\hline
		SNR & Signal to noise ratio
		\\
		\hline
		SOCP & Second order cone program
		\\
		\hline
		SON & Self-organizing network
		\\
		\hline
		TDD & Time division duplex
		\\
		\hline
		UatF & Use-and-then-forget
		\\
		\hline
		uRLLC & Ultra-reliable low-latency communications
		\\
		\hline
		WSR & weighted sum rate
		\\
		\hline
		ZF & Zero-forcing
		\\
		\hline
		\hline
	\end{tabular}
	\caption{Main acronyms used in paper.}
	\label{table:acronyms}   
	\vspace{-2em}
\end{table}

The terminology of cell-free massive MIMO systems is based on deploying a large number of access points, each serving a low number of users. Explicitly, it represents the distributed alternative for the traditional co-located massive MIMO network of~\cite{marzetta2016fundamentals, massiveMIMO5595728}, so the large number of antennas are distributed in the network instead of being co-located at a single location. The advent of massive MIMO concepts into wireless networks potentially introduces great enhancements in both spectral and energy efficiencies~\cite{massiveMIMO6798744} and is a key technology driving the fifth-generation (5G) of mobile communication networks. However, massive MIMO networks, based on the traditional cellular paradigm, do not provide uniform coverage due to its assumption of a co-located antenna system; this is especially true at the cell-edge and areas in shadow. In this regard, the theory of massive MIMO systems supports spatially distributed array deployments~\cite{massiveMIMODistributed6810508, massiveMIMO7080890, massiveMIMO6736761}, motivating researchers to investigate cell-free massive MIMO communications.
	
One distinction between user-centric cell-free MIMO and cell-free massive MIMO systems is that the former explicitly defines a serving cluster for each user, hence limiting the number of serving transmitters, while the latter assumes that, theoretically, the user can be jointly served by all the transmitters in the network~\cite{cellFreeUserCentricPower8901451}. However, in practice, even the latter implicitly defines a serving cluster because of path loss. Another distinction is the assumption of the applicability of the properties of massive MIMO channels, specifically, the theoretical applicability of favorable propagation and channel hardening, which will be discussed thoroughly later. Thus, it is not clear if the term ``massive'' should be used/dropped when referring to the user-centric cell-free scheme with a limited serving cluster.

A user-centric scheme creates a serving cluster \textit{for each user} comprising the transmitters that can contribute a useful signal~\cite{cellFreeUserCentricPower8901451}. In a nutshell, the serving cluster is constructed separately for each user based on a criterion such as serving distance~\cite{PDPUsercentricVsDisjoint8969384}, network performance~\cite{clustersbasedonNetperf7105966}, or even as a two-stage process where a base cluster is formed for each user based on large-scale fading statistics, then the clusters are optimized using scheduling or some power allocation algorithms on a per time slot basis~\cite{ammarC_RA_UC, ammarDistributed_RA_UC}. Interestingly, the user-centric scheme can be seen as a special case of the cell-free massive MIMO concept under specific power control rules.

In a cell-free scheme, the users are surrounded by serving transmitters in the user's neighborhood, thereby eliminating cell-edges, and, in turn, eliminating the traditional notion of a cell-edge user who usually suffers the worst performance. Therefore, unlike cellular networks, cells have no relevance on the acces channel. This provides a salient improvement~\cite{8768014}, which makes cell-free communication a promising technology for future mobile networks~\cite{EnergyE2020towards}. Succinctly, the key gains from this scheme are:
\begin{itemize} 
	\item Alleviated cell-edge user problem because each user is found at the effective center of its serving cluster,
	\item Enhanced signal strength and connectivity through cooperation,
	\item Suppressed interference through coordination\footnote{Coordination is different from cooperation in terms of the amount of information shared between the transmitters~\cite{DelayAware6180015}.},
	\item Improved energy efficiency due to the relative proximity of the transmitters to the users, and
	\item Augmented reliability realized through marco diversity due to distinct path loss and shadowing from each serving transmitter.
\end{itemize}
Interestingly, using cell-free massive MIMO techniques has shown to provide large improvements in the median and $95\%$-likely spectral efficiency (SE) compared to traditional networks under different scenarios~\cite{cellFreeVersusSmallCells7827017, powerControlCellFree7917284, 8886730, cellFreeStochasticGeom8972478, differentCooperationLevels8845768}. For example, the work in~\cite{cellFreeVersusSmallCells7827017} reports five-fold and ten-fold improvements over a small-cell scheme with uncorrelated and correlated shadow fading, respectively. Moreover, the user-centric cell-free scheme can outperform a cell-free massive MIMO network~\cite{cellFreeUserCentricPower8901451, UserCentricvsCellFreeBackhaul8000355, EnergyEfficiencyMmwave8676377, EECellFreeUCMMwave8292302, EECellFreeUCMMwave8516938}. It is these significant gains  - and growing interest - that motivates our effort in compiling this paper. As Fig.~\ref{fig:NofPublications} shows, academic interest in this area is growing exponentially.

As noted earlier, implementing the user-centric cell-free MIMO framework requires defining serving clusters for users. Serving all the users by all the transmitters in a large region is impractical due to many reasons. The main limitation is the capacity of the individual transmitters; in particular, serving users with distant transmitters occupies precious power and bandwidth resources but, due to high path loss, contributes little useful power at the user. Even with strict power policies that can limit the number of served users, the signaling needed to perform resource allocation would still not be scalable~\cite{cellFreeUserCentricPower8901451, ammarC_RA_UC, ammarDistributed_RA_UC}. In this regard, it is better to define serving clusters as subsets of the available transmitters. Given such an approach, a user-centric cell-free MIMO scheme becomes a promising architecture to implement cell-free MIMO communications.

The cell-free scheme under consideration is a blueprint of extensive research on architectures like massive MIMO~\cite{marzetta2016fundamentals, massiveMIMO5595728}, network MIMO~\cite{NetworkMIMO6095627}, coordinated multi-point with joint transmission (CoMP-JT)~\cite{3GPP:TS36.819}, cloud radio access network (C-RAN)~\cite{CRAN6897914}, multi-cell MIMO cooperative network~\cite{MultiCellMIMO5594708}, virtual MIMO~\cite{VirtualMIMO6601776}, and small-cell networks~\cite{smallCell6525591}. In these systems, the transmitters are called access points in the cell-free massive MIMO architecture~\cite{cellFreeVersusSmallCells7827017}, remote radio heads (RRHs) in distributed antenna systems and C-RAN~\cite{CRAN6897914}, and base stations (BSs) in small cells. Herein, we use the terms access points and distributed units (DUs) interchangeably. 

The concept of cell-free MIMO communication has a lot in common with each of the aforementioned architectures with slight differences that mainly target eliminating the cell-edges and minimizing the fronthaul traffic. For example, the network MIMO concept is based on preserving the cellular structure, where many access points share user data and channel state information (CSI) to jointly encode (in downlink or DL) and decode (in uplink or UL) the signals for the users. On the other hand, in a cell-free massive MIMO system, CSI can be acquired and processed locally at each transmitter~\cite{cellFreeVersusSmallCells7827017}, which is more scalable than network MIMO, leading to a reduced exchange of CSI.

\begin{figure}[t]
	\centering
	\begin{subfigure}{0.48\textwidth}
		\centering
		\includegraphics[width=1\linewidth]{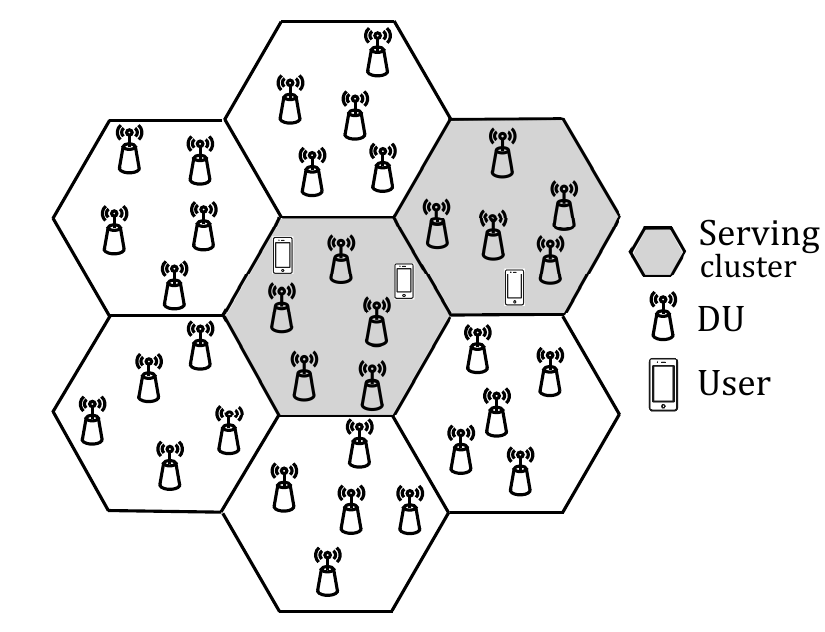}
		\caption{Cell-centric scheme.}
		\label{fig:ReviewCellCentric}
	\end{subfigure}
	\\
	\begin{subfigure}{0.48\textwidth}
		\centering
		\includegraphics[width=1\linewidth]{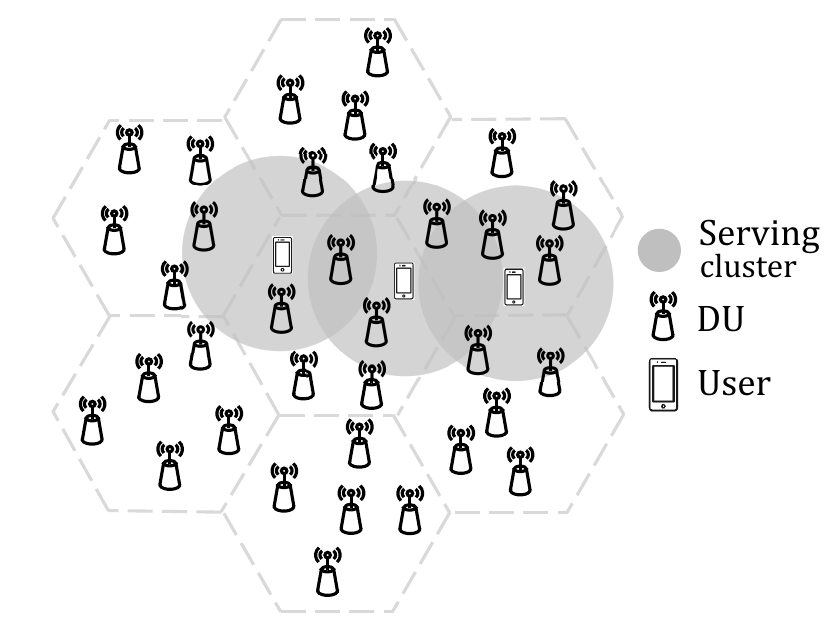}
		\caption{User-centric cell-free scheme; cells \emph{may be} applicable, \textit{but only on fronthaul}.}
		\label{fig:ReviewUserCentric}
	\end{subfigure}
	\caption{Cell-centric vs user-centric cell-free.}
	\label{fig:cellCentricVsUsercentric}
	\vspace{-1em}
\end{figure}

C-RAN is a centralized cloud-computing based network architecture that provides collaborative support and virtualization capabilities, and it shares a network architecture similar to cell-free MIMO systems, with differences in clustering of users and the centralization of some network functionalities. The downside of C-RAN is the requirement for high bandwidth fronthaul due to the use of a centralized base band unit (BBU)~\cite{CRANfronthaul8113473}. Finally, CoMP is the concept that was standardized by 3GPP in long-term evolution (LTE) release 11~\cite{3GPP:TS36.819}; it defines the modes for cooperation amongst network transmitters. 

In a cell-free, distributed network, the access points are controlled by a single or a set of central/control units (CUs), which are called different names in the literature, e.g., central processing unit~\cite{cellFreeVersusSmallCells7827017}, BBU~\cite{CRAN6897914}, C-RAN data center~\cite{perlman2015introduction}, and edge-cloud processor~\cite{burr2018ultra}. Furthermore, these CUs can have different degrees of centralization and capabilities. Our usage of the terms ``DU'' and ``CU'' is motivated by the New Radio interface standard for 5G mobile networks, which defines a distributed architecture for the next-generation NodeB (gNB) formed from a central unit (gNB-CU or simply CU) connected to distributed units (gNB-DUs or simply DUs) through the F1 interface~\cite{3GPPTR21.915}. The current standard allows for the implementation of a cell-centric scheme (also called disjoint clustering~\cite{StochasticUserCentric8449213, PDPUsercentricVsDisjoint8969384}), where the gNB-DUs under the control of each gNB-CU can serve the users within their geographical cell.

\begin{table*}[b]
	\scriptsize
	\centering
	\begin{tabular}
		{|p{0.035\textwidth}|p{0.025\textwidth}|p{0.2\textwidth}|p{0.05\textwidth}|p{0.058\textwidth}|p{0.48\textwidth}|}
		\hline
		\hline
		& \multicolumn{1}{l|}{ \textit{\textbf{Year}}}& \multicolumn{1}{l|}{ \textit{\textbf{Area}}} & \multicolumn{1}{l|}{ \textit{\textbf{Cell-free}}} & \textit{\textbf{Surveyed references}} & \multicolumn{1}{l|}{ \textit{\textbf{Main Topics}}} \\
		\hline
		\hline
		\rowcolor{anti-flashwhite}
		\!\!\!\textbf{Current work} &  --- & User-centric cell-free MIMO & \checkmark & \total{citenum} & Provide survey about the challenges, solutions and opportunities that can be provided by the user-centric cell-free massive MIMO network architecture.
		\\
		\hline
		\cite{9120231} & 2020 & Energy efficiency in C-RAN & \ding{55} & 222 & Focus on energy efficient resource allocation, softwarization and autonomous management technologies, opportunities, relevant standardization activities, open challenges, and future directions.
		\\
		\hline
		\cite{8768014} & 2019 & Cell-free massive MIMO & \checkmark & 57 & Exploit channel hardening and favorable propagation, and analyze open research challenges for network deployment and management.
		\\
		\hline
		\cite{CRANfronthaul8113473} & 2018 & C-RAN optical fronthaul & \ding{55} & 381 & Target cellular radio access network (RAN) architecture, passive optical network (PON) architectures, fiber-based common public radio interface (CPRI), variants of radio over Ethernet transport schemes, energy harvesting, power over fiber, and SON.
		\\
		\hline
		\cite{7839266} & 2017 & Clusters in CoMP & \checkmark$\!\!\!\backslash$ 
		& 128 & Browse strengths and weaknesses of the available clustering solutions which are classified as static, semi-dynamic and dynamic clustering.
		\\
		\hline
		\cite{CRAN6897914} & 2015 & C-RAN & \ding{55} & 122 & Span through the state-of-the-art literature, aspects, challenges, solutions, and possible deployment scenarios.
		\\
		\hline
		\cite{CRAN7018201} & 2015 & C-RAN & \ding{55} & 13 & Concentrate on three-layer logical structure composed of physical plane, control plane, and service plane.
		\\
		\hline
		\cite{RA_CRAN7143328} & 2015 & Resource allocation in C-RAN & \ding{55} & 15 & Briefly scan the challenges and advances in C-RAN architecture.
		\\
		\hline
		\cite{backhaulSmallCells7306536} & 2015 & Back-haul for small cells & \ding{55} & 15 & Briefly skim the RAN architecture, heterogeneous backhaul network architecture, and resource management.
		\\
		\hline
		\cite{wirelessBackhaulSmallCells7306534} & 2015 & Wireless backhaul for small cells & \ding{55} & 18 & Briefly survey flexible wireless backhaul, backhaul delay management solutions, interference management, millimeter wave backhauling, and signaling overhead.
		\\
		\hline
		\hline
	\end{tabular}
	\caption{Timeline for survey papers about distributed MIMO architectures; \checkmark, \ding{55}, or \checkmark$\!\!\!\backslash$ indicates that the topic is addressed, not addressed, or partially addressed, respectively.}
	\label{table:survey}   
\end{table*}

Fig.~\ref{fig:cellCentricVsUsercentric} illustrates the differences between a cell-centric and user-centric scheme. We emphasize that the CUs form \textit{virtual cells} only (more on this later). Unfortunately, a cell-centric approach does not provide substantial gains, because the inter-cluster interference scales with the intra-cluster signals~\cite{LimitsOfCooperation6482234}. Further, this does not solve the problem of weak signals received by cell-edge users~\cite{StochasticUserCentric8449213}. Another drawback is that cell-centric clustering produces a higher signal delay spread than the user-centric scheme~\cite{PDPUsercentricVsDisjoint8969384}. Nonetheless, although the current standard adopts the cell-centric approach, it still separates the control and user-planes making the cell-free or the user-centric scheme theoretically feasible~\cite{interdonato2019ubiquitous}. As we will explore throughout this exposition, there are also other mature technologies that can help in implementing the user-centric cell-free architecture.

\subsection{Scope, Contributions and Related Work}
The capabilities discussed in this survey paper are those concerned with implementing the cell-free architecture, which completely eliminates the cell-edges, but does not necessarily implement the CU at the network core. Our review serves as a checkpoint that articulates the challenges, methods and solutions in this area. We inspect, in detail, the state-of-the-art in this architecture and the opportunities that it can provide for future mobile networks. We also explore questions about future research directions, and we contemplate some of the challenges facing deployment. We also investigate how the user-centric cell-free MIMO architecture can benefit from other technologies like millimeter wave communication and software-defined networks (SDNs). In this context, future mobile networks need rethinking from the system and architecture levels down to the development of the physical layer, and the user-centric cell-free MIMO network can be an avenue for a revolutionized architecture.
	
\begin{figure*}[t]
	\centering
	\includegraphics[width=1\linewidth]{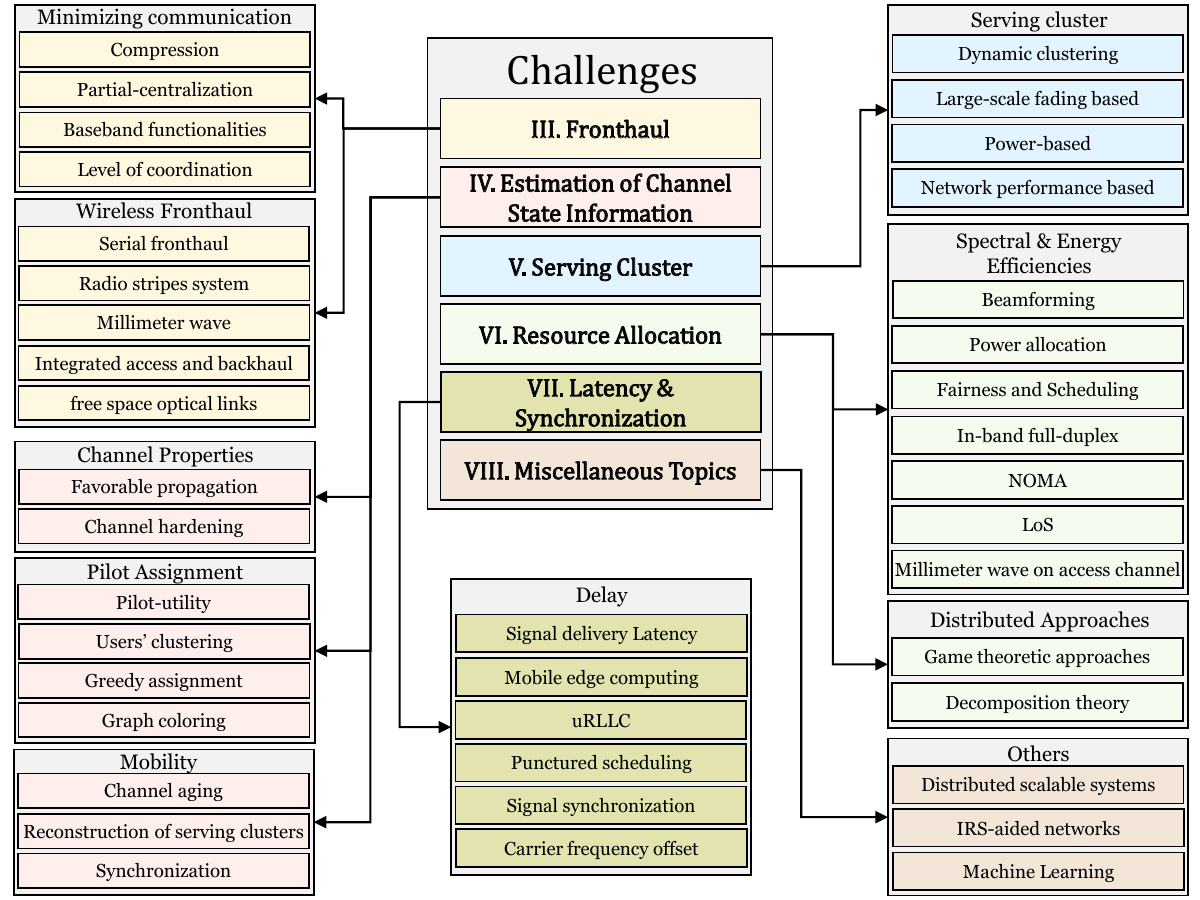}
	\caption{Main challenges surveyed in paper.}
	\label{fig:ReviewTopics}
	\vspace{-1em}
\end{figure*}
	
The contributions of this paper are summarized as follows:
\begin{enumerate}
	\item We compile a wide range of recent studies in the literature about the cell-free MIMO network scheme, with a focus on user-centric clustering, hence, providing a starting point for readers to launch further investigations.
	
	\item We identify the main physical layer challenges facing the deployment of cell-free communications. The challenges are summarized under the following titles: fronthaul capacity, CSI estimation, formation of serving cluster, resource allocation, delay, and miscellaneous topics such as scalability. Additionally, we discuss the different innovative solutions for these challenges.
	
	\item We present the important findings of, and insights into, the cell-free scheme. In this regard, we provide tables that summarize the challenges and the corresponding solutions being proposed. We also highlight the topics that are still not well-studied, hence indicating future research directions and open issues. In general, the literature shows gaps in proposing distributed resource allocation schemes that set scalability as a priority. Further, the signal delay is a weakly studied metric in the literature despite its importance for ultra-reliable low-latency communications (uRLLC). Moreover, the effect of mobility on the user-centric scheme is still vague, especially how mobility affects the re-formation of serving clusters. 
	
	More investigations for procedures and protocols to make the user-centric cell-free scheme feasible in a network with multiple CUs are still needed. We note that, in this case, the user can be served by DUs belonging to different CUs, which may not be a straightforward task. Such protocols may have to integrate autonomous management technologies like distributed SDNs to achieve a flexible and programmable network. Furthermore, we highlight the importance of some newly developed technologies like the radio stripes system.

\end{enumerate}

\begin{figure*}[t]
	\centering
	\begin{subfigure}{0.49\textwidth}
		\centering
		\includegraphics[width=1\textwidth]{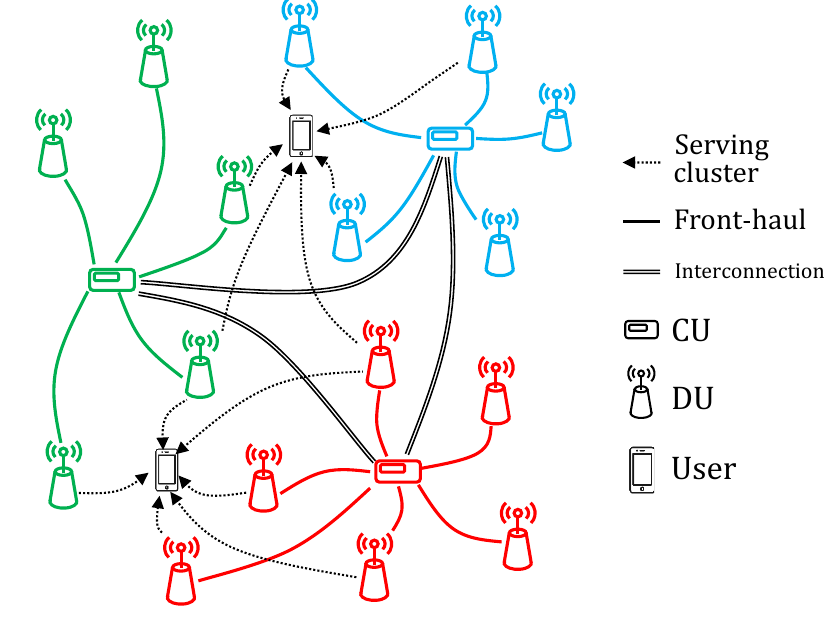}
		\captionof{figure}{With multiple CUs.}
		\label{fig:Review_CoordinatedNet_multipleCUs}
	\end{subfigure}%
	\quad
	\begin{subfigure}{0.49\textwidth}
		\centering
		\includegraphics[width=1\textwidth]{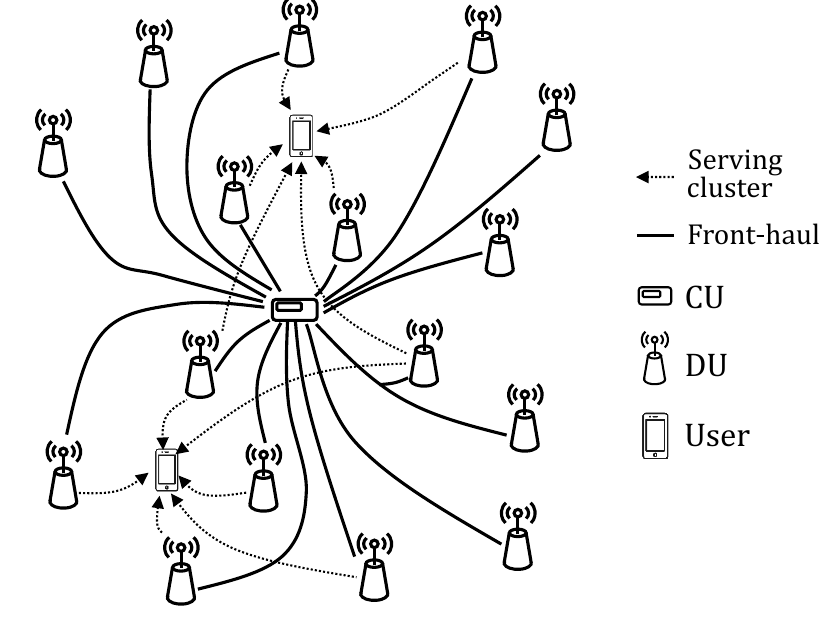}
		\captionof{figure}{With a single CU.}
		\label{fig:Review_CoordinatedNet_singleCU}
	\end{subfigure}%
	\vspace{-0.5em}
	\caption{A User-centric cell-free massive MIMO network.}
	\label{fig:Review_CoordinatedNet}
\end{figure*}

The most related survey papers to this review are those targeting coordinated distributed networks and clustering schemes. In Table~\ref{table:survey}, we summarize these papers. For the C-RAN case, the review in~\cite{CRAN6897914} addresses the advances in the C-RAN network architecture, \cite{CRANfronthaul8113473} addresses the possible C-RAN optical fronthaul technologies. Both the studies in~\cite{backhaulSmallCells7306536, wirelessBackhaulSmallCells7306534} provide a very brief review for the heterogeneous wired and wireless backhaul network architectures respectively for C-RAN. Topics like heterogeneous backhaul, in-band full-duplex, and anchor BSs are discussed, and their features are highlighted. Similarly, the survey in~\cite{9120231} handles energy efficient resource allocation in C-RAN, \cite{CRAN7018201} focuses on the physical, control, and service plane of the C-RAN, while~\cite{RA_CRAN7143328} focuses on the resource allocation schemes. 

The difference between these surveys and our paper is that they target different network architectures that are still based on the cellular structure on the access channel, which make them fundamentally different. Hence, they differ in terms of system assumptions, specifically, in terms of different definitions for the serving cluster, the existence of cell boundaries, and the CSI acquisition and sharing.

The survey paper in~\cite{7839266} targets the CoMP scheme, and it focuses on the different clustering schemes and their effect on the energy and spectral efficiencies. However, the paper is not concerned in providing cell-free communications. The investigation in~\cite{8768014} is very important as it targets cell-free massive MIMO networks. However, it does not tackle the user-centric clustering and it is not an extensive survey, where the authors focus on exploiting channel hardening and favorable propagation (more on this later). Thus, they do not focus on the topics explored herein.

In summary, the findings investigated in this survey target the user-centric cell-free MIMO architecture, and they are comprehensive compared to the surveys in the literature.
	
\subsection{Outline}
The challenges presented in this survey target critical topics. Section~\ref{sec:systemModel} briefly presents the cell-free network architecture. Section~\ref{sec:fronthaul} investigates the main bottleneck for the deployment of cell-free MIMO networks, which is the fronthaul. We analyze the solutions being studied in the literature, and we highlight some promising technologies to overcome this bottleneck. Section~\ref{sec:CSI} examines the problem of estimating the CSI, where we investigate the proposed pilot assignment (PA) policies. We discuss two important channel properties, namely channel hardening and favorable propagation, and we review what the cell-free MIMO literature states about their applicability, furthermore, we examine the topics related to mobility in cell-free networks.

\newcommand{\varClassif}{The challenges in the figure are chosen based on their impact on the performance of user-centric cell-free communications. Moreover, the topics within the challenges are based on the literature and are categorized into the general titles that can describe them.}
	
In Section~\ref{sec:servingCluster}, we investigate the feature that distinguishes user-centric cell-free MIMO from a general cell-free massive MIMO scheme, which is the formation of serving clusters. In Section~\ref{sec:resourceAllocation}, we delve deeply into the extensively studied topic of resource allocation, where we examine the optimization of different Quality of Service (QoS) requirements for future networks. In Section~\ref{sec:latsynch}, we consider topics such as latency and signal synchronization. The latency metric is less-frequently studied. This is particularly interesting, since delay is a priority for many applications, but there has been little focus on it in the user-centric cell-free MIMO literature. In Section~\ref{sec:MiscellaneousTopics}, we consider miscellaneous topics such as scalability, intelligent reflecting surfaces (IRSs)-aided networks, and the role of machine learning. We note that despite its importance, scalability of the solutions targeting the cell-free MIMO architecture is the focus of only few studies. In Section~\ref{sec:openResearchProb}, we summarize some open-research problems. Finally, in Section~\ref{sec:conclusionFutureDirections}, we compile a conclusion that states the important findings, key elements and possible future directions. Additionally, throughout the paper we present some tables that summarize some topics for easy reference.

In Fig.~\ref{fig:ReviewTopics}, we summarize the areas of the challenges discussed in this paper. \label{page:Classif}\varClassif

\begin{table*}[t]
	\footnotesize
	\centering
	\begin{tabular}
		{|p{0.19\textwidth}|p{0.1\textwidth}|p{0.12\textwidth}|p{0.16\textwidth}|p{0.14\textwidth}|p{0.12\textwidth}|}
		\hline
		\hline
		\textit{\textbf{Metric}} &  \textit{\textbf{Massive MIMO}} & \textit{\textbf{Network MIMO}} & \textit{\textbf{Cell-free massive MIMO}} &  \multicolumn{2}{c|}{ \textit{\textbf{User-centric cell-free [massive] MIMO}} } \\
		\cline{5-6}
		& & & & \textit{\textbf{Multiple CUs}} & \textit{\textbf{Single CU}}
		\\
		\hline
		Number of serving antennas & \cellcolor{anti-flashwhite} very large & \cellcolor{anti-flashwhite} medium & \cellcolor{anti-flashwhite} very large/large & large & large
		\\
		\hline
		Cost efficiency & \cellcolor{anti-flashwhite} low & \cellcolor{anti-flashwhite} low & \cellcolor{anti-flashwhite} high & moderate & high
		\\
		\hline
		Macro diversity & \cellcolor{anti-flashwhite} small & \cellcolor{anti-flashwhite} medium &  \cellcolor{anti-flashwhite}large & large & large
		\\
		\hline
		Favorable propagation* & \cellcolor{anti-flashwhite} strong & \cellcolor{anti-flashwhite} weak & \cellcolor{anti-flashwhite} moderate & moderate & moderate
		\\
		\hline
		Channel Hardening* & \cellcolor{anti-flashwhite} strong & \cellcolor{anti-flashwhite} weak & \cellcolor{anti-flashwhite} moderate & moderate & moderate
		\\
		\hline
		Uniform coverage & \cellcolor{anti-flashwhite} low & \cellcolor{anti-flashwhite} medium & \cellcolor{anti-flashwhite} high & high & high
		\\
		\hline
		Energy efficiency & \cellcolor{anti-flashwhite} large & \cellcolor{anti-flashwhite} small & \cellcolor{anti-flashwhite} very large & large & very large
		\\
		\hline
		CSI estimation & \cellcolor{anti-flashwhite} global & \cellcolor{anti-flashwhite} global & \cellcolor{anti-flashwhite} local/global & local (two degrees)/ global & local/global
		\\
		\hline
		Fronthaul capacity efficiency & \cellcolor{anti-flashwhite} high & \cellcolor{anti-flashwhite} low & \cellcolor{anti-flashwhite} medium & medium & medium
		\\
		\hline
		Computation scalability & small & medium & small/medium & small/medium/large & small/medium
		\\
		\hline
		Serving cluster & co-located & cell-centric & implicit user-centric & user-centric & user-centric
		\\
		\hline
		\hline
		\multicolumn{6}{l}{*Discussed in detail later.}
	\end{tabular}
	\caption{Qualitative comparison for different schemes. Shaded cells of this table are taken from~\cite{8768014}, however, we further add the user-centric cell-free MIMO scheme to this comparison, we also add the metrics on scalability and serving clusters.}
	\label{table:differentSchemes}   
\end{table*}

\section{Typical System Model}\label{sec:systemModel}
In Fig.~\ref{fig:Review_CoordinatedNet}, we show two typical network architectures for the user-centric cell-free MIMO network. The construction of the serving clusters is critical to minimize the load and signaling on each DU; this approach is what distinguishes the user-centric cell-free scheme from other approaches in the literature. We note that the serving cluster can be constructed based on many metrics, which will be discussed later. These metrics can be even determined ``offline'', e.g., distance-based or channel power-based. The assumption of a massive MIMO requires having number of serving antennas much greater than the number of users served on the same resources, however, this assumption may not be always adopted. As seen in the figure, the concept of cells on the access channel is not useful anymore, where the user will be surrounded by the serving DUs. However, \emph{on the fronthaul}, the concept of cells, \emph{virtual} cell identifiers, tracking and pool areas can still exist especially in a network with multiple CUs (Fig.~\ref{fig:Review_CoordinatedNet}(\subref{fig:Review_CoordinatedNet_multipleCUs})).
	
The network that adopts multiple CUs, shown in Fig.~\ref{fig:Review_CoordinatedNet}(\subref{fig:Review_CoordinatedNet_multipleCUs}), is more scalable than the one that has a single CU~\cite{8761828, bjornson2019scalable9064545}, as shown in Fig.~\ref{fig:Review_CoordinatedNet}(\subref{fig:Review_CoordinatedNet_singleCU}), because the multiple-CU architecture provides a hierarchical design with relatively low fronthaul traffic flows~\cite{mmWaveHierarchicalBackhaul7904705}. Using multiple CUs can also provide lower delay compared to a centralized CU, because the base-band signals can be collected at local CUs rather than at the core network. Moreover, such a scheme provides a smooth transition from the current cellular structure, because as noted earlier, the cells are still employed on the backhaul of the CUs.

\label{page:oldNFV}The multiple-CU architecture can be implemented using technologies such as distributed SDN~\cite{distributedSDN8187644, distributedSDN6838330} to allow for better coordination and dynamic assignment of the DUs to the CUs. This can help avoid a single point of failure and allow load balancing among the CUs~\cite{DynamicMapping8108113}. An interesting aspect in the architecture of multiple CUs is the possibility of interconnecting the CUs with serial backhaul connections. In this regard, with the help of user-centric clustering, the data for each user needs to be found only in its neighboring CUs, i.e., only neighboring CUs need be interconnected thereby decreasing the number of connections required. Alternatively, the backhaul of the CU can still be connected in a conventional way, i.e., connected directly from each CU to the core network.
	
A network architecture with a single CU, such as shown in Fig.~\ref{fig:Review_CoordinatedNet}(\subref{fig:Review_CoordinatedNet_singleCU}), is another possible deployment. This architecture is adopted in most studies about cell-free massive MIMO networks~\cite{9113273}. Unless the DUs will be processing their signals locally, such a deployment is demanding in terms of its fronthaul capacity, and most probably it should include serial fronthaul links between the DUs. Interestingly, currently there are some technologies that may allow the deployment of such a scheme, e.g., the radio stripes system~\cite{frenger2019antenna, radioStripsSystem}, which integrates the DUs and their wiring into an easy-to-deploy single adhesive tapes. We note also that despite some studies, like~\cite{differentCooperationLevels8845768}, that use a single CU per network, they still state that multiple CUs will be needed in a real deployment.

In Table~\ref{table:differentSchemes}, we summarize some of the qualitative differences between massive MIMO, network MIMO, cell-free massive MIMO and user-centric cell-free [massive] MIMO schemes. The comparison is mainly adopted from~\cite{8768014}, however, we add the user-centric cell-free [massive] MIMO scheme to the comparison with some additional metrics like the scalability and serving clusters metrics. The added metrics and comparison are based on~\cite{cellFreeUserCentricPower8901451, differentCooperationLevels8845768, EnergyEfficiency8097026, UC_CellFreOpticalFronthaule020, 9130689, FullDuplex9110914, HardwareImpairements9004558}. We also note the following: in term of cost efficiency, a multiple-CU network may require higher deployment cost than a single-CU network. For the CSI estimation, in a multiple-CU network, we can have a local DU estimation or a local CU estimation, thus we have two degrees of local CSI estimation. We also emphasize that the gain in performance afforded by a cell-free network over other architectures is highly dependent on the assumptions and the scenarios considered.

\subsection{Transmission Mode}
User-centric cell-free transmissions can inherit the transmission modes that were defined in the CoMP standard~\cite{3GPP:TS36.819}. Specifically, the network can operate in joint processing (JP) mode which is further divided into joint transmission (JT) and dynamic point selection (DPS) modes. We also have the coordinated scheduling/beamforming (CS/CB) mode. In JT, the data intended to the user is shared among the cooperating DUs and simultaneously transmitted in the network. On the other hand, in DPS, the user data is available at multiple DUs in the serving cluster which coordinate each other, but the data is transmitted from one DU (called the transmission point (TP)) at a time. Interestingly, the TP can change from a subframe to another depending on the instantaneous channel condition~\cite{3GPP:TS36.819}.

The JT and the DPS modes can be combined together to allow multiple DUs selected from the serving cluster to transmit the data. As for the CS/CB scheme (also called semi-static point selection), the data is transmitted from one DU, but user scheduling and beamforming decisions are taken with coordination between the DUs within the serving cluster. In this scheme, the data for the user needs to be available at one transmitter, hence reducing the demand on network resources, especially on the fronthaul. Thus, when the benefits of JT are outweighed by the overhead, the CS/CB mode can be used~\cite{clusters6415394}.

In the JT mode, we have two different transmission modes which achieve different data rates at the user. The first mode is coherent transmission, where the serving DUs coherently precode and send the same data symbol for each user, and hence act as a unique antenna array system. Nominally, such an approach requires phase-synchronization among the DUs~\cite{NoncoherentCRAN8482453}, which could be a challenge, especially when the DUs serving the user are under the control of different CUs. However, the use of cyclic prefix can relax this requirement, and the DUs can be assumed quasi-synchronized within a $1~{\rm km}$ or $5~{\rm km}$ radius for the LTE normal and extended cyclic prefix respectively~\cite{differentCooperationLevels8845768}. In this regard, the user-centric cell-free approach provides a significant advantage over the cell-centric scheme, because it provides a smaller signal delay spread~\cite{PDPUsercentricVsDisjoint8969384} due to centering the user at its serving cluster, hence leading to smaller serving distance variability compared to a conventional cell-edge user in a cell-centric scheme. Further, the scheme allows the user to perform coherent combining.

The second mode is non-coherent transmission, where the DUs transmit different data streams to the user, and each transmitter precodes the data independently allowing only a power gain at the receiver~\cite{3GPPTR21.915}. The data rate in such case is the summation of the individual data rates of the DUs. Generally, this mode provides smaller data rates than the coherent mode~\cite{ammarC_RA_UC} because the user requires successive interference cancellation (SIC) to decode the data streams from the serving DUs; this makes the effective received signal power to be the sum of the signal powers received from the DUs~\cite{NoncoherentCRAN8482453}. Notably, the strict phase-synchronization requirement is not necessary in such mode, making the implementation easier on the transmitter side but at the expense of a far more complex receiver that implements SIC. On the other hand, since the serving cluster size is generally small, the computational complexity of the SIC at the user may be affordable.

As expected, each of the coherent/non-coherent modes produce different expressions for the achievable data rate. Furthermore, each mode has different capacity requirements and usage model on the fronthaul. The literature of cell-free massive MIMO largely focuses on the coherent joint transmission and reception~\cite{cellFreeVersusSmallCells7827017, differentCooperationLevels8845768, 8385475, han2019sparse, 8901196, powerControlCellFree7917284, HardwareImpairements8891922, LocaPartialZFBF9069486} seen as the most probable transmission mode to be deployed, though our recent work has investigated non-coherent transmission~\cite{ammarC_RA_UC_conf, ammarC_RA_UC}.

\section{Fronthaul Issues}\label{sec:fronthaul}
The fronthaul refers to the link between a CU and a DU. The DUs need to exchange substantial signaling information with the CU, which varies depending on how much cooperation/coordination is needed. The shared messages can include CSI, beamforming vectors, power control coefficients, scheduling decisions, mobility management functions and other variables used to operate the network. The CU itself needs to at least communicate the users' data to and from the DUs. The most demanding coordination is the JT scheme where the users' data needs to be available at each DU in the serving cluster for the user. In this regard, the limited capacity of the fronthaul is a significant challenge in implementing cooperation among the DUs; this problem is especially evident when the fronthaul solution is wireless~\cite{ammarWirelessbackahulJournal}. On the bright side, user-centric clustering imposes a lower burden on the fronthaul compared to other forms of cell-free massive MIMO systems~\cite{UserCentricvsCellFreeBackhaul8000355}. This is because the serving clusters are limited and can be even optimized based on any needed metric, e.g., the fronthaul capacity.

Although the network needs to implement user-centric clustering on the access channel, the DUs must be clustered into \textit{virtual cells} with DUs within a virtual cell controlled by a single CU. Essentially, in a network with multiple CUs, a cell-like structure can still be implemented on the fronthaul side. This raises a challenge in determining how these CUs should be assigned to control the DUs, and how cooperation will be implemented. This topic is a good future research direction focused on the fronthaul.

\label{page:SDNdiscussion}
One useful direction is to use distributed SDN~\cite{distributedSDN8187644, distributedSDN6838330} as a management framework for the CUs. An SDN has a three-layer architecture composed of application, control and infrastructure (or data) layers. It uses software implementations for complex networking applications and configuration. In computer networks, SDN can easily control the network behavior, e.g., routing of data packets, through installing flow tables on the switches found in the network. What is interesting is that the SDN controllers do not need to know the details of the managing software of these switches. This turns out to be very useful to control and interconnect a large number of devices~\cite{distributedSDN8187644}.

\begin{figure}[t]
	\centering
	\includegraphics[width=1\columnwidth]{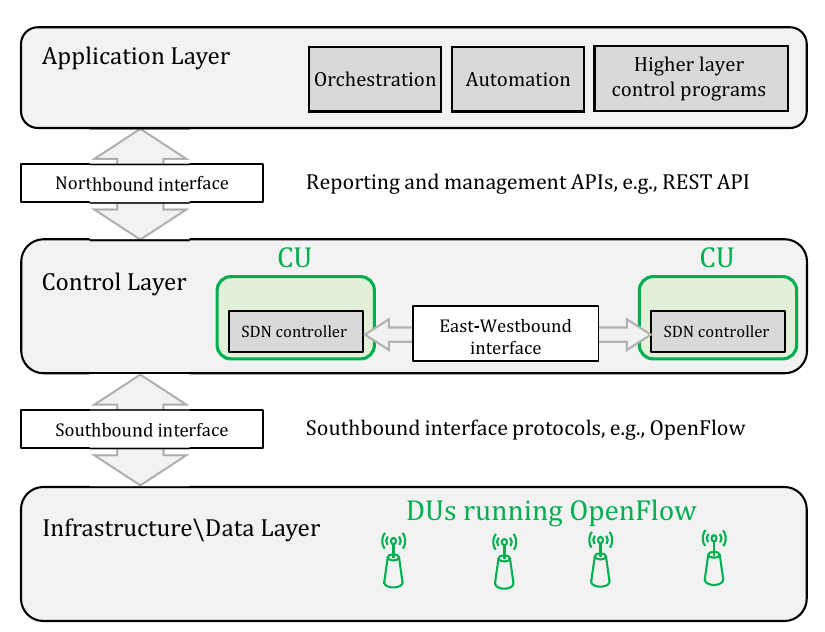}
	\caption{The three-layer communication stack of SDN~\cite{distributedSDN8187644}. We additionally assume that the SDN controllers are installed on the CUs and act as a single logical SDN controller, while the DUs run OpenFlow protocol to communicate with this logical controller.}
	\label{fig:SDN_layers}
	\vspace{-1em}
\end{figure}

An SDN implements its functionality through a set of protocols. An example of these protocols is the OpenFlow protocol~\cite{mckeown2008openflow} which allows remote administration and operates on the SDN southbound interface. In Fig.~\ref{fig:SDN_layers}, we depict the protocol architecture of the SDN. The southbound interface is the interface between the control and the infrastructure layers. It provides communication and management between the SDN controllers and the network nodes, and it pushes the instructions to the controlled nodes. In computer networks, these nodes are usually switches. On the other hand, the northbound interface provides application programming interfaces (APIs), management and reporting functionalities to the upper layer applications. In distributed SDN, we also have what can be called eastbound or westbound interfaces between the set of controllers.

Distributed SDN refers to physically distributed controllers that can be logically centralized. The importance of distributed SDN is that it avoids potential bottlenecks and single point failures. For example, if one controller fails, another controller can automatically take control of its part of the network. This is usually done through a backup master-slave configuration.

As shown in Fig.~\ref{fig:SDN_layers}, SDN can be integrated into cell-free massive MIMO systems by installing SDN controllers, such as the Floodlight controller~\cite{floodlightController}, on the CUs, and allowing the DUs to run an SDN southbound interface protocol, such as the OpenFlow protocol. This ``software upgrade'' is feasible because today's mobile networks are already packet-switched networks. The benefits of using SDN in cell-free networks are numerous, for example, using SDN allows for a dynamic assignment between the CUs and the DUs. As a consequence, the DUs can be automatically migrated from overloaded CUs to under-loaded ones. This can be a good strategy to minimize the traffic on some overloaded fronthauls. Interestingly, such an approach can even handle CU failure scenarios~\cite{DynamicMapping8108113}. Another important use is making the network programmable by allowing the CUs to use the protocols provided by SDN to easily push commands, notifications, and other control plane data to the DUs.

Distributed SDN has been widely studied in the literature of computer networks~\cite{distributedSDN8187644}, however, no one has proposed using it in cell-free networks. So, this is an open research area.

\begin{figure*}[t]
	\centering
	\includegraphics[width=1\linewidth]{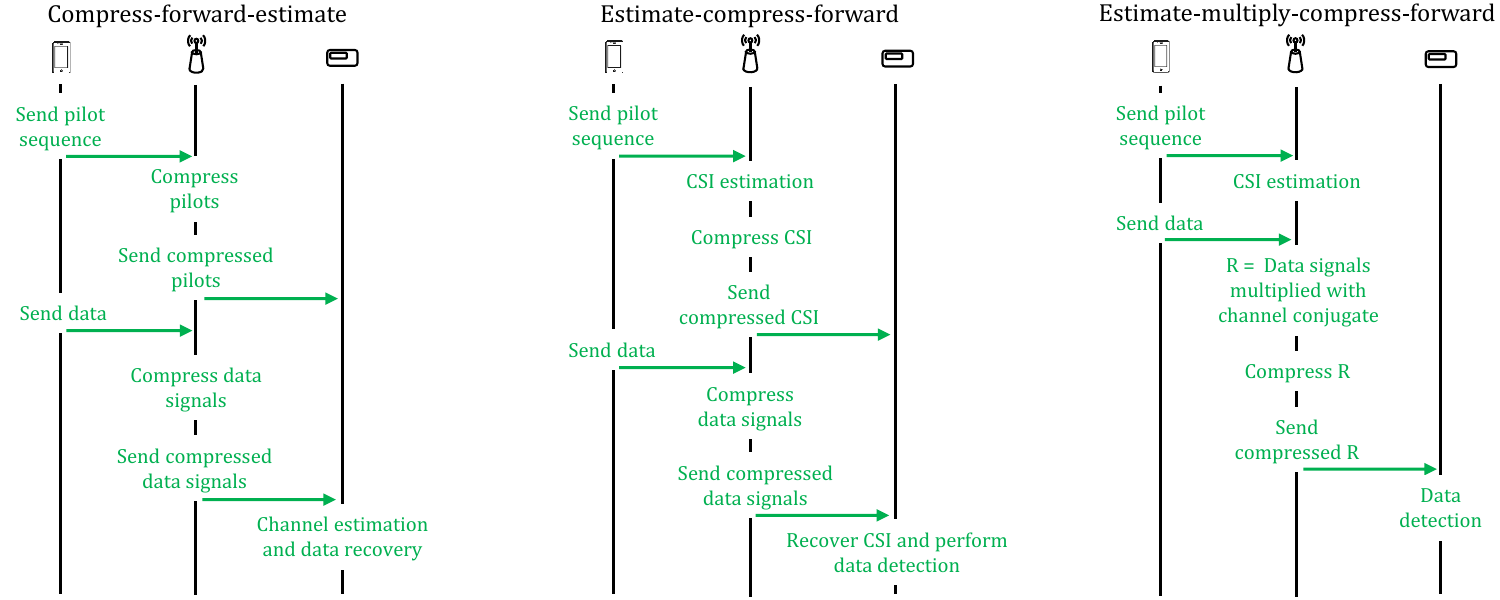}
	\caption{Three different scenarios of signals compression for uplink transmissions on fronthaul~\cite{HardwareImpairements8891922, maxMinRate8756286, DeepLearningFronthaul9110901}.}
	\label{fig:ReviewCompressScenarios}
\end{figure*}

\subsection{Minimizing Fronthaul Communication}\label{sec:MinfronthaulCom}
One way of limiting fronthaul signaling is through source coding schemes and fronthaul data compression~\cite{8693830}. Dedicated but unreliable fronthaul links have attracted the interest of many researchers, where the reliability of the fronthaul in a single cluster cooperative network has been modeled as Bernoulli~\cite{7120102,7981320}. With data compression, only a quantized version of the data and/or channel estimates are available at the CU~\cite{limitedFronthaulMmwave8678745}.
	
The work in~\cite{maxMinRate8756286} studies the effect of different quantization scenarios on the fronthaul of a centralized CU. The first case is composed of sending quantized versions of the CSI and signals, while the second scheme sends only a quantized weighted signal version. To derive the achievable rate, the authors use the use-and-then-forget (UatF) bounding technique~\cite{marzetta2016fundamentals, cellFreeVersusSmallCells7827017}, which is commonly used in massive MIMO networks and results in a tight and simple expression for the rate. The results show that the first case provides a slightly better uplink rate than the second~\cite{maxMinRate8756286}; the rate depends on the number of antennas per DU and number of users, and this difference in performance decreases as the number of antennas increases. Moreover, under the network configuration used in~\cite{maxMinRate8756286}, the results show that setting the number of quantization bits to be greater than $7$ results in performance very close to that of a perfect fronthaul.

Bussgang decomposition~\cite{5560739, bussgang1952crosscorrelation} is used to model the effect of quantization, where the nonlinear output of the quantizer is represented as a linear function that depends on the power of the quantizer input, a midrise uniform quantizer function~\cite{8730536} and some distortion noise. We note that the quantization step size is defined so that the signal to distortion noise ratio (SDNR) at the output of the quantizer is maximized~\cite{8730536}. Similarly, the study in~\cite{EnergyEfficiency8781848} uses Bussgang decomposition to model the effect of quantization.
	
The authors of~\cite{EE9212395} investigate the uplink of cell-free massive MIMO networks under different scenarios that depend on where channel estimation is performed and what information is sent back to the CU. We note that any signal to be sent to the CU needs to be quantized. Results show that quantizing the received signals and pilots then sending them to the CU to perform estimation (denoted as Quantize\&Estimate) is only marginally better in terms of performance than the case of estimating the channels at the access points and then quantizing the estimated channel and signal, and sending them to the CU (denoted as  Estimate\&Quantize). The authors study the effect of the step size of the quantizer under different scenarios. The results presented show that an appropriate choice of the number of the quantized bits can bring the performance of a limited fronthaul close to the performance of a perfect fronthaul system. A possible future work could be to adopt an alternative approach and use a non-uniform quantizer. In this regard, the quantization levels can be optimized as a function of the statistics of the signal~\cite{6226311} using algorithms like the Lloyd-Max algorithm~\cite{Compression1456239}.

The study in~\cite{ComputerAndForward7962724} employs a compute-and-forward strategy in the uplink to minimize the load on the fronthaul. The approach uses a structured lattice algorithm for physical layer network coding~\cite{4787140}. The results presented show that such an approach can decrease the traffic on the fronthaul and increase system throughput.

The investigation in~\cite{HardwareImpairements8891922} studies the uplink under a limited capacity fronthaul and hardware impairments. The authors develop a low-complexity fronthaul rate allocation scheme under three different compress-and-forward strategies between the DUs and the CU. These schemes, in increasing order of processing power required at the DU, are denoted as compress-forward-estimate, estimate-compress-forward, and estimate-multiply-compress-forward. What is particularly interesting in this work is using rate-distortion theory~\cite{cover1999elements} to relate the required fronthaul rate with the compressed signal that experiences quantization noise and needs to be transmitted over the fronthaul link. In Fig.~\ref{fig:ReviewCompressScenarios}, we summarize three different scenarios of signal compression in the fronthaul.

The authors of~\cite{6920005, 7581201} consider beamforming optimization under a preset fronthaul capacity, where the weighted \mbox{$\ell_1$-norm}, from the compressive sensing literature~\cite{CompressedSensing1614066}, is used to deal with the non-convex constraints. The work in~\cite{7287780} minimizes the total transmission power of the DUs while respecting a needed QoS constraint and the limited capacity fronthaul deployed through optical fibers. Similarly, the authors in~\cite{han2019sparse} use beamformers with a group sparse structure for cell-free massive MIMO networks with finite fronthaul capacity constraints. As shown in~\cite{han2019sparse}, and under the system assumptions, the sparse structure used can outperform benchmark solutions that use zero-forcing (ZF) precoding with serving clusters chosen based on simple metrics, e.g., fixed-size clusters consisting of DUs providing the highest signal level.

Sacrificing some performance by limiting or not sharing of CSI between the DUs, and using either instantaneous or statistical CSI, can be a key to reduce the traffic on the fronthaul~\cite{channelAcquisition5462883}. The investigation in~\cite{LocaPartialZFBF9069486} employs two distributed schemes for local beamforming that do not require instantaneous CSI exchange between the DUs and the CUs. Although the beamformers are constructed locally, the power control coefficients are optimized in a centralized fashion using long-term channel statistics. Partial-centralization~\cite{8693830} can also be used to decrease the fronthaul traffic, where the DUs integrate not only the radio function but also the baseband functions.

The work in~\cite{7996634} assumes that there are two types of DUs; a primitive one and a DU with layer-1 functions, and then formulates an objective function to determine at which site (from available ones) and how many nodes of each type are to be deployed based on a trade-off between the deployment cost and the fronthaul capacity. We note that the authors assume a fixed limited-capacity fronthaul.
	
\begin{figure*}[t]
	\centering
	\includegraphics[width=0.8\linewidth]{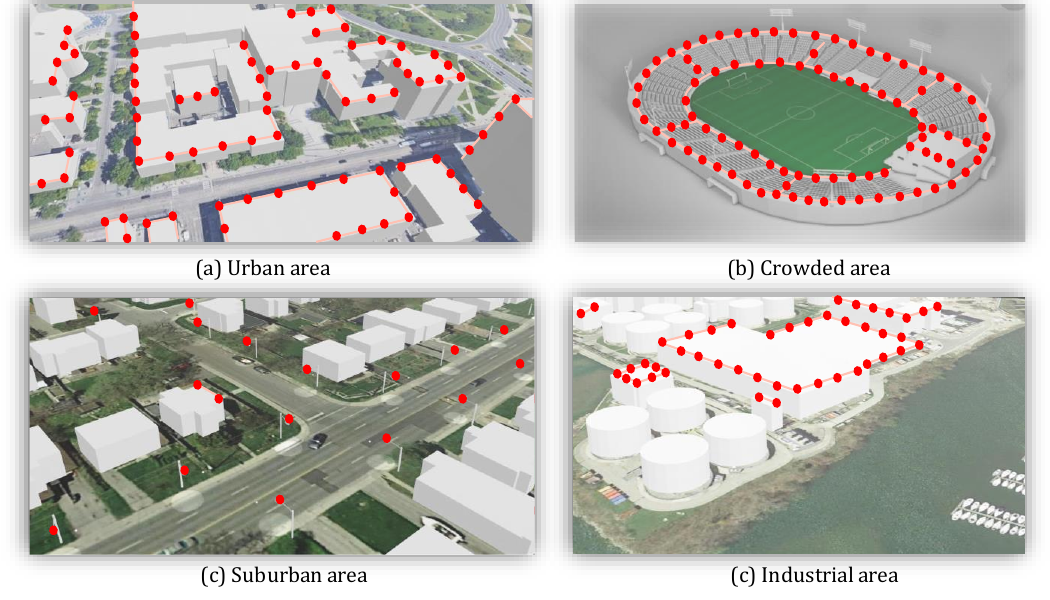}
	\caption{A dense deployment for the user-centric cell-free MIMO solution imposes logistical challenges in deploying the DUs (red dots in figure) and their fronthaul links (light red lines). The figure is not meant to reflect the density of DUs.}
	\label{fig:CoordinatedNet_review}
\end{figure*}
	
Different levels of cooperation on the uplink between the DUs and their CUs are studied in~\cite{differentCooperationLevels8845768} under a cell-free massive MIMO network architecture. In a nutshell, reduced cooperation decreases the amount of signaling that is communicated on the fronthaul. However, the DU is required to integrate more baseband functions, which leads to a less-cost effective DU deployment, while also suffering lower performance due to a less coordinated network. In~\cite{differentCooperationLevels8845768}, the authors analyze four different degrees of cooperation, under a single CU per network, described as follows:
\begin{enumerate}
	\item Fully centralized solution, where pilot and data signals received at all DUs are compiled at a single CU to perform channel estimation and data detection.\label{enum:level4}
	\item A two-stage approach that relies on the large-scale fading decoding (LSFD)~\cite{7837706, powerControlCellFree7917284, Scalable9174860, 8901196, HardwareImpairements9004558} (more on this later):\label{enum:level3}
	\begin{enumerate}
		\item Each DU locally estimates the CSI and employs an arbitrary combiner to obtain a local estimate of the user data.
		\item Data estimates with channel statistics are gathered at the CU to perform joint detection through linear processing (only channel statistics are found at the CU, no pilot signal sharing). 
	\end{enumerate}
	\item Same as in the previous scheme, but now the CU performs detection in the second stage by simply taking the average of the local estimates. So, no channel statistics are needed at the CU.\label{enum:level2}
	\item A small-cell network scheme that is fully distributed because no CSI-related information is exchanged on the fronthaul. Detection is done locally at the DUs using local CSI, and one DU serves each user.\label{enum:level1}
\end{enumerate}
The schemes are ordered from highest to lowest in terms of the fronthaul usage. In terms of performance, the fully centralized solution generally provides the highest SE, while the fully distributed scheme provides the lowest, with a crossing between the cumulative density function (CDF) of the SE for the other two schemes. In particular, this observation can differ based on the access points density, where the fully distributed scheme can outperform the third scheme listed above. Interestingly, this contradicts the results in~\cite{cellFreeVersusSmallCells7827017}, because as stated in~\cite{differentCooperationLevels8845768}, the study in~\cite{cellFreeVersusSmallCells7827017} uses maximum ratio combining (MRC) which seems to perform poorly in cell-free massive MIMO systems. However, this claim may not be valid in all scenarios.

The authors of~\cite{DeepLearningFronthaul9110901} study different uplink power control schemes, where a deep convolutional neural network (DCNN) is employed to map large-scale fading to the allocated power coefficients. The combine-quantize-and-forward scheme is used, where the access points send a quantized version of the combined received signals multiplied with the conjugate of the channel to the CU. Quantize-and-forward is also considered, where the access points send the quantized versions of both the received signals and the channel estimates to the CU, then the CU can perform MRC and ZF techniques to enhance the signal.
	
\subsection{Wired/Wireless fronthaul}
One method to limit the communication on the fronthaul would be to implement a serial fronthaul by allowing the DUs to relay the data between each other. This may be useful in some network scenarios where some DUs can act as a proxy for the CU. Nonetheless, these connections are not a scalable solution for a wide deployment unless an efficient wired solution is applied; in this case, the DUs would be connected through a wired bus network, so that the DUs do not intervene in relaying the signal to other DUs. In this regard, the radio stripes system~\cite{frenger2019antenna} could be the technology that fills this gap. The developed system is a promising architecture of cabling and internal communication between the DUs and their CUs~\cite{radioStripsSystem}, where the DUs antennas can be integrated inside single adhesive tapes providing the connections with the CU. In this cabling architecture, the radio frequency components are printed electronics on tape and the processing is done sequentially inside the cable. This means that the fronthaul is integrated with the DUs in wired connections, which is different from conventional fiber optic fronthauls. The importance of this solution also resides in its ability to be mass deployed with little impact on the environment because the connections can be integrated in the infrastructure with few logistical considerations (e.g., digging, connecting thick cables, etc).
	
In Fig.~\ref{fig:CoordinatedNet_review}, we illustrate a possible deployment of DUs in different environments, denoted as urban, crowded, suburban, and industrial areas. Planning for a dense deployment of DUs shows that each environment requires different strategies to map the DUs on the structures found in the area. The figure also shows that deploying a dense fronthaul infrastructure is challenging, requiring flexible links with little impact on the environment, in addition to the burden of maintenance.

The deployed locations of the DUs play a crucial role in maximizing the access rate~\cite{ammarWirelessbackahulJournal, ammarWirelessbackahulConf, Arin8529184}. Using a wireless fronthaul provides flexibility in deploying the DUs more than that provided by traditional fiber-optic links. This is because the DUs can be easily deployed in the locations that they are mostly needed rather than in the locations where the fronthaul is available or can be installed. In this context, the optimal placement of the DUs may not coincide with the locations of the wired fronthaul connections. Not to mention the prohibitive cost, geographical limitation and logistical considerations required to deploy wired connections. Furthermore, wireless connections provide the ability to quickly build up/tear down a network. Wireless fronthaul can also be used to support a wired fronthaul. On the other hand, a wireless fronthaul deployment requires careful tuning of the network resources~\cite{ammarWirelessbackahulJournal} and may require advanced interference cancellation techniques to be successful. Wireless fronthaul provides a high flexibility, but it may not be feasible for some network configurations, e.g., a large number of users served on the same channel resources through beamforming~\cite{ammarWirelessbackahulJournal}.

Wireless fronthauls may be point-to-point links, making millimeter wave communication an attractive solution~\cite{5GBackhaulNetworks6963798}. However, generally, studies about cell-free networks have used millimeter wave on the access channel~\cite{EnergyEfficiencyMmwave8676377, EECellFreeUCMMwave8292302, EECellFreeUCMMwave8516938, 9130689, DeepLearning8815888}, but, the effect of using this technology on the fronthaul is still not studied. Thus, the effect of deploying millimeter wave on the fronthaul and optimizing its usage for cell-free communications is an open research topic.
	
Another possible solution for a wireless backhaul is the integrated access and backhaul (IAB) technology~\cite{3GPPTR38.874} which targets mainly millimeter wave communication. IAB allows part of the wireless spectrum in the access channel to be used in the backhaul or the fronthaul. The IAB node may serve as a first hop or second hop node, so it can allow a fraction of the CUs to use wireless connections to other CUs (IAB-donors) that already have a wired connection to the core network~\cite{IAB9040265}. In this regard, it can serve as a plug-and-play solution to deploy the fronthaul for some DUs. The use of a full-duplex in-band backhaul under user-centric cell-free MIMO network is investigated in~\cite{WSRjointAccessBack8786917}, where the authors maximize the network weighted sum rate (WSR) through joint access-backhaul beamforming. The authors solve the problem using successive lower-bound maximization by introducing a concave lower-bound approximation for the rate expression based on signal to interference and noise ratio (SINR) convexification, which seems to outperform mean squared error (MSE)-based approaches as shown in the numerical results presented in~\cite{WSRjointAccessBack8786917}.
	
The uplink performance of both cell-free massive MIMO and user-centric cell-free MIMO networks is investigated in~\cite{UC_CellFreOpticalFronthaule020}. The authors assume a two-level design for the fronthaul, wherein free space optics (FSO) links connect the access points to some aggregation units, then fiber optics is used to connect the aggregated units to a single CU. Two hardware models are considered for the FSO link, which are denoted as the clipping model and the hardware impairment model. FSO links~\cite{6844864} provide an advantage compared to fiber optics in the sense that they do not require digging and installation, however, the technology requires a Line of Sight (LoS) connection, and it is sensitive to atmospheric conditions like snow and fog. The results presented show that under different scenarios, the user-centric approach outperforms the cell-free massive MIMO scheme with a huge gap in the sum SE due to lower interference~\cite{UC_CellFreOpticalFronthaule020}. Moreover, the hardware impairment model seems to provide better performance compared to the clipping model.

\subsection{Lessons Learned}\label{sec:fronthaul_LL}
\begin{table}[t!]
	\footnotesize
	\centering
	\begin{tabular}{|>{\raggedright\arraybackslash}p{0.08\textwidth}|>{\raggedright\arraybackslash}p{0.16\textwidth}|>{\raggedright\arraybackslash}p{0.1\textwidth}|>{\raggedright\arraybackslash}p{0.05\textwidth}|}
		\hline
		\hline
		\multicolumn{1}{|l|}{ \textit{\textbf{Solution}}} & \multicolumn{1}{l|}{ \textit{\textbf{Pros}}} & \multicolumn{1}{l|}{ \textit{\textbf{Cons}}} & \multicolumn{1}{l|}{\textit{\textbf{Ref.}}}\\
		\hline
		\hline
		Different degrees of cooperation or function centralization & Flexible functional split, adaptive to network conditions, low-complexity schemes exist, can decrease/increase costs of access points & Can sacrifice performance due to limited cooperation & \cite{differentCooperationLevels8845768, bjornson2019scalable9064545, maxMinRate8756286}
		\\
		\hline
		Signal compression scenarios & Minimize load on fronthaul, different scenarios for different needs exist (Fig.~\ref{fig:ReviewCompressScenarios}), lower fronthaul blocking probability, step-size of quantizer (or number of quantization bits) can be optimized based on the SDNR, quantization distortion is uncorrelated with the quantizer output & Quantization error introduces additional noise & \cite{maxMinRate8756286, differentCooperationLevels8845768, 8693830, HardwareImpairements8891922, EE9212395}
		\\
		\hline
		Millimeter wave-based wireless fronthaul & Large bandwidth, flexible deployment of DUs, Lower deployment costs & Requires LoS, suffers from blockages & \cite{EnergyEfficiencyMmwave8676377, 9130689}
		\\
		\hline
		Turn on/off access points & Dynamic control for fronthaul load based on network status & Affect performance due to smaller number of serving access points, waste resources that are put in standby & \cite{powerAllocEnergyEfficiency9136914}
		\\
		\hline
		Multiple CUs & Hierarchical fronthaul design & Could affect delay for users served by multiple CUs & \cite{ammarDistributed_RA_UC, LocaPartialZFBF9069486, UC_CellFreOpticalFronthaule020}
		\\
		\hline
		Effective wired solutions such as the radio stripes system & Can be integrated within environment for dense deployment, few logistical considerations & Could be costly in terms of deployment and maintenance & \cite{frenger2019antenna, radioStripsSystem}
		\\
		\hline
	\end{tabular}
	\caption{Solutions related to fronthaul.}
	\label{table:prosCons_fronthaul}
	\vspace{-2em}
\end{table}
It seems that a centralized implementation for the control functionalities using a single CU requires using a wired fronthaul. This puts a huge burden on developing an effective and flexible wired fronthaul that camouflage well within the environment, such as the radio stripes system~\cite{frenger2019antenna}, to allow for a dense deployment of DUs. Additionally, deploying a dense fronthaul infrastructure is challenging, requiring flexible links with little impact on the environment, in addition to the burden of maintenance. For a distributed implementation of the control functionalities of the network using multiple-CU network, wireless fronthaul can be used, but still with some dependence on a wired fronthaul. Furthermore, millimeter wave communication and IAB could be exploited to provide large bandwidth for the fronthaul.

Limiting cooperation and partial centralization are useful strategies to handle the limited capacity of the fronthaul. In this context, local beamforming techniques and optimized beamforming techniques with limited CSI sharing among the DUs are needed. The price to be paid is a possible loss in performance and less cost-effective DUs.

Distributed SDN can play a major role in managing the traffic on the fronthaul and other crucial functionalities such as a dynamic assignment of the DUs to the CUs. Signal compression is a milestone for communication on the fronthaul. In this regard, many different scenarios can be used such as those detailed in Fig.~\ref{fig:ReviewCompressScenarios}. These schemes, in increasing order of processing power required at the DU, are denoted as compress-forward-estimate, estimate-compress-forward, and estimate-multiply-compress-forward~\cite{HardwareImpairements8891922}. There seem to be no particular compression scenarios preferred for cell-free communications, so choosing the scenario could be a preference or constrained by the different processing power required at the DU, processing load and fronthaul communication. Studying the usage of non-uniform quantizers on the fronthaul is still a novel topic, and it would be interesting to see studies on this topic in the~future.

\newcommand{\varTableDes}{advantages and \emph{current drawbacks}}
Finally, in Table~\ref{table:prosCons_fronthaul} we summarize some \varTableDes\ of selected solutions targeting fronthaul~links.

\section{Estimation of Channel State Information}\label{sec:CSI}
Much of capabilities potentially provided by the distributed user-centric, cell-free, architecture is only possible if accurate CSI is available. Serving users with many DUs requires timely CSI to be available, which introduces large overhead for channel estimation. In this regard, in the research community, time division duplex (TDD) systems are far more common than the frequency division duplex (FDD) counterpart. This is because of the lower feedback overhead~\cite{FeedbackOverhead6449246} required to share the CSI; in TDD systems channel reciprocity can be exploited to use a single pilot training phase (in the uplink).

\begin{figure}[t]
	\centering
	\includegraphics[width=1\columnwidth]{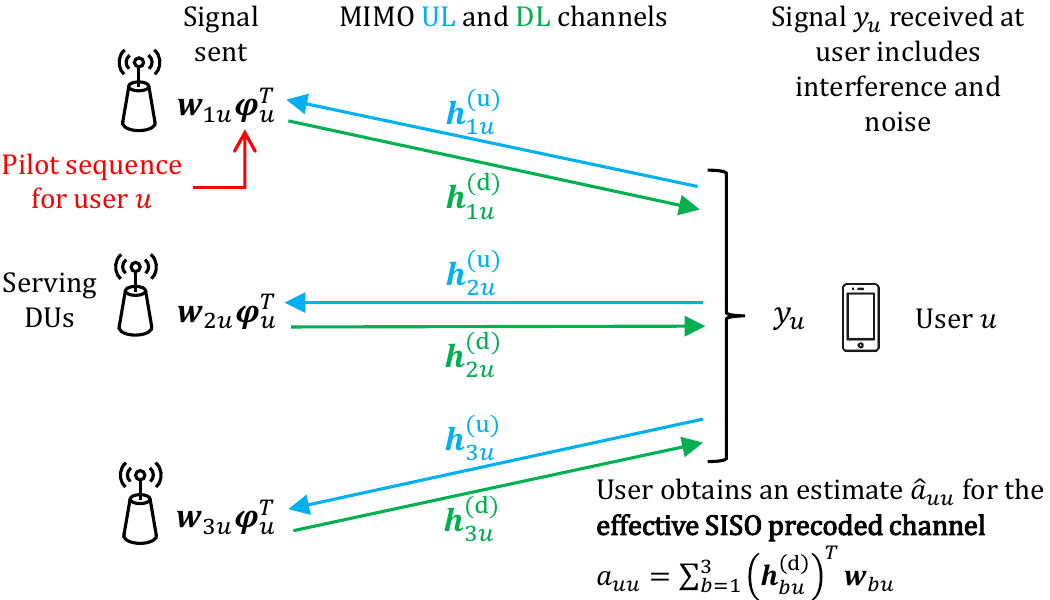}
	\caption{Downlink training with the number of pilot sequences independent of the number of DUs. The user can use linear minimum mean square error (MMSE) estimation to estimate the effective SISO precoded channel; ${\bf h}_{bu}$: MIMO channel between DU $b$ and user $u$, ${\bf w}_{bu}$: beamformer used by DU $b$ to serve user $u$; ${\bm \varphi}_u$: pilot sequence assigned for user $u$.}
	\label{fig:Review_DL_Training}
	\vspace{-1em}
\end{figure}

Although a cell-free scheme can benefit from downlink pilot training~\cite{DLTrainingCSI8799031, Rician9099874} in TDD mode to estimate an effective downlink channel~\cite{kay1993fundamentals}, especially for multi-antenna users, still using downlink pilots will result in an additional overhead that may be dependent on the number of DUs. Hence, in real-world deployments such a scheme may not be practically feasible. One way to make the downlink pilot sequence independent from the number of DUs, is to let the DUs use the pilots assigned to the users (so no dedicated pilots are assigned to each DU). Based on this, each user estimates the effective single-input single-output (SISO) precoded channel ($a_{uu}$ in Fig.~\ref{fig:Review_DL_Training}) from all the serving DUs~\cite{DLTrainingCSI8799031}, where $a_{uu}$ is the accumulated precoded channels from the serving DUs. So, technically, the user does not know the individual channels to the serving DUs. In Fig.~\ref{fig:Review_DL_Training}, we show an illustration for such downlink pilot training scheme, where each DU sends the precoded pilot assigned to each user $u$, i.e., sends ${\bf w}_{bu} {\bm \varphi}^T_u$.

The problem of CSI estimation becomes more critical for FDD systems or for scenarios that introduce channel non-reciprocity~\cite{ChannelNonReciprocity8910383}, such as the existence of antenna calibration errors~\cite{SLNRnonIdealReciprocity6589033}. Some antenna calibration methods have been proposed to compensate for the phase mismatch on the feeds of the antenna arrangement~\cite{o2017mimo, stayton2011systems}. If the antenna calibration errors greatly affect the system performance, even for TDD systems, both downlink and uplink pilots would be needed.

Some studies~\cite{PowerClusteringCellFreeFDD8968400} investigate the use of FDD for cell-free MIMO. A key component in these studies is exploiting the angle-reciprocity of the multipath components, so that the required CSI acquisition overhead scales with only the number of served users. Results demonstrate that the proposed estimation technique outperforms conventional subspace-based~\cite{32276} and gradient-descent based~\cite{8446023} techniques. We note that, in FDD systems, downlink CSI estimation can still benefit from angle reciprocity when the uplink and downlink carrier frequencies being used are relatively close to each other~\cite{7524027}.

The study in~\cite{cellFreeFDD9014542} proposes a feedback reduction technique that exploits angle reciprocity in FDD cell-free MIMO systems. The technique is based on feeding back the path gain information (PGI) of a few selected dominant paths between the BS and the user and obtaining the angles of departures (AoDs) directly from the uplink pilot signal by exploiting angle reciprocity. Their results show that the proposed ``dominating PGI feedback'' scheme achieves greater than $60\%$ reduction in the feedback overhead compared to the conventional scheme relying on CSI feedback. This is a promising result that warrants further investigation.

In terms of managing CSI estimation, the approaches proposed in the literature can be divided based on different levels of centralization used. We can have a centralized approach in which the pilot signals received by the DUs are gathered at the CU, then the CU performs channel estimation, and jointly processes the UL and DL data signals~\cite{bjornson2019scalable9064545}. Alternatively, we can have a decentralized scheme where each DU locally estimates the channels of its associated users and uses this information to locally process data signals~\cite{differentCooperationLevels8845768}. Hence, only the decoding and encoding of data signals is carried out at the CU. For a minimum mean square error (MMSE) precoding scheme, each DU can either estimate the channels to all the users in the network within some boundary, or estimate the channels to only its served users, and use a statistical approximation for the channels for the other users~\cite{CoherentCRAN8606433}. The second approach is more scalable, because the number of channels to be estimated only grows with the number of users in the serving cluster of the DUs, and is less affected by the number of the users in a large network area.

The authors of~\cite{DeepLearning8815888} target the high computational complexity of CSI estimation when using millimeter waves in cell-free massive MIMO networks by proposing a deep learning framework. The fast and flexible denoising convolutional neural network (FFDNet)~\cite{FFDNet8365806} is used because of its ability to both reduce training and testing latency and deal with different noise levels using the same neural network. The simulation results presented indicate that FFDNet is a promising solution, where it provides faster CSI estimation than benchmark CSI estimators without introducing any additional error. The results also show that FFDNet outperforms the existing deep denoising convolutional neural network (DnCNN)-based method.

\subsection{Favorable Propagation and Channel Hardening}\label{sec:FavorableP_ChannelH}
In this subsection, we discuss two decisive properties that can significantly help in alleviating the CSI acquisition problem. These properties are useful for signal detection using the channel statistics, and they are believed to be achievable in massive MIMO systems~\cite{marzetta2016fundamentals}.

For any two mutually independent $N \times 1$ random vectors ${\bf u}$ and ${\bf v}$ whose elements are independent and identically distributed (i.i.d) zero-mean random variables with respective variances $\sigma_v^2$ and $\sigma_u^2$, we almost surely ($\overset{{\rm a.s.}}{\longrightarrow}$) have~\cite{EAsymptotic6457363}
\begin{align}\label{eq:favorableProp1}
	\frac{1}{N} {\bf u}^H {\bf u} \overset{{\rm a.s.}}{\longrightarrow} \sigma_u^2,
	\ \text{and}\
	\frac{1}{N} {\bf u}^H {\bf v} \overset{{\rm a.s.}}{\longrightarrow} 0,
	\quad \text{as}\ N \rightarrow \infty 
\end{align}
Furthermore, using the Lindeberg-Levy central limit theorem, the term $\frac{1}{\sqrt{N}} {\bf u}^H {\bf v}$ converges in distribution ($\overset{{\rm d}}{\longrightarrow}$) as
\begin{align}
\frac{1}{\sqrt{N}} {\bf u}^H {\bf v} \overset{{\rm d}}{\longrightarrow} \mathcal{CN} \left(0, \sigma_u^2 \sigma_v^2 \right), \quad \text{as}\ N \rightarrow \infty
\end{align}
The implication is that if the vectors ${\bf u}$ and ${\bf v}$ are the serving channels for any two users, where $N$ would be the number of serving antennas, and $N$ is large, we can assume that these channels are orthogonal, which will cancel inter-user interference. This crucial property is called \emph{favorable propagation}~\cite{FavProp7903703}.
	
Another influential property is \emph{channel hardening}, which is a direct consequence of the law of large numbers. Channel hardening occurs when the effective channel gain is close to its mean, i.e., the channel gain is not random~\cite{ChannelHardening1327795} and, hence, using the mean of the gain works well for signal detection. For example, for a channel ${\bf H}_b \in \mathbb{C}^{N \times |\mathcal{E}_b|}$, between a BS $b$ and a set $\mathcal{E}_b$ of $K$ served users who are experiencing small-scale fading ${\bf G}_b$ with vectors $\{ {\bf g}_{bu} = [ {\bf G}_b ]_{u} \sim \mathcal{CN} \left({\bf 0}, {\bf I}_N \right) : u \in \mathcal{E}_b\}$, we state that, for a fixed $K$,
\begin{align}
	\frac{ {\bf H}_b^H {\bf H}_b }{N}
	=
	{\bf D}_b^{1/2} \frac{ {\bf G}_b^H {\bf G}_b }{N} {\bf D}_b^{1/2}
	\simeq
	{\bf D}_b
	,
	\quad \text{as}\ N \rightarrow \infty, 
\end{align}
where ${\bf D}_b = {\rm diag}\left(\{\beta_{bu}\}_{u \in \mathcal{E}_b}\right)$ is a diagonal matrix representing the large-scale fading experienced by the users $u \in \mathcal{E}_b$. This shows that the small-scale fading between the BS and each user acts in a deterministic fashion overall.

Channel hardening requires that the norms of the channel vectors fluctuate only a little, so that the following condition must be valid~\cite{ChannelHardeningDegree7880691}
\begin{align}
	\frac{\|{\bf g}_{bu}\|^2}{ \mathbb{E}\left\{ \|{\bf g}_{bu}\|^2 \right\} } \rightarrow 1,
	\quad \text{as}\ M \rightarrow \infty
\end{align} 
where $\mathbb{E} \left\{\cdot\right\}$ is the expectation operator. This condition follows from the Chebyshev's inequality, where for any $\epsilon \ge 0$, we have
\begin{align}
	&{\rm Pr} \left\{
	\left| \frac{\|{\bf g}_{bu}\|^2}{ \mathbb{E}\left\{ \|{\bf g}_{bu}\|^2 \right\} } - 1 \right|^2 \le \epsilon \right\} 
	\nonumber \\
	&\quad =
	1 - {\rm Pr} \left\{
	\left| \frac{\|{\bf g}_{bu}\|^2}{ \mathbb{E}\left\{ \|{\bf g}_{bu}\|^2 \right\} } - 1 \right|^2 > \epsilon \right\}
	\nonumber \\
	&
	\quad \ge
	1 - \frac{1}{\epsilon} \frac{{\rm Var}\left\{ \|{\bf g}_{bu}\|^2 \right\} }
	{ \left(\mathbb{E}\left\{ \|{\bf g}_{bu}\|^2 \right\} \right)^2 }
\end{align}
where ${\rm Pr} \left\{\cdot\right\}$ is the probability operator, and ${\rm Var}\left\{\cdot\right\}$ is the variance operator. Based on this analysis, we can say a channel is hardened for a user $u$ if~\cite{ChannelHardeningDegree7880691} 
\begin{align}\label{eq:channelHardening}
	\frac{{\rm Var}\left\{ \|{\bf g}_{bu}\|^2 \right\} }
	{ \left(\mathbb{E}\left\{ \|{\bf g}_{bu}\|^2 \right\} \right)^2 }
	\rightarrow 0,\quad N \rightarrow \infty
\end{align}

As discussed in~\cite{ChannelHardeningDegree7880691}, massive MIMO channels may not always harden, which raises a doubt as to whether this property will be valid in user-centric cell-free MIMO networks, especially, in scenarios where a more distributed antenna system is preferred~\cite{ammarWirelessbackahulJournal}. Additionally, because the cell-free MIMO scheme has still not been deployed, no field trials seem to have been conducted to prove such a strong claim. The stochastic geometry based study performed in~\cite{cellFreeStochasticGeometry8379438} concludes that for cell-free massive MIMO networks channel hardening is limited to very special scenarios, like communication under a small path-loss exponent (PLE). Hence, having a LoS channel component may increase the probability of channel hardening. Despite this, the work also reports that the use of 5-10 antennas per DU can substantially improve channel hardening and favorable propagation. 

We note that the assumption of channel hardening and favorable propagation is used in many studies in the literature of cell-free massive MIMO~\cite{EnergyEfficiency8097026, DLTrainingCSI8799031}. All in all, it seems that it remains a point of contention as to whether channel hardening and favorable propagation can be assumed in user-centric cell-free networks. Accordingly, a robust model should at least evaluate the conditions in~\eqref{eq:favorableProp1} and~\eqref{eq:channelHardening} for the network configuration under consideration before assuming the validity of these properties.

\begin{figure}[t]
	\centering
	\includegraphics[width=1\columnwidth]{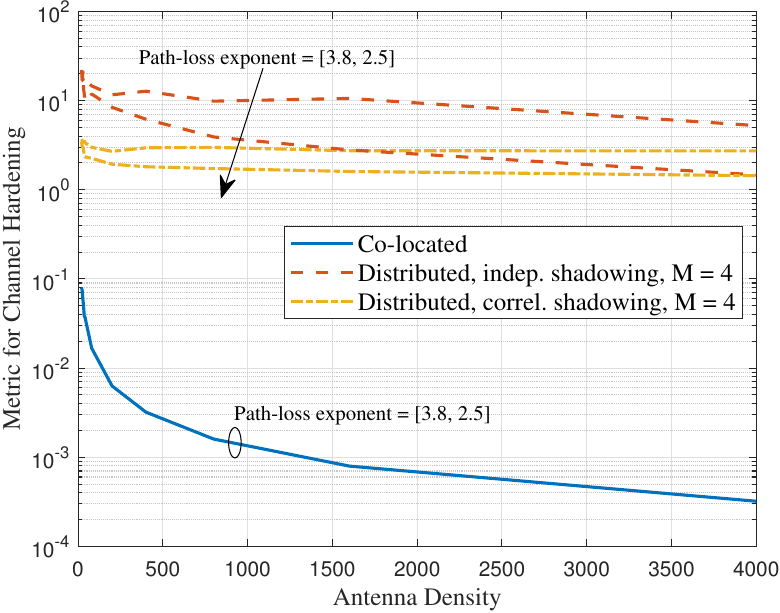}
	\caption{Metric Shown in~\eqref{eq:channelHardening} as a function of antenna density.}
	\label{fig:ChannelHardening_metric}
	\vspace{-1em}
\end{figure}

In Fig.~\ref{fig:ChannelHardening_metric}, we plot the result of the expression in~\eqref{eq:channelHardening} as a function of the antenna density. The antenna density is simply the number of serving DUs in one square~kilometer (${\rm km}^2$) area multiplied by the number of antennas $M$ per DU. We simulate a circular area of radius $500$ meters with a typical user found at the center, and we generate the locations of the serving DUs according to Poisson Point Process (PPP) with an exclusion region of $20$ meters around the typical user located at the origin. 

We plot three cases for the antennas: co-located (i.e., all antennas are found at one location), distributed with independent shadowing and distributed with distance-based two-component correlated shadowing model that was validated in~\cite{CorrelatedShadowing4357088, CorrelatedShadowing104090} and used for cell-free MIMO systems in many studies such as~\cite{cellFreeVersusSmallCells7827017, cellFreeUserCentricPower8901451, differentCooperationLevels8845768}. All results are averaged using Monte Carlo simulations over $100$ network realizations each with $10^4$ channel realizations. Each DU is equipped with $M=4$ antennas in the case of distributed deployment, while they are assumed to be located at a single transmitter for the co-located case. Furthermore, we use typical network environment, Rayleigh fading, and path loss parameters suitable for transmissions with short serving distances. The considered environment results in a PLE of value $3.8$. We also plot the results for a manually modified PLE of $2.5$ to check the impact of the PLE.

The results under the described scenario show that the co-located deployment provides the highest profile of channel hardening followed by the case of distributed deployment with correlated shadowing then by that of distributed deployment with independent shadowing. Moreover, the results show that a lower PLE provides better channel hardening for a distributed deployment of DUs, which is consistent with the conclusions in~\cite{cellFreeStochasticGeometry8379438}. However, as discussed previously, we note that different network configurations may produce different profiles for channel hardening. 

\subsection{Pilot Assignment}
An effective PA policy is crucial when the number of users served or requesting access across the network at a specific carrier is larger than the length $\tau_p$ of the pilot sequence. Introducing PA policies in cell-free networks can help in keeping co-pilot users as mutually distant as possible. It can also determine which users necessitate a dedicated pilot that is not used by nearby users in order to minimize pilot contamination.

The authors of~\cite{randomVsStructuredPilots8403508} report that in a typical environment in a cell-free network, a well-structured PA policy can reduce the required length of the pilot sequence by a factor of $3$–$3.75$ while achieving negligible contamination. In this regard, it is assumed that a $3\%$ decrease in average SE is the definition for a negligible contamination. For many reasons, PA is a serious issue for user-scheduling optimization problems~\cite{ammarC_RA_UC, ammarC_RA_UC_conf}. In this case, the number of users is much larger than the available resources, and the DUs need to select the users to be served. In such problem types, it is impractical to optimize the PA based on a rate-based utility because the users to be scheduled are still not chosen. Hence, an instantaneous achievable rate cannot be used in practice. Thus, unless statistical approaches are used to assign the pilots, the PA must be based on some offline metric, e.g., the locations of the users. 

A disadvantage of rate-based approaches is the large overhead and computing load needed just to assign the pilots, which makes the scheme hard to deploy in a real network scenario. It is worth noting that reassigning pilots and then redoing channel estimation after is not a viable option, as it will leave little time for actual data transmission during the limited channel coherence time. Based on this, despite the fact that a location-based PA will not produce the optimal solution, it can be a viable choice for non-orthogonal PA.

The study in~\cite{8385475} proposes a downlink orthogonal pilot assignment strategy for cell-free massive MIMO. The users are divided into two groups based on a defined user pilot-utility. The first group is assigned orthogonal pilots, while the second decodes data using only statistical CSI. The pilot-utility is a function of the Doppler spread, channel hardening degree, and some prioritization weights. A drawback in this scheme is that the number of antennas serving the users may not be large enough to guarantee channel hardening; this is needed to efficiently use statistical CSI. Further, to avoid data-to-pilot interference, the users belonging to the second group are still not allowed to receive their data during the pilot training phase of the first group.

\begin{figure}[t]
	\centering
	\includegraphics[width=0.8\columnwidth]{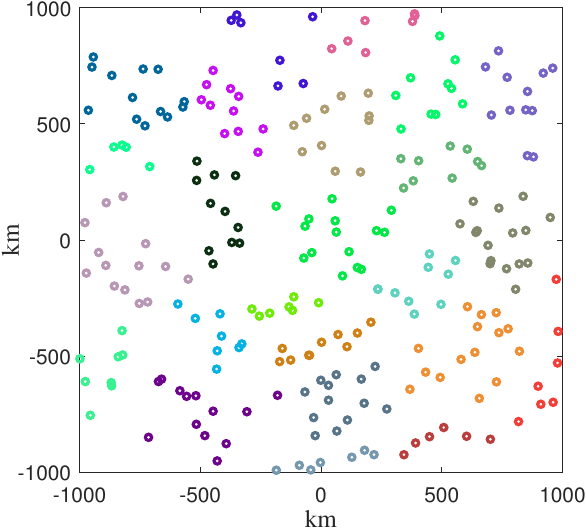}
	\caption{Users clustered based on their location using the HAC algorithm. Users within each cluster (markers with same color) use orthogonal pilots. The assigned of pilots inside each group can be further optimized.}
	\label{fig:ClusteredUsers}
	\vspace{-1em}
\end{figure}

The work presented in~\cite{Scalable9174860} proposes to use a user-group clustering or a K-means clustering for pilot assignment in cell-free network. Similarly, the study in~\cite{pilotAssign9178782} optimizes PA through a location-based rate maximization. The proposed algorithm clusters the users based on their location, then a combinatorial search using the Hungarian algorithm~\cite{kuhn1955hungarian} is performed to assign the pilots to the users within each cluster. The studies in~\cite{ammarC_RA_UC, ammarC_RA_UC_conf} employ the hierarchical agglomerative clustering (HAC)~\cite{bien2011hierarchical} to cluster the users into groups each containing fewer users than the number of available orthogonal pilots. The orthogonal pilots are then assigned to the users inside each group to prevent co-pilot sequences within each group. The HAC algorithm has been shown to be more consistent than algorithms like the K-means and Gaussian mixture models, because it is insensitive to the choice of the distance-metric used to construct the cluster~\cite{karypis2000comparison}; however, it has a polynomial computational complexity compared to a linear complexity for the aforementioned algorithms. In Fig.~\ref{fig:ClusteredUsers}, we show an example of clustering the users to perform~PA.

Non-orthogonal PA has also been studied in massive MIMO networks~\cite{GraphCodePilotContamin8487005}. The system assumes a ``crowd scenario'', where the number of users is very large and the users have intermittent access behavior, hence making random access a natural choice. This scenario is often found in internet of things (IoT) networks. The study in~\cite{masoumi2020cell} uses greedy~\cite{cellFreeVersusSmallCells7827017} and graph coloring-based~\cite{GraphColoringPilot9110802, PiltAssigGraphColoring7217795} algorithms to assign pilot sequences among different users to control the resultant pilot contamination in a coexistence scenario of cell-free massive MIMO underlaid with device-to-device (D2D) communication. The greedy method is based on reassigning the pilot for the user experiencing the lowest data rate to minimize its pilot contamination. The procedure is repeated for a specific number of times to enhance performance. Nonetheless, achieving optimal performance would require trying all possible combinations, which is not an practical solution. The graph coloring method is based on selecting co-pilot users one-by-one such that the potential interference is minimized.

A graph-coloring theoretic approach is also used in~\cite{graphColoringUsercentric7080877} to solve the PA problem in a network with user-centric clustering. A similar scheme based on the worst performing user is proposed in~\cite{ashikhmin2017pilot}, where different stopping criteria for the pilot reassignment procedure are used. The framework in~\cite{pilotPowerControl8450041} proposes to minimize the pilot contamination by optimizing the pilot transmission power by formulating a min-max optimization problem. The problem minimizes the users' largest normalized MSE and is solved using a sequential convex approximation method. Interestingly, the authors in~\cite{8914726} perform pilot assignment based on the Tabu search algorithm, which can escape from local optima by recording the solutions of the last few iterations~\cite{TabuSearch5957382}.

The study in~\cite{CSI6415397} employs a covariance-aided CSI estimation using a Bayesian approach. Specifically, the covariance matrix of the channel vectors is used to obtain information about the mean and the spread of the multipath angles of arrival at the BS. The authors then exploit this technique to design a coordination protocol that assigns the pilot sequences to the users. The study in~\cite{uplinkPilotcellFreeMassiveMIMO2020} minimizes the MSE of the channel estimation in cell-free massive MIMO networks with a large number of users accounting for channel correlation. The optimization of the MSE is based on proposing an adjustable phase-shift pilot set allocation scheme. Active users are identified through non-orthogonal pilot phase shift hopping patterns and non-orthogonal adjustable phase shift pilots. Such an approach is useful when the users are sporadically active~\cite{mMTC8053903}, which is the case in massive machine-type communications~(mMTC).

Power control for the reused pilots potentially provides an opportunity to improve CSI estimation. Possible approaches can be based on the bit error rate or the MSE of the channel as evaluation metrics~\cite{PowerControlConventional7801046}. Finally, semi-blind CSI estimation techniques can be a key to mitigate pilot contamination and minimize the complexity of CSI estimation. This area is not investigated for cell-free MIMO systems and is, thus, an important future research direction.

\subsection{High-mobility Users}
\begin{table}[t!]
	\footnotesize
	\centering
	\begin{tabular}{|>{\raggedright\arraybackslash}p{0.08\textwidth}|>{\raggedright\arraybackslash}p{0.16\textwidth}|>{\raggedright\arraybackslash}p{0.1\textwidth}|>{\raggedright\arraybackslash}p{0.05\textwidth}|}
		\hline
		\hline
		\multicolumn{1}{|l|}{ \textit{\textbf{Solution}}} & \multicolumn{1}{l|}{ \textit{\textbf{Pros}}} & \multicolumn{1}{l|}{ \textit{\textbf{Cons}}} & \multicolumn{1}{l|}{\textit{\textbf{Ref.}}}\\
		\hline
		\hline
		Greedy and graph coloring-based algorithms & Reduce pilot contamination, applied in sequential pattern, can escape from local optima by recording the solutions of last few iterations & Some algorithms have high computational complexity & \cite{masoumi2020cell, cellFreeVersusSmallCells7827017, GraphColoringPilot9110802, PiltAssigGraphColoring7217795, GraphCodePilotContamin8487005, graphColoringUsercentric7080877, ashikhmin2017pilot} 
		\\
		\hline
		Location-based PA & Easy to implement, low signaling requirements, suitable for user scheduling problems & Neglect shadowing and other network performance metrics & \cite{ammarC_RA_UC, randomVsStructuredPilots8403508, Scalable9174860, pilotAssign9178782, bien2011hierarchical}
		\\
		\hline
		Rate-based PA & Advanced design, high performance & High signaling requirements, not suitable for user scheduling problems & \cite{8385475, pilotAssign9178782}
		\\
		\hline
		Large-scale fading based resource allocation & Lower CSI estimation load &  May require channel hardening & \cite{DLTrainingCSI8799031}
		\\
		\hline
		Rely on channel hardening to decode data symbols & No downlink pilot training is needed & Channel hardening may not apply & \cite{DLTrainingCSI8799031, cellFreeVersusSmallCells7827017}
		\\
		\hline
		Bayesian estimation & Exploitation of the side-information lying in the second-order statistics
		of the channels & Sensitive to the degree of overlap between the subspaces of covariance matrices for the channels of desired signal and interference & \cite{RicianFadingPhaseShifts8809413, CSI6415397}
		\\
		\hline
		PGI feedback for FDD systems & Reduction in the feedback overhead compared to the conventional scheme relying on CSI feedback (more than 60\% reduction), exploitation of angle reciprocity for the uplink and downlink channels, number of bits required for the channel vector quantization scales linearly with the number of dominating paths (not the number of transmit antennas) & Spatial domain channel needs to be represented by a small
		number of multi-path components
		& \cite{cellFreeFDD9014542, PowerClusteringCellFreeFDD8968400}
		\\
		\hline
		AoA information aided CSI estimation & Works for FDD and for multi-tap propagation channel paths & Still require feed back for the PGI & \cite{cellFreeFDD9014542, 8446023, PowerClusteringCellFreeFDD8968400}
		\\
		\hline
		Non-reciprocity calibration and CSI estimation based on FFDNet and deep neural networks & Faster CSI estimation than conventional schemes, recover the downlink CSI across the entire Bandwidth from the uplink CSI obtained at a small number of pilot subcarriers (interpolation) & Requirement for labeled data to train the classifier network & \cite{nonReciprocityCellFreeDNN9098852, FFDNet8365806}
		\\
		\hline
		Antenna calibration methods & Ensure channel reciprocity, do not use downlink pilots training phase & Additional signal processing overhead & \cite{o2017mimo, stayton2011systems, ArinNonReciprocity8421240}
		\\
		\hline
	\end{tabular}
	\caption{Solutions related to CSI.}
	\label{table:prosCons_CSI}
	\vspace{-2em}
\end{table}
Serving high-mobility users under the user-centric cell-free scheme is an open research topic that is still not well investigated. The channel between high-speed users and the DUs changes very quickly, hence leading to a mismatch between the CSI at the time the channel was estimated and at the time it is used for data transmission. This mismatch is referred to \emph{channel aging}~\cite{CSIAging6608213}. Though not well-investigated in the literature, channel aging can affect the fading statistics that are used to estimate the instantaneous channel response. 

Incorporating channel aging effects requires using a time-varying model for the channel that relates the temporal autocorrelation function of the channel with the propagation geometry, velocity, frequency, and antenna characteristics~\cite{1512123}. In this regard, the most famous models are the Jakes' model~\cite{jakes1994microwave} (also called the Clarke-Gans model) and the autoregressive process~\cite{CSIAging6608213}. Channel aging is a serious problem for conventional cellular networks as well. Works targeting this topic are mostly for the co-located massive MIMO model~\cite{channelAgingMassiveMIMO7307172, channelAgingMassiveMIMO7473866, channelAgingMassiveMIMO8122014}. As shown in~\cite{channelAgingMassiveMIMO8122014}, optimally selecting the frame duration can alleviate channel aging.

In a user-centric cell-free MIMO network, a high mobility profile requires the user to update its serving clusters very often. This will produce a lot of signaling overhead and delays which will affect the effective throughput. The effect of this recurrent change on the performance is still not known. In conventional fourth generation LTE networks, the handover process has a latency on the order of about $45$-$50$ ms~\cite{3GPP:TR36.881, zhang2020seamless}, however, this duration could be longer for cell-free networks because the user can be served by many transmitters, i.e., DUs, which are themselves controlled by many CUs.

In general, mobility dependent handover schemes that involve identifying high-mobility users~\cite{504082} could be a candidate to improve the handover process in user-centric cell-free networks. To model the mobility of the users, mobility models used in conventional networks are helpful; trace-based and synthetic models, such as the random waypoint (RWP)~\cite{johnson1996dynamic}, can be used to emulate mobile users~\cite{roy2011handbook, 5343061}. Metrics such as, handoff rate~\cite{Ravi7006787}, sojourn time~\cite{6477064}, and handoff probability~\cite{6477064}, which were developed initially for conventional networks can be redefined for cell-free networks. We note that the handoff probability can be interpreted as the probability that the serving BS does not remain the best candidate in one user's movement period.

A study of the SE is performed in~\cite{CSIAgingzheng2020cell} under the assumptions of channel aging and a cell-free massive MIMO system. Interestingly, using a cell-free massive MIMO framework provides better performance than the small-cell network architecture in both static and mobile scenarios. The authors also employ some fractional power control to improve the system performance. The system model uses a time-multiplexed pilot scheme to maintain the orthogonality of the pilots in the presence of channel aging, however, it is still not known if better pilot design can be obtained. Furthermore, the study in~\cite{RicianFadingPhaseShifts8809413} partially addresses mobility, because it studies the performance assuming Rician fading with phase shift of the LoS component. We note that the a phase shift of the LoS component of the channel can change due to user mobility and hardware effects such as phase noise.

Cell-free MIMO networks can benefit from studies developed for other systems like massive MIMO networks. From the literature of co-located massive MIMO BSs, channel prediction~\cite{machineLearningChannelAging8979256} is used to minimize the effect of CSI aging. However, such approaches need further investigation to evaluate their performance in user-centric clustering cell-free MIMO systems to determine if they can be implemented when the channel hardening and favorable propagation does not apply. Another issue that can be experienced by high-mobility users is synchronization errors for the received signals because of the Doppler shifts~\cite{Synchronization4287203}, which also still needs further investigation.

Finally, in Table~\ref{table:prosCons_CSI} we summarize some \varTableDes\ of selected solutions targeting CSI acquisition.

\subsection{Lessons Learned}\label{sec:CSI_LL}
Antenna calibration methods in user-centric cell-free settings is still not well studied. While the literature has focused on TDD systems, more studies are still needed for FDD systems, because both TDD and FDD systems are usually standardized in mobile networks. Current studies for FDD in cell-free networks have shown that more than 60\% reduction in the uplink feedback overhead (for the CSI estimated at the user, i.e., in the downlink) can be achieved compared to the conventional scheme relying on CSI feedback~\cite{cellFreeFDD9014542}. In this context, exploitation of angle reciprocity in FDD systems may be less effective in millimeter wave communication because the uplink and downlink carrier frequencies could be relatively far from each other due to the availability of large bandwidth.

For downlink CSI estimation schemes, the user can estimate the effective SISO precoded channel. If the user needs to estimate the precoded channel from each serving DU, the pilot sequence used in the downlink will be dependent on the number of DUs. This could be a problem when the number of DUs is very large. Thus, unless some alternate CSI scheme is developed, only the effective SISO precoded channel could be estimated by the user. Additionally, FFDNet and DnCNN can be used to reduce the pilot training period~\cite{DeepLearning8815888, FFDNet8365806}. This area still needs more investigation.

Pilot assignment policies can keep co-pilot users as far as possible from each other and hence decrease pilot contamination. For example, a well-structured PA policy can reduce the required length of the pilot sequence by a factor of $3$–$3.75$ while achieving negligible contamination~\cite{randomVsStructuredPilots8403508}. PA policies should take into consideration the overhead and computation complexity, otherwise the policy may not be feasible for an actual deployment. The reason for this is that if the PA policy can decrease the length of the pilot sequence but still wastes considerable time to be constructed, then there is no point in decreasing the pilot sequence in the first place. In actual deployment, the time wasted in optimizing the PA policy will be added as a communication overhead, which has the same effect of using a longer pilot training period (i.e., longer pilot sequence). Non-centralized PA policies could be also one way of making the policy feasible to be deployed, also, non-orthogonal PA is a promising scheme in ``crowd scenarios''.

Power control for the pilots is another direction for optimizing the performance of CSI estimation. Users with high mobility will suffer from channel aging that will increase the MSE of channel estimation. Incorporating channel aging effects requires using a time-varying model for the channel that relates the temporal autocorrelation function of the channel with the propagation geometry, velocity, frequency, and antenna characteristics~\cite{1512123}. Trace-based and synthetic models, such as the RWP~\cite{johnson1996dynamic}, can be used to emulate the movement of users~\cite{roy2011handbook, 5343061}. Moreover, high-mobility requires the users to frequently update their serving cluster, this becomes a problem in cell-free networks that usually includes a dense deployment of DUs. High mobility could also result in some synchronization errors.

\section{Formation of Serving Cluster}\label{sec:servingCluster} 
There are many reasons to construct serving clusters for the users. Serving clusters are needed because of the limited capacity of the DUs, the high complexity of signal processing and computational load needed by each DU when serving all users, the high fronthaul load, and the CSI estimation overhead. Furthermore, due to path loss, it is useless to serve the users by all the DUs in the network. Next, we identify two main frameworks to construct serving clusters in a user-centric clustering scheme.

\subsection{Utility-based Clustering}
Metrics used to construct the serving clusters can provide different degrees of complexity and performance optimality. Thus, a trade-off between these two metrics can be constructed based on the network scenario and the type of service provided.

It is important to note that, both users served by many DUs and conventional users may exist in the same network, especially in locations where dense deployment of DUs is not feasible. For example, users that are experiencing a good signal from a close-by BS may not need to be served by many DUs. In this context, if the signal level is above the threshold required by the user's application, increasing the size of the serving cluster will not introduce any enhancements but may, rather, introduce delay. Hence, a method to determine when the user will be a conventional one or not, should be defined. In this context, network slicing~\cite{NetworkSlicing8320765} may play a role in determining the size of the serving cluster.

One simple idea would be to exclude the DUs that cannot provide a signal power comparable to that provided by the nearest DUs~\cite{8491240}. The authors of~\cite{8281464} develop an iterative elimination algorithm for power control to remove ineffective BSs for each user.

Another topic is how to divide the resources between the conventional users and the users served by a cluster of DUs. The survey in~\cite{7839266} analyzes different methods to construct the serving cluster in cooperative networks. As stated, optimal clustering is a key challenge in cooperative networks. Additionally, the work identifies three main clustering schemes; static, semi-dynamic and dynamic clustering. Each scheme provides a different trade-off between overhead and interference cancellation capabilities. Even in terms of performance, some optimized clustering techniques can maximize the SE but degrade the energy efficiency (EE).

In general, user-centric clustering is a more dynamic clustering scheme than the cell-centric approach, because it can adopt different clustering strategies, however, the clustering process is more complex~\cite{6920005}. The work in~\cite{ammarC_RA_UC} defines the serving cluster in two stages. In the first stage the cluster for each users is defined based on a large-scale fading threshold. Then, this cluster is further optimized during the optimization of user-scheduling based on a weighted sum rate utility. Two serving cluster formation techniques are proposed in~\cite{EnergyEfficiency8097026} which depends on either the received power or the large-scale fading statistics.

The study in~\cite{clusters6415394} considers BSs clustering under the user-centric clustering scheme, where the cluster size is controlled through a single penalty parameter introduced in a regularized weighted MSE minimization problem. Reference~\cite{6328484} proposes a greedy joint clustering selection algorithm based on the WSR. The authors assume the existence of a set of candidate clusters based on the large-scale fading statistics, which can be either based on an absolute threshold or a relative one, i.e., with respect to an anchor BS.

\begin{figure*}[t]
	\centering
	\includegraphics[width=1\textwidth]{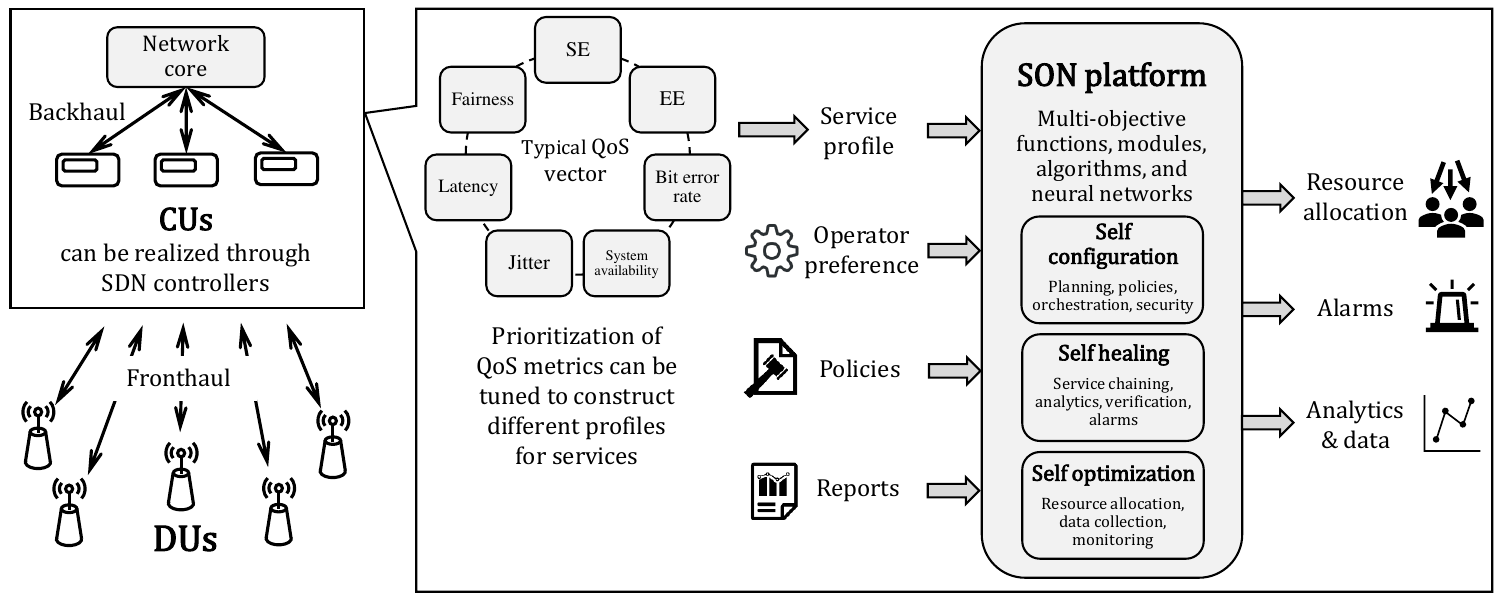}
	\caption{Platforms such as self-organizing networks (SON) used as a network management platform for user-centric cell-free MIMO networks. SDN also can be employed to support dynamic CU-DU assignment and flexible population of control plane from the CUs to the DUs. We also show a typical Quality of Service (QoS) vector that quantifies the network performance and can be used to construct service profiles for users.}
	\label{fig:QoSVector}
	\vspace{-1em}
\end{figure*}

\subsection{Lessons Learned}\label{sec:ServingCluster_LL}
An explicit construction of serving clusters is what distinguishes a user-centric scheme from a general cell-free network. The construction of the serving clusters limits the number of users served by each DU, hence limits the computational complexity, CSI estimation and load on the DU. It also decreases the load on the fronthaul and removes the DUs that cannot contribute to the users' desired signal due to a large path loss.

The construction of the serving clusters can be based on either offline or online parameters. Additionally, static, semi-dynamic and dynamic clustering can be adopted to prioritize some metric. Utility-based optimization frameworks can be built for the users under different scenarios.

Finally, in Table~\ref{table:prosCons_ServingClusters} we summarize some \varTableDes\ of selected solutions targeting the formation of serving clusters.

\begin{table}[t!]
	\footnotesize
	\centering
	\begin{tabular}{|>{\raggedright\arraybackslash}p{0.08\textwidth}|>{\raggedright\arraybackslash}p{0.14\textwidth}|>{\raggedright\arraybackslash}p{0.12\textwidth}|>{\raggedright\arraybackslash}p{0.05\textwidth}|}
		\hline
		\hline
		\multicolumn{1}{|l|}{ \textit{\textbf{Solution}}} & \multicolumn{1}{l|}{ \textit{\textbf{Pros}}} & \multicolumn{1}{l|}{ \textit{\textbf{Cons}}} & \multicolumn{1}{l|}{\textit{\textbf{Ref.}}}\\
		\hline
		\hline
		Weighted sum-rate utility & Consider fairness among users & Most problems are NP-hard & \cite{ammarC_RA_UC} 
		\\
		\hline
		Combinatorial search & Does not require smooth utility function or convexity & No guarantee to find an optimal solution, polynomial-time complexity is only for limited scenarios, only works when the feasible set of the problem is discrete & \cite{Ahmad9084256, ammarDistributed_RA_UC, TabuSearch5957382} 
		\\
		\hline
		Static clustering & Low computational complexity & No consideration for optimality & \cite{7839266}
		\\
		\hline
	\end{tabular}
	\caption{Solutions related to formation of serving clusters. Please check~\cite{7839266} for an extensive survey on clustering schemes.}
	\label{table:prosCons_ServingClusters}
	\vspace{-2em}
\end{table}



\section{Resource Allocation}\label{sec:resourceAllocation}
Optimizing resource allocation is a preeminent tool to achieve high performance in cell-free MIMO networks. The optimization can target different QoS metrics, including boosting the SE of the network to maximizing the EE, or minimizing the transmit power under a QoS constraint, etc. This is usually described through maximizing a utility or minimizing a cost function under some constraints. A typical QoS vector capturing various facets of the optimization problem is shown in Fig.~\ref{fig:QoSVector}. We also integrate our discussion on the distributed SDN and management platforms in the same figure, where \emph{optimized} resource allocation is seen as a main component in the network.

\subsection{Spectral Efficiency (SE)}
SE is the main metric targeted in any QoS vector, as it is closely related to the capacity of the communications. The studies in~\cite{PrecodingDistrib2020Atzeni, 6151868, WSRMSE8307115, 6920005} optimize beamforming strategies through minimizing the weighted sum MSE, which is an objective function that is easier to work with than the WSR, but it yields a minor penalty in terms of sum-rate performance~\cite{PrecodingDistrib2020Atzeni}. The analysis in~\cite{differentCooperationLevels8845768} focuses on receiver cooperation on the uplink using MMSE-based schemes under different levels of cooperation between the DUs and their CU. An important conclusion is that MMSE-based schemes outperform conjugate beamforming in cell-free massive MIMO networks irrespective of the level of cooperation among the access points and their number of antennas. Moreover, \cite{bjornson2019scalable9064545} studies beamforming with dynamic clustering including CSI estimation and addressing scalability.

The WSR is an interesting utility function, because it describes the network performance while accounting for other metrics, such as fairness among the users represented as weights for the rates. We note the fairness metric can be implemented through different methods, e.g., proportional fair rate allocation~\cite{yu2011adaptive}. Moreover, implementing the fairness amongst users is a milestone in scheduling problems~\cite{ammarC_RA_UC, ammarDistributed_RA_UC}. A WSR problem is usually of the form 
\begin{subequations}\label{eq:weightedSumRate}
	\begin{align}
		\stepcounter{probNum}
		(\mathrm{P}\arabic{probNum})\quad
		\underset{{\bf x}}{\mathrm{maximize}} \quad & \sum_{u \in \mathcal{U}} \delta_{u} \log\left( 1 + 
		\gamma_{u}\left({\bf x}\right)
		\right) & 
		\label{eq:weightedSumRate_obj}
		\\
		\mathrm{subject\ to}\quad 
		& {\bf x} \in \mathcal{X},
		&
		\label{eq:weightedSumRate_constraints}
	\end{align}
\end{subequations}
where $\mathcal{U}$ is the set of users, $\delta_{u}$ is the fairness weight for user $u$, $\gamma_{u}$ is the SINR of user $u$, ${\bf x}$ is vector of the optimization or decision variables, and $\mathcal{X}$ is a set of constraints. The decision variables often include transmit and receive beamformers, user-scheduling, and allocated power (which can be related to user-scheduling as well).

The investigation in~\cite{Cell-freeWSR2005.12331} employs the WSR to optimize the beamforming by approximating the problem by a conic-quadratic program based on the inner approximation framework~\cite{marks1978general}, where the authors use a lower bound for the logarithm function to obtain a local optimum. The authors also report the globally optimal solution using a branch-reduce-and-bound optimization framework. Unfortunately, this approach mostly has exponential complexity with respect to the problem size~\cite{tuy2005monotonic}.

An alternative utility can use the MSE covariance between the transmitted complex data symbol $s_{u}$ for user $u$ and the estimated one $\hat{s}_{u}$, which is denoted as ${\rm e}_{u} \triangleq \mathbb{E}_{s_u}\left[ | \hat{s}_{u} - s_{u} |^2 \right]$. This metric can be used to construct an MSE-based optimization problem that is similar to~\eqref{eq:weightedSumRate} and is defined as~\cite{6920005}
\begin{subequations}\label{eq:MSE}
	\begin{align}
		\stepcounter{probNum}
		(\mathrm{P \arabic{probNum}})\quad
		\underset{{\bf x}}{\mathrm{maximize}} \quad & \sum_{u \in \mathcal{U}} \delta_{u} \left( \rho_u {\rm e}_{u}\left({\bf x}\right) - \log \left(\rho_u\right)
		\right) & 
		\label{eq:MSE_obj}
		\\
		\mathrm{subject\ to}\quad 
		& {\bf x} \in \mathcal{X},
		&
		\label{eq:MSE_constraints}
	\end{align}
\end{subequations}
where $\rho_u$ is an introduced MSE weight. In most cases, the problem in~\eqref{eq:MSE} can be formulated as a second order cone program (SOCP)~\cite{6920005}, and it can be solved using solvers such as CVX~\cite{grant2014cvx}.

Alternative problems to optimize the SE can also include max-min fractional problems of the form $\max_{\bf x} \min_n A_n({\bf x})/B_n({\bf x})$, sum-of-ratios problems ($\max_{\bf x} \sum_{n} A_n({\bf x})/B_n({\bf x})$), product-of-ratios problems ($\max_{\bf x} \prod_{n} A_n({\bf x})/B_n({\bf x})$), and weighted product of SINRs ($\max_{\bf x} \prod_{u \in \mathcal{U}} \delta_{u}	\gamma_{u}\left({\bf x}\right)$). We emphasize that this list is not meant to be exhaustive. More details on problem types and optimized metrics that could be targeted can be found in~\cite{bjornson2013optimal}.


\begin{table*}[t!]
	\footnotesize
	\centering
	\begin{tabular}{|p{0.15\textwidth}|p{0.3\textwidth}|p{0.3\textwidth}|p{0.1\textwidth}|}
		\hline
		\hline
		\multicolumn{1}{|l|}{ \textit{\textbf{Topic}}} & \multicolumn{1}{l|}{ \textit{\textbf{Challenges}}} & \multicolumn{1}{l|}{ \textit{\textbf{Proposed Solutions}}} & \multicolumn{1}{l|}{\textit{\textbf{Reference}}}\\
		\hline
		\hline
		\multirow{16}{*}{Fronthaul/backhaul} & \multirow{16}{*}{Limited capacity, scalability} & Radio stripes system, serial fronthaul & \cite{frenger2019antenna, interdonato2019ubiquitous} 
		\\
		\cline{3-4}
		& & Joint-access fronthaul design, Integrated access and backhaul (IAB) & \cite{WSRjointAccessBack8786917, IABStochastic2019}
		\\
		\cline{3-4}
		& & Partial-centralization: Different levels of coordination between the DUs and CUs, e.g., DUs having different baseband functionalities, DUs with different coordination levels & \cite{differentCooperationLevels8845768, 8693830}
		\\
		\cline{3-4}
		& & Conjugate beamforming (downlink) and matched filtering (uplink) when channel hardening and favorable propagation applies &
		\cite{cellFreeVersusSmallCells7827017}
		\\
		\cline{3-4}
		& & Local beamforming with a second layer decoding at the CU & \cite{RicianFadingPhaseShifts8809413}
		\\
		\cline{3-4}
		& & Data compression schemes and source coding& \cite{maxMinRate8756286, 8693830, 8730536, EE9212395, HardwareImpairements8891922, limitedFronthaulMmwave8678745}
		\\
		\cline{3-4}
		& & Distributed SDN in multiple-CU network & 
		\cite{distributedSDN8187644, distributedSDN6838330}
		\\
		\cline{3-4}
		& & Free space optical (FSO) links & 
		\cite{UC_CellFreOpticalFronthaule020}
		\\	
		\hline
		\multirow{12}{*}{CSI estimation} & \multirow{9}{*}{Non-orthogonal pilots, pilot overhead} & Pilot assignment policies using users' clustering algorithms, orthogonality of channels, Bayesian approach, graph coloring, and combinatorial search & \cite{ashikhmin2017pilot, pilotAssign9178782, randomVsStructuredPilots8403508, bjornson2019scalable9064545, CSI6415397, graphColoringUsercentric7080877}
		\\
		\cline{3-4}
		&
		&
		Channel reciprocity for TDD system, angle reciprocity for FDD system, feed back for the path gain information
		&
		\cite{PowerClusteringCellFreeFDD8968400, cellFreeFDD9014542}
		\\
		\cline{3-4}
		& & Large-scale fading decoding (LSFD) & \cite{powerControlCellFree7917284, Scalable9174860, 8901196, HardwareImpairements9004558, RicianFadingPhaseShifts8809413, differentCooperationLevels8845768}
		\\
		\cline{2-4}
		&
		Hardware impairments
		&
		Second layer decoding at the CU, different forward strategies to the CU
		&
		\cite{HardwareImpairements8891922, HardwareImpairements8476516, HardwareImpairements9004558}
		\\
		\cline{2-4}
		& Computational complexity & Deep learning & \cite{DeepLearning8815888, FFDNet8365806}
		\\
		\hline
		\multirow{11}{*}{Pilot Assignment} & \multirow{8}{*}{Pilot reuse may be unavoidable} & Structured PA policies: clustering algorithms for users to reuse non-orthogonal pilots, graph coloring-based algorithms, power control for the reused pilots, heap-structure based strategy.& \cite{randomVsStructuredPilots8403508, ashikhmin2017pilot, pilotPowerControl8450041, PiltAssigGraphColoring7217795, PowerControlConventional7801046, masoumi2020cell, uplinkPilotcellFreeMassiveMIMO2020, ammarC_RA_UC, ammarDistributed_RA_UC, 8360138, cellFreeVersusSmallCells7827017, GraphCodePilotContamin8487005, pilotAssign9178782, Scalable9174860, GraphColoringPilot9110802, FullDuplex9110914}
		\\
		\cline{3-4}
		& &
		Based on a pilot utility metric, use statistical CSI for some users without assigning them pilots. 
		&
		\cite{8385475, pilotAssign9178782}
		\\
		\cline{2-4}
		& Need for centralized control & --- & \cite{randomVsStructuredPilots8403508}
		\\
		\hline
		Resource allocation 
		& Optimizing beamforming, power, scheduling, etc & \emph{Addressed in a separate table} &
		\\
		\hline
		\multirow{9}{*}{Signal synchronization} & \multirow{9}{*}{\parbox{0.3\textwidth}{Stringent synchronization across a widely distributed network which may include consumer-grade equipment}} & Coarse synchronization using Precision Time Protocol (PTP) (IEEE 1588v2) which is better than Network Time Protocol (NTP) & \cite{IEEE1588_7949184}
		\\
		\cline{3-4}
		& & Complex equalization and advanced synchronization techniques between the DUs, quasi-synchronous systems & \cite{Synchronization4287203}
		\\
		\cline{3-4}
		& & Over-the-air synchronization and calibration protocols & \cite{synchronizationCalibration6760595}
		\\
		\cline{3-4}
		& & Deep neural networks (DNN) & \cite{nonReciprocityCellFreeDNN9098852}
		\\
		\hline
		\multirow{11}{*}{Latency} & \multirow{11}{*}{\parbox{0.3\textwidth}{Study delay-aware scenarios and communications with deadlines, control the communications delay, synchronization issues}} & Lyapunov optimization techniques,	queueing theory, optimize data to be admitted for transmission queues  & \cite{8901196}
		\\
		\cline{3-4}
		& &
		Partially observable Markov decision process 
		& \cite{DelayAware6180015}
		\\
		\cline{3-4}
		& & Discrete-time Markov model, Markov fluid model &
		\cite{9130689}
		\\
		\cline{3-4}
		& & Mobile edge computing (MEC) &
		\cite{8974591, mobileEdgeComputing7901477, 7498684}
		\\
		\cline{3-4}
		& & Punctured scheduling &
		\cite{puncturedScheduling9011578}
		\\
		\cline{3-4}
		& & Protocols such as Precision Time Protocol (PTP) &
		\cite{IEEE1588_7949184, synchronizationCalibration6760595}
		\\
		\hline
		\multirow{5}{*}{Scalability} & \multirow{5}{*}{\parbox{0.3\textwidth}{Scaling of complexity with respect to system size}} & Local resource allocation, local partial ZF (PZF), local protective partial ZF (PPZF) & \cite{Scalable9174860, bjornson2019scalable9064545, LocaPartialZFBF9069486}
		\\
		\cline{3-4}
		& &
		Statistical approaches
		&
		\cite{7837706, powerControlCellFree7917284, 8901196, HardwareImpairements9004558}
		\\
		\cline{3-4}
		& &
		Distributed schemes
		&
		\cite{ammarDistributed_RA_UC, differentCooperationLevels8845768, PowerAlloc8630677, maxMinRate8756286, LocaPartialZFBF9069486, DistribResourceAllo7676375, EEdistributedAntenna7482747}
		\\
		\hline
		\multirow{3}{*}{Mobility} & \multirow{3}{*}{\parbox{0.3\textwidth}{Channel aging, serving cluster reformation, synchronization errors}}
		& Fractional power control, channel prediction~\cite{machineLearningChannelAging8979256} & \cite{CSIAgingzheng2020cell, RicianFadingPhaseShifts8809413}
		\\
		\cline{3-4}
		& &
		Optimal selection of the frame duration
		&
		\cite{channelAgingMassiveMIMO8122014}
		\\
		\hline
		\hline
	\end{tabular}
	\caption{Synopsis of the challenges with proposed solutions.}
	\label{table:challenges}  
	\vspace{-1em} 
\end{table*}


\textit{Power control} algorithms are needed to balance fairness, latency, and network throughput. The analysis in~\cite{cellFreeUserCentricPower8901451} performs power allocation by formulating both maximization of a lower-bound for sum rate and minimum-rate problems. We note that the min-rate maximization is very aggressive in terms of user fairness when compared to WSR problems; it does not provide the flexibility to control the fairness, so the optimal solution, if feasible, provides equal throughput to all the users in the network. Often the max-min rate problem can be re-written as a convex program allowing us to obtain a solution for the problem. Notably, the work shows that the user-centric clustering scheme outperforms the cell-free massive MIMO deployment in which no serving clusters are defined. The authors also report that the latter scheme is not scalable and the power allocation routine requires many iterations to converge, because the number of variables to be optimized is larger than the user-centric cell-free scheme. In addition, the study in~\cite{masoumi2020cell} solves two power allocation problems for cell-free massive MIMO underlaid with D2D communications using Geometric Programming (GP) to maximize either the sum rate or the weighted product of SINRs. The investigation in~\cite{maxMinRate8756286} maximizes the minimum rate through optimizing beamforming. 

The authors of~\cite{8901196} study joint power control and scheduling in the uplink of both co-located and cell-free massive MIMO networks. Additionally, the investigation in~\cite{powerControlCellFree7917284} considers a near-optimal power control algorithm that is simpler than the max–min power control problem for cell-free massive MIMO networks. Hence, implementing a cell-free distributed computation of the power control coefficients comes at the cost of reduced performance. The authors use LSFD that achieves a two-fold gain over conjugate beamforming in terms of $95\%$-likely per-user SE.

\textit{The LSFD scheme} is also studied in~\cite{HardwareImpairements9004558} under hardware impairments. Other studies, e.g.,~\cite{RicianFadingPhaseShifts8809413}, also employed LSFD as a second layer decoding that is performed at a central CU to mitigate inter-user interference. The scheme mainly depends on applying conjugate beamforming at the DUs, then forwarding the resulting signal to the CU, where a weighted combining is performed using some optimized weights. Mathematically, the signal on the uplink is jointly decoded at CU $c$ as follows
\begin{align}\label{eq:JointDetection}
	\bar{y}_{cu} = \sum_{b \in \mathcal{C}_u} a_{cbu}^* {\bf \hat{h}}_{bu}^T {\bf y}_{bu}
\end{align}
where $\mathcal{C}_u$ is the serving cluster of user $u$ (i.e., the serving DUs), ${\bf y}_{bu} \in \mathbb{C}^{M \times 1}$ is the uplink signal received at DU $b$ from user $u$, with $M$ representing the number of antennas per DU, ${\bf \hat{h}}_{bu} \in \mathbb{C}_{M \times 1}$ is the estimated channel between the two peers, and $a_{cbu}$ is a combining weight that can be optimized at the CU to enhance the efficacy of joint detection, e.g., to maximize the effective SINR. As can be seen in~\eqref{eq:JointDetection}, the serving cluster of the user is assumed to be under the control of a single CU, thus~\eqref{eq:JointDetection} corresponds to a network with a single CU. Hence, it is not clear how such an approach can be implemented for a network with multiple CUs. The results in~\cite{RicianFadingPhaseShifts8809413} show that LSFD improves the uplink SE under different considered LoS phase-aware channel estimators that include the phase-aware MMSE, the linear MMSE (LMMSE), and the least-square (LS) estimator. Altogether, LSFD seems a promising scheme for the uplink of cell-free massive MIMO networks.

The authors of~\cite{8281464} study power control and load balancing in the uplink by solving three problems; minimization of power consumption, maximization of minimum QoS, and maximization of sum SE. The authors use either conjugate or ZF beamforming and develop an iterative elimination algorithm to remove ineffective serving DUs for each user. Their results show that the developed method provides better performance than both maximum signal to noise ratio (SNR) association and full-set joint transmission, especially in the high QoS regime. Similarly, \cite{9236635} optimizes max-min power allocation under correlated and uncorrelated channel fading. Interestingly, the CSI estimation in the uplink is performed under limited channel covariance knowledge by using element-wise MMSE~\cite{8094949}.

The authors of~\cite{clusteringOverlappingCoordinationClusters5594709} use a numerical linear soft interference nulling technique that can outperform or at least provide comparable performance to ZF with full network coordination at moderate SNRs. However, the technique requires sharing CSI between the DUs.

\textit{Non-orthogonal multiple-access (NOMA)} schemes have also been studied for cell-free communication. This solution employs power domain multiplexing and SIC to support multiple users with the same time-frequency resources~\cite{NOMA6736749}, hence boosting the capacity of the system at the expense of added interference. We remind the reader that SIC has also been suggested for non-coherent transmissions from multiple DUs to a single user. The study in~\cite{NOMACellFree8368267} shows that a NOMA scheme with cell-free MIMO networks allows the system to serve more users overall compared to an orthogonal (i.e., OMA) counterpart. However, the achievable sum rate using NOMA in cell-free networks can be lower than that in an orthogonal scheme even in the regime of low number of users due to both intra-cluster pilot contamination and error propagation of imperfect SIC.

The authors of~\cite{NOMA8895763} show that the switching between NOMA and OMA can enhance system performance. The need for switching depends on both the length of the coherence time of the channel and the total number of users. The study in~\cite{CellFreeNOMA9024101} investigates the performance of NOMA in cell-free massive MIMO networks using different types of linear precoders such as conjugate beamforming, full-pilot zero-forcing (fpZF) and modified regularized ZF (mRZF) precoders. We note that mRZF and fpZF allow each DU to construct the beamformers using only its local CSI. The results show that with perfect SIC, the mRZF and fpZF precoders significantly outperform conjugate beamforming in both NOMA and OMA deployments.

In contrast, the authors in~\cite{CellFreeNOMA9130101} study the uplink performance with optimal fronthaul combining by formulating a max-min power allocation problem. Further, the investigation in~\cite{CellFreeNOMA8957510} derives the achievable rate in a primary massive MIMO network and an underlaid cell-free massive MIMO network (i.e., secondary network) both implementing the NOMA scheme. The authors mention that the intuition from such an underlay deployment is that massive MIMO concepts are expected to be widely deployed in 5G mobile networks, hence in some scenarios it can be exploited when deploying the cell-free scheme.

If the studied problem is \textit{combinatorial}, intelligent scheduling algorithms can be used to partition time, frequency and spatial resources. In this regard, combinatorial search tools like mixed integer programming (MIP)~\cite{pochet2006production} and constraint programming (CP)~\cite{rossi2006handbook} can be useful. Importantly, these tools can, sometimes, efficiently solve combinatorial problems that are, usually, computationally hard. This is because the problem consists of a set of variables that can take a corresponding set of possible values (called the domain), a set of constraints, and some sort of cost function. The solution is found by an efficient systematic search, where the variables are labeled and the constraints actively prune the domains of the unassigned variables during the search. Many techniques are employed to perform an efficient search, which include backtracking, domain filtering, constraint propagation, advanced search strategies like branch-and-bound, and the concept of global constraints~\cite{hooker2007integrated}.

While not an optimal approach in, e.g., sum-rate sense, \textit{conjugate beamforming} seems to be used in the literature to enhance the system scalability, cope with single-antenna DUs, make complex analysis easier, or limit the load on the fronthaul~\cite{cellFreeVersusSmallCells7827017}. Importantly, conjugate beamforming can be optimal for massive MIMO scenarios~\cite{6415388}, when channel hardening applies. However, conjugate beamforming cannot suppress interference well~\cite{RicianFadingPhaseShifts8809413}, hence other local schemes like the LSFD~\cite{7837706, powerControlCellFree7917284, Scalable9174860, 8901196, HardwareImpairements9004558}, local MMSE combining~\cite{differentCooperationLevels8845768}, and the weighted MMSE~\cite{MSE5756489, ammarC_RA_UC} could be a better candidate for beamforming in cell-free MIMO networks. In this regard, in the weighted MMSE approaches~\cite{MSE5756489}, the beamformers are constructed using the channels between each DU and the users in the network, and it seems to provide superior performance~\cite{ammarC_RA_UC} compared to local ZF and conjugate beamforming. The authors of~\cite{8599043} present a modified conjugate beamforming scheme for cell-free massive MIMO networks. The scheme eliminates the self-interference without using matrix inversions at the expense of fading-rate coordination among the DUs.

Other schemes have also been considered: the high density of the deployed DUs in cell-free massive MIMO networks raises the possibility of \textit{LoS communication}. Based on this insight, the authors in~\cite{201114573} analyze uplink performance using a probabilistic LoS channel model~\cite{ITU:P.1410_5, 8859647}. Their results show that increasing the access point density from $128$ to $1024$ per ${\rm km}^2$ dramatically increases the LoS probability from $40\%$ to $95\%$. The authors also study the system performance under either joint or stream-wise processing of user data at the CU, where with the availability of accurate CSI, the stream-wise combining can approach the performance of joint decoding.

Based on channel hardening, the study in~\cite{multicastCellFree8017421} investigates \textit{multigroup multicasting }in cell-free massive MIMO networks using conjugate beamforming based on a short-term power constraint. The scheme seems to be effective only when the number of served users is small. The study in~\cite{FullDuplex9110914} analyzes the deployment of in-band \textit{full-duplex} in cell-free massive MIMO networks. As stated by the authors, the deployment of full-duplex communication has a limitation on self-interference cancellation, which makes it advisable for short-range communications provided by cell-free massive MIMO systems. The authors optimize power control and access point-user association to optimize the SE and EE. The results show that the cell-free scheme outperforms massive MIMO and small-cell network schemes. Similarly, the authors in~\cite{FullDuplex8943119} study in-band full-duplex in cell-free massive MIMO where MMSE joint detection and joint scheduling are proposed to mitigate the self interference.

\begin{table*}[t]
	\centering
	\begin{tabular}{|p{0.15\textwidth}|p{0.3\textwidth}|p{0.3\textwidth}|p{0.1\textwidth}|}
		\hline
		\hline
		\multicolumn{1}{|l|}{ \textit{\textbf{Optimized metric}}} & \multicolumn{1}{l|}{ \textit{\textbf{Problem Type}}} & \multicolumn{1}{l|}{ \textit{\textbf{Approach}}} & \multicolumn{1}{l|}{\textit{\textbf{Reference}}}\\
		\hline
		\hline
		\multirow{2}{*}{}
		$\!\!\!\!\!\!$
		\multirow{9}{*}{\parbox{0.15\textwidth}{\small Beamforming and precoding}} & \multirow{7}{*}{\small Weighted Sum Rate (WSR)} & Minimize the weighted sum mean square error (MSE) & \cite{6151868, PrecodingDistrib2020Atzeni, 6920005, 7581201, EE8777141}
		\\
		\cline{3-4}
		& & SINR convexification based concave lower-bound approximation & \cite{WSRjointAccessBack8786917}
		\\
		\cline{3-4}
		& & Approximate the problem by a conic-quadratic program based on the inner approximation framework~\cite{marks1978general} & \cite{Cell-freeWSR2005.12331}\\
		\cline{2-4}
		& {\small Minimum-rate maximization} & Geometric programming (GP) & \cite{cellFreeUserCentricPower8901451, PowerAlloc8630677, maxMinRate8756286, powerControlCellFree7917284}
		\\
		\cline{2-4}
		& {\small Sum-rate maximization} & Sequential optimization framework &
		\cite{cellFreeUserCentricPower8901451, clusteringOverlappingCoordinationClusters5594709}
		\\
		\hline
		\multirow{10}{*}{\parbox{0.15\textwidth}{\small Power*}} & \multirow{10}{*}{\parbox{0.3\textwidth}{\small Sum rate, weighted max product of SINRs, max-min SINR, max-min SE, target SINR-constrained
		}}
		& GP & \cite{masoumi2020cell, LocaPartialZFBF9069486, HardwareImpairements8891922, 8281464, Rician9099874}\\
		\cline{3-4}
		&
		& Bisection search, second-order cone program & \cite{cellFreeVersusSmallCells7827017, cellfreeIofTPilots2020, 9236635, uplinkPilotcellFreeMassiveMIMO2020, NOMA8895763, 9079911}\\
		\cline{3-4}
		& & Sequential convex approximation & \cite{DLTrainingCSI8799031, Rician9099874, CellFreeNOMA9130101}
		\\
		\cline{3-4}
		& & Successive lower-bound (upper-bound) maximization (minimization) method~\cite{SuccessiveLowerBoundRazaviyayn2013unified, cellFreeUserCentricPower8901451} &
		\cite{EnergyEfficiencyMmwave8676377}
		\\
		\cline{3-4}
		& & Standard semi-definite programming (SDP)  & \cite{PowerClusteringCellFreeFDD8968400}
		\\
		\cline{3-4}
		& & Path-following algorithm that invokes one simple convex quadratic program at each iteration. & \cite{8360138}
		\\
		\hline
		\multirow{5}{*}{\parbox{0.15\textwidth}{\small User-scheduling }} & \multirow{3}{*}{\parbox{0.3\textwidth}{\small Energy efficiency maximization}} & Hierarchical decomposition technique, iterative successive convex approximation, combinatorial search & \cite{EE8695055, FullDuplex9110914}
		\\
		\cline{2-4}
		& \multirow{2}{*}{\parbox{0.3\textwidth}{\small WSR}} & MSE, weighted $\ell_1$-norm, fractional programming &
		\cite{6920005, ammarC_RA_UC, ammarC_RA_UC_conf}
		\\
		\hline
		\multirow{10}{*}{\parbox{0.15\textwidth}{\small Energy efficiency}} & \multirow{10}{*}{\parbox{0.3\textwidth}{\small Max-min SINR, target SINR-constrained, power-budget minimization subject to QoS constraints}} & Bisection search, second-order cone program, successive convex approximation, GP & \cite{EnergyE2020towards, EnergyEfficiency8097026, powerAllocEnergyEfficiency9136914, EnergyEfficiency8689095, EE7900388, cellfreeIofTPilots2020, EnergyEfficiency8781848, EE9212395, EE8695055, EnergyEfficiencyMmwave8676377, EECellFreeUCMMwave8292302, EECellFreeUCMMwave8516938, UC_CellFreOpticalFronthaule020, Rician9099874}
		\\
		\cline{3-4}
		& & Dual-decomposition-based method, bisection search method, fractional programming~\cite{zappone2015energy}, Dinkelbach method~\cite{dinkelbach1967nonlinear} &  \cite{EEdistributedAntenna7482747}
		\\
		\cline{3-4}
		& & Gradient projection method, bounds of the Lambert function~\cite{LambertFuncthoorfar2008inequalities}, Karush–Kuhn–Tucker (KKT) conditions based &
		\cite{EE8777141}
		\\
		\cline{3-4}
		& & IRS-aided cell-free network & \cite{IRS9352948, IRS9363171, IRS8910627, IRS9308937, IRS9298843}
		\\
		\hline
		\multirow{9}{*}{\parbox{0.15\textwidth}{\small Cluster formation}} & {\small Sum SE} & Power control & \cite{6920005, 8281464, bjornson2019scalable9064545}
		\\
		\cline{2-4}
		& \multirow{2}{*}{\parbox{0.3\textwidth}{\small WSR}} & Scheduling, fractional programming, weighted MSE minimization & \cite{ammarC_RA_UC, clusters6415394, ammarC_RA_UC_conf}
		\\
		\cline{2-4}
		& \multirow{3}{*}{\parbox{0.3\textwidth}{\small Management platform}} & self-organizing network (SON) platform, dynamic clustering algorithms, Self organizing CoMP, hierarchical clustering via minimax linkage & \cite{3GPP:TR36.902}
		\\
		\cline{2-4}
		& \multirow{3}{*}{\parbox{0.3\textwidth}{---}} &
		Based on offline parameter (e.g., large-scale fading), static, semi-dynamic and dynamic clustering to optimize a specific QoS
		& \cite{limitedFronthaulMmwave8678745, 8491240, clustersbasedonNetperf7105966, 6151868, 6328484}
		\\
		\hline
		\hline
		\multicolumn{4}{l}{*Corresponds to works that specifically addresses transmit power. Note that power can be implicitly optimized using beamforming weights.}
	\end{tabular}
	\caption{Resource allocation problems.}
	\label{table:ResourceAllocation}   
	\vspace{-1em} 
\end{table*}

\subsection{Energy Efficiency (EE)}
Cell-free massive MIMO networks outperform their collocated counterpart in terms of EE under a QoS constraint~\cite{EnergyEfficiency8097026}. The optimization of EE is necessary due to many reasons:
\begin{itemize}
	\item The increase in the data rates required by the next-generation mobile systems should not require a proportional increase in the transmission power. Without optimizing the EE, the demand in transmitted power may become unmanageable~\cite{alsharif2019energy}.
	\item 5G subscriptions are expected to reach $3.5$ billion subscriptions in the year $2026$~\cite{EricssonMobilityReport}. This exponential increase of connected wireless devices and the accompanying rapid expansion of wireless networks, causes environmental concerns, like increased carbon-dioxide ($\mathrm{CO}_2$) emissions and electromagnetic pollution~\cite{footprintComm5978416}.
	\item The additional economic expenses for deploying a densely distributed network that could consume high power and hence produce high electricity bills for the network operators~\cite{alsharif2019energy}.
\end{itemize}

The EE is computed in $\mathrm{bits/joules}$, and it is defined as a the ratio of the data rate ($\mathrm{bits/s}$) to the power consumption ($\mathrm{watts = joules/s}$). Both the radiated and the circuit power are usually included in the calculation of the power consumption~\cite{6815733} (else, the optimal EE usually occurs at zero power i.e., vanishingly small transmission rates). Resource allocation problems that maximize the EE can be naturally cast as a fractional program, hence making fractional programming~\cite{zappone2015energy, FR8314727, FR8310563} and the Dinkelbach method~\cite{dinkelbach1967nonlinear} suitable tools to solve such problems. 

The problem can be defined as follows~\cite{EnergyEfficiencyMmwave8676377}
\begin{subequations}\label{eq:EE}
	\begin{align}
		\stepcounter{probNum}
		(\mathrm{P \arabic{probNum}})\quad
		\underset{{\bf x}}{\mathrm{maximize}} \quad & \frac{ \sum_{u \in \mathcal{U}} \log\left( 1 + 
			\gamma_{u}\left({\bf x}\right)
			\right)}{ \sum_{b \in \mathcal{B}} D_b\left({\bf x}\right)} & 
		\label{eq:EE_obj}
		\\
		\mathrm{subject\ to}\quad 
		& {\bf x} \in \mathcal{X},
		&
		\label{eq:EE_constraints}
	\end{align}
\end{subequations}
where $D_b\left({\bf x}\right)$ is a power consumption function corresponding to DU $b$, and $\mathcal{B}$ is the set of the DUs. We note that there are many alternatives for the EE problem in~\eqref{eq:EE}, where some studies define the problem per access point~\cite{EnergyEfficiency8097026}. The power consumption accounts for the transmission power and other power consumption for the hardware components. For power consumption models, the readers can refer to~\cite{7437385, 6056688, 7456325, 7031971}.

Notably, the study in~\cite{EnergyE2020towards} maximizes the \textit{EE per unit area} (i.e., $\mathrm{bits/joules/km^2}$) by optimizing the pilot reuse factor and DU density. The results show that wisely choosing a pilot reuse factor enables lower interference, which seems to increase the EE per unit area up to a specific value. The difference between EE and EE per unit area is that, in the latter, an area-based power consumption model is used. Importantly, as indicated by the authors in~\cite{EnergyE2020towards}, using a cell-free approach necessitates the use of EE per unit area, because each user receives a joint transmission from many DUs. This is in contrast to conventional networks where a user is served by a single BS. Consequently, the definition of area SE, in which the received user rate is multiplied with the BS density does not hold in the cell-free scenario.

The work in~\cite{EnergyEfficiency8097026} optimizes EE under both a per-user SE and per-access point power constraint. Power allocation is also considered in~\cite{powerAllocEnergyEfficiency9136914, 8970501} with the choice of turning off some DUs to foster EE. The investigation in~\cite{EE7900388} maximizes EE in cell-free MIMO networks by optimizing the power control coefficients under ZF beamforming. Additionally, the work in~\cite{cellfreeIofTPilots2020} optimizes power allocation for an IoT network that is based on the cell-free massive MIMO framework. The developed scheme allows infrequent transmit power adaptation and provides a $40\%$ improvement in uplink and downlink rates and a $95\%$ in EE compared to a non-optimized cell-free scheme. Energy harvesting~\cite{EnergyHarvesting9201540} and sleep modes~\cite{9149862} are also investigated for cell-free massive MIMO.

The studies in~\cite{EnergyEfficiencyMmwave8676377, EECellFreeUCMMwave8292302, EECellFreeUCMMwave8516938} analyze the performance of both cell-free massive MIMO network and the user-centric architecture based on \textit{millimeter wave communication}. The authors optimize the power allocation so that the EE is maximized. We note that the combination of both user-centric cell-free networks and millimeter wave communication is promising because cooperative networks can alleviate the problem of poor connectivity suffered by millimeter wave communication~\cite{mmWaveConnectivity8579590, mmwave6732923}. The short range and large bandwidth of millimeter wave communication may also help in deploying a very dense cell-free network. This harmonization is simply based on the fact that millimeter wave communication requires a dense deployment of transmitters due to the high path loss, and a dense deployment requires coordination to prevent interference from getting out of control.


EE maximization seems to be extensively researched under different scenarios and optimization variables like power constrained problems that optimize both power and beams~\cite{EE8777141}, power-rate constrained problems that optimize both transmit covariance matrix and select active DUs~\cite{EEdistributedAntenna7482747}, and power-rate-capacity constrained problems that optimize both power allocation, antenna activation, and DU-user association~\cite{EE8695055}. Furthermore, the investigation in~\cite{EnergyEfficiency8781848} maximizes the EE under per-user power, fronthaul capacity and throughput requirement constraints.

Generally speaking, EE in cellular networks is a well researched topic~\cite{7446253}, and many of the techniques, e.g., environmental and RF energy harvesting~\cite{EnergyHarvestignEnvironmental7010878, EnergyHarvestignRF6951347} and offloading techniques~\cite{localCaching7150324}, developed for cellular networks, can benefit user-centric cell-free networks as well. Furthermore, due to the new architecture provided by the cell-free network, we now deal with far lower service distances and BS transmission power budgets than in conventional networks. On the other hand, distributing the available number of antennas on many DUs increases the circuit power consumption due to deploying more hardware. Hence, we must establish the implied trade-off through a robust study for the effect of distributing (or co-locating) the available antennas on more~(fewer)~DUs~on~the~EE. 

\begin{table*}
	\scriptsize
	\centering
	\begin{tabular}
		{|p{0.03\textwidth}|p{0.09\textwidth}|p{0.12\textwidth}|p{0.07\textwidth}|p{0.06\textwidth}|p{0.01\textwidth}|p{0.04\textwidth}|p{0.04\textwidth}|p{0.04\textwidth}|p{0.2\textwidth}|}
		\hline
		\hline
		& \multicolumn{1}{l|}{ \textit{\textbf{Model}}} & \multicolumn{1}{l|}{ \textit{\textbf{Topic}}} & \multicolumn{1}{l|}{\textit{\textbf{Control}}} & \multicolumn{1}{l|}{\textit{\textbf{Direction}}} &
		\multicolumn{1}{l|}{\textit{\textbf{Duplex}}} &
		\multicolumn{1}{l|}{\textit{\textbf{CU}}} &
		\multicolumn{1}{l|}{\textit{\textbf{LoS}}} &
		\multicolumn{1}{l|}{\textit{\textbf{Topology}}} & \multicolumn{1}{l|}{\textit{\textbf{Additional notes}}}\\
		\hline
		\hline
		\cite{masoumi2020cell} & Mathematical optimization  & Pilot-assignment, power allocation & Centralized & UL & TDD & Single & No & Random & 
		Cell-free massive MIMO underlaid with D2D communication, low resolution analog-to-digital converters (ADCs), reduce pilot contamination through greedy and graph coloring-based algorithms, GP.\\ 
		\hline
		\cite{cellfreeIofTPilots2020} & Mathematical optimization& Power allocation, EE & Centralized & DL, UL & TDD & Single & No & Random & IoT network organized as cell-free massive MIMO, non-orthogonal pilots, infrequent transmit power adaptation.
		\\
		\hline
		\cite{EnergyEfficiency8097026} & Mathematical optimization & EE & Centralized & DL & TDD & Single & No & Random & Conjugate beamforming, serving clusters.
		\\ 
		\hline
		\cite{DLTrainingCSI8799031} & Mathematical optimization & Power allocation & Centralized & DL, UL & TDD & Single & No & Random & Conjugate beamforming, global knowledge of the large-scale fading (statistical CSI), computation is performed at the large-scale fading time scale, max-min SINR problem.\\
		\hline
		\cite{PrecodingDistrib2020Atzeni} & Mathematical optimization & Beamforming & Semi-distributed through feedback channels & DL & TDD & Single, non-existent & No & Grid & Over-the-air signaling between BSs, multi-receive antennas, weighted sum MSE minimization problem.\\
		\hline
		\cite{StochasticUserCentric8449213} & Stochastic geometry & Beamforming & ---  & DL, UL & TDD & Non-existent & No & 2D PPP & Both cell-centric and user-centric are analyzed, approximation of the direct and interfering channel strength by a gamma distribution.\\ 
		\hline
		\cite{clusters6415394} & Mathematical optimization & Clustering, beamforming & Centralized & DL & --- & Single & No & Random & Joint design of clustering and beamforming, low-latency fronthaul network, group-sparse structure of beamformers\\ 
		\hline
		\cite{ammarC_RA_UC} & Mathematical optimization & Beamforming, user-scheduling, pilot assignment & Centralized & DL & TDD & Single & No & Random & Formation of serving clusters, both coherent and non-coherent transmission modes are studied,  weight sum rate, fractional programming, coordinated descent, pilot assignment is performed using HAC.\\ 
		\hline
		\cite{maxMinRate8756286}
		& Statistical analysis, mathematical optimization & Beamforming & Centralized & UL & TDD & Single & No & Random & Minimum rate maximization, quantized version of the signal on the CU, limited-capacity fronthaul, usage of GP, problem decomposition into two sub-problems
		\\ 
		\hline
		\cite{EnergyE2020towards} & Stochastic geometry, mathematical optimization & EE & Centralized & DL & TDD & Single & No & 2D PPP & Perfect fronthaul, conjugate beamforming, estimated CSI with pilot contamination \\ 
		\hline
		\cite{EnergyEfficiencyMmwave8676377} & Mathematical optimization & EE, power allocation & Centralized & DL, UL & TDD & Single & Probab. & Random & Both cell-free massive MIMO and user-centric are addressed, millimeter wave communication, ZF beamforming
		\\ 
		\hline
		\hline
	\end{tabular}
	\caption{Qualitative comparison and classification for selected references, Part 1 of 4.}
	\label{table:References}   
	\vspace{-1em}
\end{table*}

\begin{table*}
	\scriptsize
	\centering
	\begin{tabular}
		{|p{0.03\textwidth}|p{0.09\textwidth}|p{0.12\textwidth}|p{0.07\textwidth}|p{0.06\textwidth}|p{0.01\textwidth}|p{0.04\textwidth}|p{0.04\textwidth}|p{0.04\textwidth}|p{0.2\textwidth}|}
		\hline
		\hline
		& \multicolumn{1}{l|}{ \textit{\textbf{Model}}} & \multicolumn{1}{l|}{ \textit{\textbf{Topic}}} & \multicolumn{1}{l|}{\textit{\textbf{Control}}} & \multicolumn{1}{l|}{\textit{\textbf{Direction}}} &
		\multicolumn{1}{l|}{\textit{\textbf{Duplex}}} &
		\multicolumn{1}{l|}{\textit{\textbf{CU}}} &
		\multicolumn{1}{l|}{\textit{\textbf{LoS}}} &
		\multicolumn{1}{l|}{\textit{\textbf{Topology}}} & \multicolumn{1}{l|}{\textit{\textbf{Additional notes}}}\\
		\hline
		\hline
		\cite{cellFreeFDD9014542} & Statistical analysis, mathematical optimization & AoA information aided CSI estimation, path gain information, power allocation & Centralized & DL & FDD & Single & No & Random & Feed back the path gain information to the BS, geometric channel model with multi-tap propagation path.
		\\
		\hline
		\cite{EE9212395} & Statistical analysis, mathematical optimization & Fronthaul, EE, SE & Centralized, semi-distributed & UL & TDD & Single & No & Random & Limited fronthaul, Bussgang decomposition to model the quantization of signals, different scenarios for communication on fronthaul
		\\ 
		\hline
		\cite{cellFreeUserCentricPower8901451} & Mathematical optimization & Power allocation & Centralized & DL, UL & TDD & Single & No & Random & Multi-antenna users, correlated shadowing, use-and-then-forget (UatF) bounding technique, zero-forcing beamforming scheme that does not require CSI at the user, unlimited fronthaul.
		\\
		\hline
		\cite{cellFreeVersusSmallCells7827017} & Statistical analysis, mathematical optimization & Power allocation, pilot assignment & Centralized & DL, UL & TDD & Single & No & Random & Conjugate beamforming on the downlink and matched filtering on the uplink, no sharing of instantaneous CSI, random vs greedy assignment of pilots, unlimited fronthaul, rely on channel hardening, correlated shadowing.
		\\
		\hline
		\cite{powerAllocEnergyEfficiency9136914} & Mathematical optimization & Power allocation & Centralized & DL & TDD & Single & No & Random & Unlimited fronthaul, DUs can be turned off, power consumption is composed of transmit power and power dissipation in the transceiver hardware
		\\
		\hline 
		\cite{8901196} & Queueing theory, mathematical optimization & Power allocation, user-scheduling & Centralized & UL & TDD & Single & No & Random & LSFD, Lyapunov optimization, time slots, data packets are  generated according to a stationary and ergodic stochastic process, transmission queues.
		\\
		\hline
		\cite{HardwareImpairements8891922} & Statistical analysis, mathematical optimization & Power allocation, hardware impairments, fronthaul & Centralized & UL & TDD & Single & No & Random & Limited capacity fronthaul, hardware impairments, three different compress-and-forward strategies, signal quantization, single-antenna DUs.
		\\
		\hline
		\cite{HardwareImpairements8476516} & Statistical analysis, mathematical optimization & Power allocation, hardware impairments & Centralized & DL, UL & TDD & Single & No & Random & Classical hardware distortion models, single-antenna DUs, unlimited fronthaul capacity.
		\\
		\hline
		\cite{HardwareImpairements9004558} & Statistical analysis & Hardware Impairments & Centralized & UL & TDD & Single & No & Random &
		Low-complexity receiver cooperation schemes, local channel estimation that is sent to the CPU for joint detection
		\\
		\hline
		\cite{RicianFadingPhaseShifts8809413} & Statistical analysis, Bayesian estimation & SE, Line of Sight (LoS) phase-aware channel estimation & Centralized & DL, UL & TDD & Single & Yes & Random & Rician fading, phase shift of LoS component, knowledge of the phase shift at access point, three channel estimators based on the availability of priori information are considered (phase-aware MMSE, non-aware linear MMSE (LMMSE), and least-square (LS) estimator), error-free fronthaul, coherent and non-coherent, single-antenna access points, second layer decoding using LSFD performed at the CU to mitigate the inter-user interference.
		\\
		\hline
		\hline
	\end{tabular}
	\caption{Qualitative comparison and classification for selected references, Part 2 of 4.}
	\label{table:References2}   
\end{table*}

\begin{table*}
	\scriptsize
	\centering
	\begin{tabular}
		{|p{0.03\textwidth}|p{0.09\textwidth}|p{0.12\textwidth}|p{0.07\textwidth}|p{0.06\textwidth}|p{0.01\textwidth}|p{0.04\textwidth}|p{0.04\textwidth}|p{0.04\textwidth}|p{0.2\textwidth}|}
		\hline
		\hline
		& \multicolumn{1}{l|}{ \textit{\textbf{Model}}} & \multicolumn{1}{l|}{ \textit{\textbf{Topic}}} & \multicolumn{1}{l|}{\textit{\textbf{Control}}} & \multicolumn{1}{l|}{\textit{\textbf{Direction}}} &
		\multicolumn{1}{l|}{\textit{\textbf{Duplex}}} &
		\multicolumn{1}{l|}{\textit{\textbf{CU}}} &
		\multicolumn{1}{l|}{\textit{\textbf{LoS}}} &
		\multicolumn{1}{l|}{\textit{\textbf{Topology}}} & \multicolumn{1}{l|}{\textit{\textbf{Additional notes}}}\\
		\hline
		\hline
		\cite{cellFreeStochasticGeom8972478} & Stochastic geometry & Coverage, SE & Centralized & DL & TDD & Single & No & 2D PPP & Unlimited-capacity error-free fronthaul, conjugate beamforming.
		\\
		\hline
		\cite{bjornson2019scalable9064545} & Mathematical optimization & Scalability, initial access, pilot assignment, cluster formation & Centralized, Decentralized with some feedback & DU, UL & TDD & Single & No & Multiple & Different degrees of cooperation among the access points, Two deployments are analyzed (centralized and decentralized with some feedback to CU).
		\\
		\hline
		\cite{NOMA8895763} & Statistical analysis, mathematical optimization & Power allocation, NOMA/OMA mode switching & Centralized & DL & TDD & Single & No & Random & Max-min bandwidth efficiency, SIC, switching between orthogonal multiple access and NOMA, pilot contamination, imperfect SIC, conjugate beamforming.
		\\
		\hline
		\cite{9130689} & Queueing theory, mathematical optimization & Staistical delay, error-rate buonded QoS & Centralized & DL & TDD & Single & No & Random & Millimeterwave, hybrid automatic repeat request with incremental redundancy (HARQ-IR) and the finite blocklength coding (FBC), Hybrid digital-analog precoders, large number of access point are simulated ($1000$), large number of antennas ($[100, 800]$).
		\\
		\hline
		\cite{uplinkPilotcellFreeMassiveMIMO2020} & Mathematical optimization & Power allocation, pilot assignment & Centralized & UL & TDD & Single & No & Random & High-density for users, spatially correlated channels, adjustable phase shift pilot set allocation scheme, space-frequency and angle-delay domain channels, directional antennas, error-free fronthaul network, OFDM, detailed transmission frame, two arrays of antennas per access point (each has 100 antennas).
		\\
		\hline
		\cite{LocaPartialZFBF9069486} & Statistical analysis, mathematical optimization & local beamforming, power allocation & Distributed for beamforming, centralized for power control & DL & TDD & Multiple & No & Random & Local partial ZF (PZF) and the other uses local protective partial zero-forcing (PPZF) beamforming, distributed scheme uses long-term channel statistics, infinite capacity of the fronthaul, power control performed centrally using long-term channel statistics.
		\\
		\hline
		\cite{PowerClusteringCellFreeFDD8968400} & Statistical analysis, mathematical optimization & AoA information aided CSI estimation, power allocation & Centralized & DL, UL & FDD & Single & No & Random & Multipath component estimation for the AoA, max-min power control, standard SDP, geometric channel model with multi-tap propagation path, power control is performed at the CU at the angle-coherence time-scale.
		\\
		\hline
		\hline
	\end{tabular}
	\caption{Qualitative comparison and classification for selected references, Part 3 of 4.}
	\label{table:References3}  
	\vspace{-1em}  
\end{table*}

\begin{table*}
	\scriptsize
	\centering
	\begin{tabular}
		{|p{0.03\textwidth}|p{0.09\textwidth}|p{0.12\textwidth}|p{0.07\textwidth}|p{0.06\textwidth}|p{0.01\textwidth}|p{0.04\textwidth}|p{0.04\textwidth}|p{0.04\textwidth}|p{0.2\textwidth}|}
		\hline
		\hline
		& \multicolumn{1}{l|}{ \textit{\textbf{Model}}} & \multicolumn{1}{l|}{ \textit{\textbf{Topic}}} & \multicolumn{1}{l|}{\textit{\textbf{Control}}} & \multicolumn{1}{l|}{\textit{\textbf{Direction}}} &
		\multicolumn{1}{l|}{\textit{\textbf{Duplex}}} &
		\multicolumn{1}{l|}{\textit{\textbf{CU}}} &
		\multicolumn{1}{l|}{\textit{\textbf{LoS}}} &
		\multicolumn{1}{l|}{\textit{\textbf{Topology}}} & \multicolumn{1}{l|}{\textit{\textbf{Additional notes}}}\\
		\hline
		\hline
		\cite{ammarDistributed_RA_UC} & Mathematical optimization & User scheduling, beamforming, cluster formation, scalability & Distributed & DL & TDD & Multiple & No & Random, uniform or with hotspots & Formation of serving clusters, weight sum rate, fractional programming, coordinate descent, pilot assignment is performed using HAC.
		\\
		\hline
		\cite{nonReciprocityCellFreeDNN9098852} & Deep neural networks & Non-reciprocity calibration & Centralized & DL & TDD & --- & No & Random & Radio frequency chains are Linear Time Invariant (LTI), orthogonal frequency division multiplexing (OFDM), channels with different Power Delay Profile (PDP), cascaded deep neural networks.
		\\
		\hline
		\cite{differentCooperationLevels8845768} & Statistical analysis & Degrees of cooperation & Different degrees of centralization & UL & TDD or FDD & Single & No & Grid and Random & Four different levels of cooperation among the access points (from fully centralized till fully distributed), spatially correlated fading, users transmit with equal powers, optimized receive combining, arbitrary assignment for pilots.
		\\
		\hline
		\cite{8360138} & Statistical analysis, mathematical optimization & Power allocation, security aspect, pilot assignment & Centralized & DL & TDD & Single & No & Random & Pilot spoofing attack~\cite{6151778}, path-following algorithms, active eavesdropping, unlimited-capacity error-free fronthaul, usage of achievable secrecy rate, derivation of upper-bound and lower-bound of data rate.
		\\
		\hline
		\cite{UC_CellFreOpticalFronthaule020} & Mathematical optimization & EE, Power allocation & Centralized & UL & --- & Single with aggregation units & No (access), yes (fronthaul) & Random & Two-level fronthaul, free space optical (FSO) links, aggregation nodes, use-and-then-forget (UatF), two types of hardware models are considered: clipping and hardware impairment models
		\\
		\hline
		\cite{201114573} & Statistical analysis & SE bounds & Centralized & UL & TDD & Single & Probab. & Random & Perfect CSI, conjugate beamforming with joint or stream-wise processing of user data at the CU, error-free fronthaul network
		\\
		\hline
		\cite{Rician9099874} & Statistical analysis, mathematical optimization & Power allocation, EE, SE & Centralized & DL & TDD & Single & Probab. & Random & Conjugate beamforming for pilots sent in downlink, ideal fronthaul
		\\
		\hline
		\cite{powerControlCellFree7917284} & Mathematical optimization & Power allocation & Centralized & DL & TDD & Single & No & Random & Single-antenna access points, zero-forcing, conjugate beamforming, comparison with small-cell architecture, conventional COST Hata path-loss model.
		\\
		\hline
		\hline
	\end{tabular}
	\caption{Qualitative comparison and classification for selected references, Part 4 of 4.}
	\label{table:References4}  
	\vspace{-1em} 
\end{table*}


\subsection{Distributed Approaches}\label{sec:distributedRA}
Distributed resource allocation schemes help scale down the complexity of optimized resource allocation. When using a distributed approach, network operations are not implemented at a single point in the network; hence, as the system size increases, the newly available resources can be used to compensate for the increased complexity. Another incentive for distributed resource allocation is the increased density for the deployed DUs, which yields a large load on the fronthaul. In addition, the strict low-latency requirements for future mobile networks~\cite{she2020tutorial} necessitates implementing intelligence at the network access. The attractive features like self-organization and optimization discussed previously also motivates such approach. In this section, we examine numerous frameworks and studies that can achieve distributed resource allocation for cell-free communications.

\emph{Game theoretic approaches} are powerful tools to model the interactions between self-interested users and predict their choice of strategies. These tools can be categorized under areas like general Game Theory, Auction Theory, and Stable Matching Theory. The use of these tools comes as a natural choice for distributed resource allocation schemes, because they define agents that maximize their own utility in a distributed fashion. Using game theoretic approaches to study distributed resource allocation requires defining separate logical entities that coordinate/cooperate/compete between each other to achieve the objective, which is usually to minimize a cost function or to maximize a utility function.

Examples of these studies include interference-pricing~\cite{interferencePricing1626432}, which is performed by allowing users to announce the compensation paid by other users for their interference. Then resource allocation, e.g., power control, can be performed by treating this problem as a supermodular game which refers to games characterized by strategic complementarities, that include situations where when one agent changes its strategy, the others want to do the same~\cite{levin2003supermodular}. Another example is auction-based problems, where the users compete for the resources through two stages~\cite{bertsekas1979distributed}: the first one is a biding phase, where the users bid on the resources, while the second one is an assignment stage where the resource is assigned to the highest bidder.

The shortcomings of game theoretic approaches is that in some cases the model requires implementing feedback channels between the agents, which is not always feasible and results in a large delay. Moreover, the convergence of the game theoretic approaches are toward a Nash equilibrium which can be far from the sum-rate maximum. Moreover, most resource allocation problems need to be simplified or relaxed before they can be formulated in a game theoretic framework.

In a non-game theoretic approach, \emph{decomposition theory} divides the problem into a master problem and many derivative problems. The decomposition method can obtain the optimal solution when the original problem is convex and its feasible set can be separated~\cite{DistribResourceAllo7676375}. The studies in~\cite{PowerAlloc8630677, maxMinRate8756286} investigate the uplink max-min SINR problem through decoupling the problem into two sub-problems. The first problem designs the receiver filter through a generalized eigenvalue problem, while the second maximizes the efficiency of power allocation through a standard GP~\cite{OptPowerGP4275017}.

The study in~\cite{LocaPartialZFBF9069486} employs two distributed schemes that provide no additional overhead on the fronthaul; one that uses what is called local partial ZF (PZF) and the other uses local protective partial zero-forcing (PPZF) beamforming. Briefly, these beamformers are locally constructed at the DUs and are based on the principle of suppressing the interference caused to the users with the largest channel gain. Their results show that the proposed schemes achieve a performance that is comparable to regularized zero-forcing (RZF) and better than conjugate and ZF beamforming. Constructing the beamformers locally does not require exchanging the CSI over the fronthaul, and it allows for schemes like ZF to construct the beamformers using a channel matrix with a small dimension, hence leading to lower complexity in calculating the pseudo-inverse matrix needed by ZF.

The study in~\cite{EEdistributedAntenna7482747} divides the problem of maximizing the EE into three subproblems, namely, rate maximization, EE maximization without a rate constraint, and a power minimization problem, and each subproblem can be efficiently solved. Specifically, the first problem maximizes the rate, then if the maximum rate satisfies the rate requirement, the second problem is solved by maximizing the EE without a rate constraint, which can be solved by fractional programming~\cite{zappone2015energy, FR8314727, FR8310563}. If the solution obtained satisfies the rate requirement, it is the optimal solution. Otherwise, a third problem which minimizes the power used under the rate constraint is solved using bisection.

The study in~\cite{FederatedLearning9124715} uses the cell-free massive MIMO network scheme to implement the federated learning model~\cite{federatedLearning9084352, mcmahan2017communication} for\textit{ machine learning}, which allows for distributed data training at the users through iterative data sharing and aggregation with the CU. This iterative process terminates upon reaching a learning accuracy level for the model. The authors mention that the presented scheme reduces the training time up to $55\%$ over baseline approaches. However, the effect of latency due to the communication overhead required in their the framework is not studied. Still, the federated learning framework looks interesting, especially with the growing use of on-mobile artificial intelligence (AI)~\cite{edgeAI9134426, edgeIntelligence8736011}.

Finally, the study in~\cite{ammarDistributed_RA_UC} uses a novel leakage and interference metric to implement two fully distributed user scheduling and beamforming schemes. The first one can be implemented at the DUs, while the second can be implemented at the CUs in user-centric cell-free MIMO networks with multiple CUs. The results presented show that, compared to a centralized resource allocation solution, the schemes significantly decrease the computational complexity, while achieving 72\% and 90\% of the performance of the centralized solution for the DU-distributed and CU-distributed systems, respectively. The results also show that the CU-distributed scheme provides $1.3$- to $1.8$-fold improvements in network data rate compared to the DU-distributed scheme, which highlights the importance of deploying a multiple-CU cell-free network.


\subsection{Lessons Learned}\label{sec:RA_LL}
Conjugate beamforming is used in the literature for analytical tractability, scalability, fronthauls with limited capacity, or for some cases such as single-antenna DUs. However, conjugate beamforming does not explicitly suppress interference~\cite{cellFreeVersusSmallCells7827017}. Hence, in practice, especially in the absence of channel hardening, other local schemes such as the LSFD~\cite{7837706, powerControlCellFree7917284, Scalable9174860, 8901196, HardwareImpairements9004558}, local MMSE combining~\cite{differentCooperationLevels8845768}, and the weighted MMSE~\cite{MSE5756489, ammarC_RA_UC} could be better candidates for beamforming in cell-free MIMO networks irrespective of the level of cooperation~\cite{differentCooperationLevels8845768}. One interesting scheme is the weighted MMSE approach developed initially for the cellular networks~\cite{MSE5756489}. CSI estimation and scalability should be accounted for in the proposed resource allocation schemes for cell-free networks.

The high density of the deployed DUs in cell-free massive MIMO networks raises the possibility of having LoS communication. Based on this insight, a probabilistic LoS channel model~\cite{ITU:P.1410_5, 8859647} can be optionally used in the studies. The study in~\cite{201114573} shows that increasing the access point density from 128 to 1024 per ${\rm km}^2$ dramatically increases the probability of LoS communication from $40\%$ to $95\%$.

Power control algorithms can be used to balance fairness, latency, and throughput. A tradeoff between latency and data rate is still not addressed. EE maximization seems to be extensively researched under different scenarios and optimization variables for cell-free communications~\cite{EE8777141, EEdistributedAntenna7482747,EE8695055,EnergyEfficiency8781848}.
	
Fractional programming~\cite{zappone2015energy, FR8314727, FR8310563} and the Dinkelbach method~\cite{dinkelbach1967nonlinear} are suitable tools to solve such EE optimization problems. Transmit power adaptation, energy harvesting~\cite{EnergyHarvesting9201540} and sleep modes~\cite{9149862} can foster EE in cell-free MIMO networks. LSFD seems a promising scheme for the uplink of cell-free massive MIMO networks~\cite{powerControlCellFree7917284, HardwareImpairements9004558}. However, it has been studied only for a single-CU network.

The combination of both user-centric cell-free networks and millimeter wave communication is promising because cooperative networks can alleviate the problem of poor connectivity suffered by millimeter wave communication~\cite{mmWaveConnectivity8579590, mmwave6732923}. The short range and large bandwidth of millimeter wave communication may also help in deploying a very dense cell-free network. This harmonization is simply based on the fact that millimeter wave communication requires a dense deployment of transmitters due to the high path loss, and a dense deployment requires coordination to prevent interference from getting out of control.

Distributed resource allocation schemes provide a main playground for cell-free communications, because they can scale down the computational complexity per node and prioritize scalability. Game theory, auction theory, stable matching theory, decomposition theory, and advanced optimization problems can be used as tools to implement a distributed resource allocation scheme~\cite{interferencePricing1626432, DistribResourceAllo7676375}. A possible drawback of some of these tools, such as game theory and auction theory for example, is that the system model and actual problem needs to be simplified before using these tools. Unfortunately, it is not guaranteed that these simplifications will not affect the accuracy of the conclusions derived.

Finally, in Table~\ref{table:prosCons_RA} we summarize some \varTableDes\ of selected solutions targeting resource allocation.

\newcommand{\varEmptyDrawback}{\emph{Needs more investigations to identify}}

\begin{table}[t!]
	\footnotesize
	\centering
	\begin{tabular}{|>{\raggedright\arraybackslash}p{0.08\textwidth}|>{\raggedright\arraybackslash}p{0.16\textwidth}|>{\raggedright\arraybackslash}p{0.1\textwidth}|>{\raggedright\arraybackslash}p{0.05\textwidth}|}
		\hline
		\hline
		\multicolumn{1}{|l|}{ \textit{\textbf{Solution}}} & \multicolumn{1}{l|}{ \textit{\textbf{Pros}}} & \multicolumn{1}{l|}{ \textit{\textbf{Cons}}} & \multicolumn{1}{l|}{\textit{\textbf{Ref.}}}\\
		\hline
		\hline
		Infrequent transmit power adaptation & More stable decisions for the network, lower computation load, less strict time constraint for the network decisions than instantaneous adaptations & Sacrifice performance, may require properties like channel hardening & \cite{cellfreeIofTPilots2020, DLTrainingCSI8799031}
		\\
		\hline
		Conjugate beamforming & Enhanced system scalability, cope with single-antenna DUs, mathematical tractability, limit the load on the fronthaul, low signaling and computation overhead because only the CSI for serving channel is needed & Requires favorable propagation because it lacks interference nulling capabilities & \cite{EnergyEfficiency8097026, DLTrainingCSI8799031, EnergyE2020towards, cellFreeStochasticGeom8972478, NOMA8895763, 201114573}
		\\
		\hline
		Weighted MMSE-based and local MMSE Beamforming & High performance, feasible for user-centric clustering, can be implemented in a distributed fashion & May require DUs to estimate CSI for non-served users & \cite{ammarC_RA_UC, ammarDistributed_RA_UC, WMMSE5756489, differentCooperationLevels8845768}
		\\
		\hline
		Local partial ZF (PZF) and local protective partial zero-forcing (PPZF) beamforming & Do not require exchanging the CSI over the fronthaul, use a channel matrix with a small dimension, low complexity in calculating the pseudo-inverse matrix needed by ZF, outperform conjugate and ZF beamforming, can be used for long-term power control strategies & Sacrifice performance due to limited cooperation & \cite{LocaPartialZFBF9069486}
		\\
		\hline
		LSFD & A promising scheme for uplink, can achieve about two-fold gain over conjugate beamforming in terms of $95\%$-likely per-user SE & No clear directions on how to implement it on multiple-CU networks & \cite{powerControlCellFree7917284, 7837706, Scalable9174860, 8901196, HardwareImpairements9004558, RicianFadingPhaseShifts8809413}
		\\
		\hline
		(Millimeter wave + cell-free) communication for access channel & Alleviates the problem of poor connectivity for millimeter wave communication; allows for dense deployment of access points due to coordination, short communication range and large bandwidth & Requires LoS, even with dense deployment of access points blockages could still exist & \cite{EnergyEfficiencyMmwave8676377, EECellFreeUCMMwave8292302, EECellFreeUCMMwave8516938}
		\\
		\hline
		Distributed approaches & Highly scalable, lower complexity per node (DU/CU), Low overhead on the fronthaul & May sacrifice performance, hard to eliminate the dependence of decisions between the access points due to the coupled nature of performance metrics (e.g., SINR) & \cite{ammarDistributed_RA_UC, differentCooperationLevels8845768, bjornson2019scalable9064545, LocaPartialZFBF9069486, EE9212395, PrecodingDistrib2020Atzeni}
		\\
		\hline
	\end{tabular}
	\caption{Solutions related to resource allocation.}
	\label{table:prosCons_RA}
	\vspace{-2em}
\end{table}


\section{Latency and Synchronization}\label{sec:latsynch}
\subsection{Latency}
An important metric that is rarely visited in the literature on cell-free massive MIMO networks is the latency of data delivery. The authors of~\cite{DelayAware6180015} tackle delay-aware BS-discontinuous transmission control and user scheduling to reduce inter-cluster interference for downlink coordinated MIMO schemes with energy harvesting capability. The proposed transmission is adaptive, on long timescales, to both the queue state information (QSI) and energy state information (ESI); it also allows for the management of inter-cluster interference through bursty data arrivals which are generally delay-sensitive. We note that intermittent transmissions are very useful for mMTC, which can support high densities of users~\cite{mMTC7565189}. The authors model the problem as a Markov decision process (MDP), which is solved using distributed stochastic learning and approximate dynamic programming~\cite{bertsekas1995dynamic}.

The study in~\cite{8901196} employs Lyapunov optimization techniques~\cite{4455486} to optimize joint power control and scheduling in the uplink of both co-located and cell-free massive MIMO. The algorithm developed in this publication optimizes the amount of data that can be admitted to the transmission queues and allocates the throughput to each user on a per time slot basis, hence reducing the average delay of the network.

Controlling the delay is fundamental for many applications, such as tactile internet~\cite{tactileInternet7470948, tactileInternet8605315} and Telehealth services~\cite{Telehealthdorsey2016state}. Lyapunov optimization~\cite{4455486} is an important tool to study delay-aware scenarios~\cite{7891004} and communication with deadlines~\cite{7010886}. In this regard, queueing theory is another useful tool, because queueing greatly affects the delay time of data packets. Thus, studying the delay of the communication in cell-free systems helps in designing the layout, capacity and control of the network.

uRLLC is one of the three service categories supported by 5G networks through the features introduced by the New Radio interface~\cite{3GPPTR21.915}. To achieve ultra-reliability and low-latency, short-packet data communication~\cite{uRLLC8403963} is proposed which may affect the data rate, however, this is not a serious problem because the required throughput is usually not stringent. As discussed in~\cite{uRLLC8403963}, reliability can be improved without violating the latency requirement by utilizing resources in the frequency, antenna, and spatial domain.

Emphasizing the spatial domain, cell-free communication can be a key to boost the reliability constraint for uRLLC applications. The study in~\cite{9130689} focuses on statistical delay for massive uRLLC (muRLLC) in user-centric cell-free MIMO networks using millimeter waves. What is notable is the integration of hybrid automatic repeat request with incremental redundancy (HARQ-IR)~\cite{HARQIR6954538} and finite blocklength coding (FBC)~\cite{FBC7134725} in the analysis. The first technology is different from the traditional ARQ scheme in the sense that it can adaptively control the transmission rate based on decoding feedback to make it convenient for low-latency applications. The latter technology uses short codeword block-lengths to enable reliable communications while allowing for time-sensitive data flows.

Mobile edge computing (MEC) is one of the proposed frameworks for latency critical applications~\cite{mobileEdgeComputing7901477}. In this solution, the cloud computing capabilities are delegated among some network edge servers~\cite{7498684}. This means that cell-free architecture with multiple CUs provides an opportunity to implement MEC. The authors of~\cite{8974591} study the performance in a user-centric cell-free MIMO network implementing MEC, where the network is assumed to contain a single CU and MEC servers associated with each access point. Tools like stochastic geometry and queueing theory are implemented to derive successful edge computing probability (SECP), and successful communication and computation probabilities for a target computation latency. Notable results show that for a target SECP, a more spatially distributed antenna system (less antennas per access point) provides better EE compared to a less distributed one (i.e., more co-located antenna system). However, due to deployment cost efficiency, considering a multiple-CU network looks more logical than assuming that each access point includes an MEC server, hence the considered metrics for a multiple-CU network still need to be investigated.

Finally, punctured scheduling~\cite{puncturedScheduling9011578} is another interesting topic to be studied for the user-centric cell-free MIMO scheme. In this type of scheduling, the aim is to allow uRLLC traffic to overwrite longer ongoing enhanced Mobile Broadband (eMBB) transmissions. However, the uRLLC traffic should not degrade the performance of eMBB communications. Herein, we can spot a tradeoff between the latency of uRLLC and the rate loss for eMBB traffic. Hence, a \emph{joint scheduling} solution needs to be established which considers these two different types of metrics (latency and rate loss). Several solutions are being considered such as optimized re-transmissions, signal space diversity and opportunistic spatial preemptive scheduling~\cite{8322772, 8538471, 8408793}. This topic has still not been investigated for cell-free communications.

\subsection{Signal Synchronization, Inter-DU Carrier Frequency Offset}
The need to serve users with multiple DUs can increase the delay of signal delivery. This is critical when the DUs are under the control of different CUs and the network lacks accurate synchronization. Moreover, the geographically distributed DUs can introduce unavoidable differences in the signals' time-of-arrival. Coupled with synchronization issues, the signals received at the users can incur higher signal delay spread and inter-DUs carrier frequency offset.

The data precoding and decoding for the users needs to be executed at multiple CUs simultaneously. Hence, data synchronization errors can accumulate and affect the power delay profile (PDP) by increasing the signal delay spread~\cite{PDPUsercentricVsDisjoint8969384}. Thus, a synchronization process needs to be employed at regular intervals~\cite{interdonato2019ubiquitous}. Moreover, to allow coherent transmission, the DUs need to maintain relative signal synchronization on both time and phase between their transmissions~\cite{synchronizationCalibration6760595}. Importantly, the cyclic prefix in orthogonal frequency division multiplexing (OFDM) systems can help in tolerating this requirement. In this context, the DUs can be assumed quasi-synchronized within a $1~{\rm km}$ or $5~{\rm km}$ radius for the LTE normal and extended cyclic prefix respectively~\cite{differentCooperationLevels8845768}. Furthermore, the user-centric cell-free system coupled with a small cluster size helps overcome this problem and provides an advantage over cell-centric cooperative schemes~\cite{PDPUsercentricVsDisjoint8969384}.

Quasi-synchronous systems that employ complex equalization and advanced synchronization techniques in OFDM systems are discussed in~\cite{Synchronization4287203}. Doppler shifts and oscillator instabilities can cause a carrier frequency offset (CFO) between the received carrier and the local sinusoidal signal used for demodulation, which, in turn, results in a loss of mutual orthogonality among the subcarriers causing inter-carrier interference. Furthermore, a wrong placement position for the discrete Fourier Transform window can inflict a timing error. These impurities lead to a loss in the achievable SNR.

The hardware of the transmitter and the receiver are not reciprocal in the sense that they can introduce different amplitude scaling and phase shifts on the uplink and downlink channels~\cite{ArinNonReciprocity8421240}, respectively. In addition, the carrier frequency offset between the DUs can result from imperfect local oscillators, and this can also produce inter-carrier interference~\cite{5910719}. Generally speaking, solving synchronization problems requires two phases; the first estimates the different timing and frequency errors, and the second implements the estimated parameters on multiple DUs to do correction for the synchronization~\cite{davydov2016timing}.

The investigation in~\cite{HardwareImpairements8476516} studies power control in cell-free massive MIMO networks with hardware impairments. Hardware distortion can be characterized through models~\cite{schenk2008rf, valkama201115} that account for impairments like carrier-frequency and sampling-rate offsets, in-phase/quadrature-phase imbalance, phase-noise, inter-carrier interference, and non-linearity of analog devices~\cite{residualImpairments7106472}. As claimed by the authors, the results show that a cell-free massive MIMO system can tolerate hardware impairments without performance reduction. Similarly,~\cite{HardwareImpairements8891922} studies the uplink of cell-free massive MIMO systems under limited fronthaul, hardware impairments, and different signal/CSI quantization and transmission techniques. A notable statement in~\cite{HardwareImpairements8476516}, is that the distortion introduced by hardware impairment makes the channel estimate and estimation error non-Gaussian distributed, which prevents using the standard capacity lower bound in~\cite{6415388}. However, since the fact that the capacity expression used is a lower-bound, coupled with its analytic tractability, makes the Gaussian capacity expression, with an appropriate accounting for the reciprocity errors, very useful. 

The paper in~\cite{HardwareImpairements9004558} studies the uplink performance including the effect of hardware impairments under different receiver cooperation schemes, that include LSFD, simple LSFD and simple centralized decoding. Their results show that the LSFD scheme provides the largest SE, and that a cell-free MIMO network outperforms a small-cell network under the scenario studied. Reference~\cite{RicianFadingPhaseShifts8809413} studies the uplink SE of cell-free massive MIMO network assuming Rician fading with phase shift of the LoS component, i.e., non-static phase for the LoS component. However, to perform channel estimation, the authors assume knowledge of the phase shifts at the access point although it changes per coherence block. Hence, in practice how this phase shift can be known is not clear and may be an area of future research.

The study in~\cite{synchronizationCalibration6760595} develops over-the-air synchronization and calibration protocols for distributed MIMO systems. Their results show that the developed system provides sufficient accuracy for satisfactory performance. However, the system is developed for cell-centric clustering or small cells, hence, it may not be be directly applicable for user-centric cell-free networks. AI-based advanced prediction tools can be employed to enhance the signal synchronization between the DUs. Deep neural networks (DNNs) are also considered in~\cite{nonReciprocityCellFreeDNN9098852} as a non-traditional calibration method for the radio frequency front ends and CSI interpolation.

In terms of protocols, coarse synchronization using the IEEE 1588v2 Precision Time Protocol (PTP)~\cite{IEEE1588_7949184} is a possible solution. PTP is a packet-based timing technology which was originally designed to provide precise timing for critical industrial automation applications. Importantly, the PTP provides accuracy in the nanosecond range which is better than the Network Time Protocol (NTP). However, no studies exists on using this protocol for synchronizing the DUs in cell-free communications.

In general, synchronization systems can include channel equalization, usage of cyclic prefix that exceeds the channel impulse response duration, reference blocks that produces coarse estimates of the synchronization parameters, and frequency tracking between CFO acquisition phase.

\subsection{Lessons Learned}\label{sec:latsynch_LL}
uRLLC traffic corresponds to latency critical applications. This kind of communication is more sensitive to delay than to data rate, thus cell-free communication should provide solutions for this use case.

Topics being proposed to decrease the delay are MEC~\cite{mobileEdgeComputing7901477}, punctured scheduling~\cite{puncturedScheduling9011578}, short-packet data communication~\cite{uRLLC8403963}, hybrid coding schemes such as HARQ-IR~\cite{HARQIR6954538} and FBC~\cite{FBC7134725}, and a better utilization for the resources in the frequency, antenna, and spatial domains~\cite{uRLLC8403963}. Furthermore, limiting the size of the serving cluster could be another solution to minimize the delay in user-centric cell-free networks.

Queueing theory, MDP, dynamic programming, and Lyapunov optimization techniques~\cite{4455486} seem to be useful tools to study latency in coordinated networks~\cite{DelayAware6180015}. QSI can be as an indication of congestion that increases the latency, thus the amount of data admitted to the transmission queues can be optimized based on QSI to reduce the average delay of the network~\cite{8901196}.

\begin{table}[t!]
	\footnotesize
	\centering
	\begin{tabular}{|>{\raggedright\arraybackslash}p{0.08\textwidth}|>{\raggedright\arraybackslash}p{0.14\textwidth}|>{\raggedright\arraybackslash}p{0.12\textwidth}|>{\raggedright\arraybackslash}p{0.05\textwidth}|}
		\hline
		\hline
		\multicolumn{1}{|l|}{ \textit{\textbf{Solution}}} & \multicolumn{1}{l|}{ \textit{\textbf{Pros}}} & \multicolumn{1}{l|}{ \textit{\textbf{Cons}}} & \multicolumn{1}{l|}{\textit{\textbf{Ref.}}}\\
		\hline
		Optimize the amount of data admitted to the transmission queues & Minimize data transmission latency due to congestion, makes data scheduling more efficient & Transmission queues of users are coupled making the problem hard to tackle, queues can become unstable leading to infinite delay & \cite{8901196, 4455486}
		\\
		\hline
		HARQ-IR with FBC & Suitable for uRLLC, HARQ-IR adaptively control the transmission rate based on decoding feedback, FBC use short codeword block lengths, statistical delay constraint guarantees, more efficient for delay-sensitive applications than traditional ARQ & \varEmptyDrawback & \cite{9130689, HARQIR6954538, FBC7134725}
		\\
		\hline
		Short-packet data communication & Prioritize low-latency, Small delay overhead for re-transmitted packets & Negatively affect the data rate due to large packet overhead  & \cite{uRLLC8403963}
		\\
		\hline
		MEC & Improve performance of latency critical applications, can be supported through multiple-CU networks & May increase the cost of network end devices due to pushing cloud computing capabilities to the network edge hence precluding dense deployment, may decrease EE & \cite{8974591, mobileEdgeComputing7901477, 7498684}
		\\
		\hline
		Punctured scheduling & Reduce latency of uRLLC traffic, allow uRLLC traffic to overwrite longer ongoing eMBB transmissions, tradeoff between latency of uRLLC and rate loss of eMBB & Potential rate loss to eMBB traffic & \cite{puncturedScheduling9011578, 8322772, 8538471, 8408793}
		\\
		\hline
	\end{tabular}
	\caption{Solutions related to latency and synchronization.}
	\label{table:prosCons_lat}
	\vspace{-2em}
\end{table}


The punctured scheduling concept seems to be similar to the cognitive radio concept. In punctured scheduling, we are trying to schedule uRLLC traffic within longer ongoing eMBB traffic, while in cognitive radio, secondary users are trying to gain opportunistic access to the spectrum within the spectrum gaps between the transmissions of the primary users. Based on this observation, the partially observed Markov decision process (POMDP) may be a suitable tool to study the punctured scheduling problem, because POMDP has been successfully applied for cognitive radio to provide opportunistic access policies~\cite{POMDP7895211}. However, definitely, uRLLC differs in terms of priority when compared to the opportunistic access in cognitive radio, thus some modifications for the proposed POMDP model will be needed. We note that a POMDP can be used to account for uncertainty in the states of the studied problem, and it allows deriving strategies to optimize the concerned parameters.

Signal synchronization and PDP are other areas that can be investigated for cell-free communications. Geographically distributed DUs can introduce unavoidable differences in the signals' time-of-arrivals. Coupled with synchronization issues, the signals received at the users can incur higher signal delay spread and inter-DUs carrier frequency offset.

Coherent transmission is affected by lack of synchronization more than the non-coherent mode, so the DUs need to maintain relative signal synchronization of both time and phase between their transmissions~\cite{synchronizationCalibration6760595}. Quasi-synchronous systems, complex equalization and advanced synchronization techniques can be further studied. Usage of cyclic prefix and synchronization protocols such as the IEEE 1588v2 PTP~\cite{IEEE1588_7949184} can also minimize the effect of non-accurate synchronization.

Finally, in Table~\ref{table:prosCons_lat} we summarize some \varTableDes\ of selected solutions targeting latency and synchronization.

\section{Miscellaneous Topics}\label{sec:MiscellaneousTopics}

\subsection{Scalability}
Scalability is defined as the ability of the system to accommodate a growing amount of work gracefully~\cite{bondi2000Scalabilitycharacteristics}. Despite its importance, scalability is a poorly defined and understood term~\cite{duboc2010framework}. A precise and rigorous definition for scalability does not exist, in this regard, we quote the following statement from~\cite{hill1990scalability}: ``I encourage the technical community to either rigorously define scalability or stop using it to describe systems''.

In its general understanding, scalability targets a dimension of interest like response time, processing overhead, space, memory, etc. In~\cite{bjornson2019scalable9064545}, a cell-free based scheme is considered scalable if the tasks of signal processing for channel estimation and data reception, fronthaul signalling and power control per DU can be kept within finite complexity as the number of users served goes to infinity. A different formulation for scalability is to define two metrics; a scalability metric $m(k)$ which measures a quality property, e.g., response time, and a criterion $z(k)$ which defines a target for $m(k)$. We note that $k$ is the scale of parameter of interest, e.g., number of users served on the same channel. Then, the scalability is defined based on a relation between $m(k)$ and $z(k)$, e.g., linear or asymptotic relation.

The work in~\cite{Scalable9174860} proposes an initial access algorithm and pilot assignment scheme. For the decoding, the authors use a suboptimal but scalable LSFD, while for pilot assignment they propose to use two alternatives that are based on a user-group clustering and a K-means clustering, respectively. Nonetheless, the work assumes deploying one CU in the network which, by becoming the bottleneck, could jeopardize scalability - the CU, especially the fronthaul links, can become overloaded. Similarly, the work in~\cite{powerControlCellFree7917284} uses LSFD that can achieve two-fold gains over conjugate beamforming in terms of $95^{\rm th}$ percentile per-user SE.

One form of implementing scalability is to deploy a distributed resource allocation system for the user-centric cell-free MIMO network. In general, a distributed resource allocation scheme scales better than a centralized one, since, as the system size grows, more resources become available to handle the additional complexity. Interestingly, such an approach complies with the definition of scalability found in~\cite{weinstock2006system}, which states that it is the ability to handle increased workload by repeatedly applying a cost-effective strategy for extending the capacity of the system. This, again, motivates deploying a distributed system of CUs rather than focusing all the load at a single one. The study in~\cite{ammarDistributed_RA_UC} targets scalability through two distributed resource allocation schemes and through the use of statistical CSI and traffic distribution to scale down the required  CSI estimation.


\subsection{Intelligent Reflecting Surfaces}\label{sec:IRS}
Intelligent reflecting surfaces (IRSs) or reconfigurable intelligent surfaces is an emerging technology that aims to enhance network coverage, data rate and EE with a low deployment cost. An IRS is a planar passive array structure that is controlled by an IRS controller. The array elements can superpose the received signals and reflect them coherently toward specific positions to achieve passive beamforming and enhance the signal coverage~\cite{IRS8910627}. This structure can be placed on objects, e.g., buildings and walls, to create an active scattering environment. This is very useful in areas with poor coverage and blockages, including indoor areas. What makes this technology special is that it targets the channel quality instead of targeting the transmitter or the receiver as most technologies do. This makes IRS, at least as a concept, an idea with a new perspective.

\begin{figure}[t]
	\centering
	\includegraphics[width=1\columnwidth]{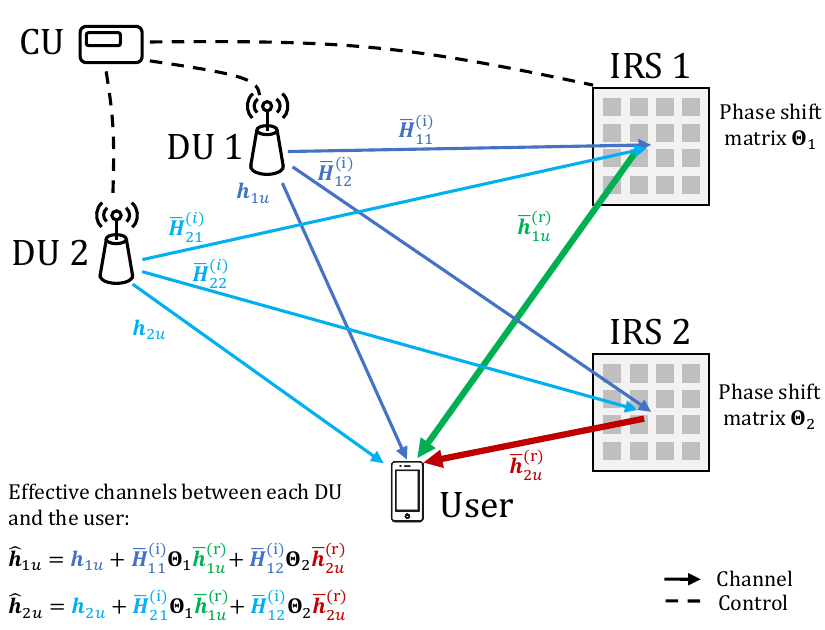}
	\caption{IRS-aided communications.}
	\label{fig:IRSaided}
	\vspace{-1em}
\end{figure}

An IRS is different from active relay systems that employ decode and forward or amplify and forward because it does not possess complex signal processing capabilities, but relies on passively reflecting the received signals through controlling the signals' phase shifts on each array element of the IRS~\cite{IRS9308937}. As indicated by~\cite{IRS9308937, IRS9363171, IRS9298843}, an IRS can be very useful for cell-free massive MIMO because the deployment cost of IRSs is much less than that for access points. In Fig.~\ref{fig:IRSaided}, we show an IRS-aided network. With the reflection support from the IRSs, we can construct a better effective channels ${\bf \hat{h}}_{bu}$ between the DUs and the user. As can be seen from this figure, these channels are composed of the concatenation of the channel from the DU (${\bf h}_{bu}$) and the reflected accumulated channels through the IRSs (${\bf \bar{H}}_{bi}^{\rm (i)} {\bm \Theta}_i {\bf \bar{H}}_{iu}^{\rm (r)}$). Furthermore, the reflected channel form each IRS can be controlled through a phase shift matrix ${\bm \Theta}_i$. This phase shift can be optimized to enhance the performance.
	
The authors in~\cite{IRS9363171} show that using IRS in cell-free MIMO networks enhances the EE. As stated in~\cite{IRS9363171}, the motivation to use IRS in cell-free networks is their low cost and power consumption. In detail, the study in~\cite{IRS9363171} jointly optimizes the transmit beamformers at the DUs and the reflection coefficients of the IRS to maximize the EE under a limited-capacity fronthaul constraint. However, the problem seems to be intractable due to the coupling between the optimization variables, thus approximations are applied and local optimum is obtained. As such, there is still a lot of space for improvement in future studies for both the system model and design criteria.

The authors of~\cite{IRS9352948} mention that IRSs can be used to create favorable propagation conditions with low operating cost due to the higher EE provided by such deployment compared to a denser network of DUs. Similar to~\cite{IRS9363171}, the authors also optimize both the beamformers of the DUs and the reflecting coefficients of the IRSs so that EE is maximized. In this regard, the impact of transmit power, density and size of the IRSs is investigated. The results of~\cite{IRS9352948} show that more than $2$-fold gain in the EE is observed when part of the DUs are replaced by IRSs in a typical network configuration.

Reference~\cite{IRS9298843} similarly optimizes the beamformers and the reflection coefficients of the IRS in an IRS-aided cell-free communications through a decentralized design. Both low and high resolution phase shifts are considered for the IRS. To do so, the authors transform the original problem into a sequence of majorized subproblems. As expected, the system that includes IRSs can outperform a counterpart without IRSs.

\subsection{Role of Machine Learning}\label{sec:AI}
Machine learning is a very powerful technology proposed for wireless communications. However, only very few studies seem to target user-centric cell-free networks, thus, studying the implication of machine learning on user-centric cell-free networks is still an open issue.

The authors of~\cite{DeepLearningFronthaul9110901} use DCNN to study different uplink power control schemes. The DCNN seems to succeed in mapping the large-scale fading to the allocated power coefficients. The authors of~\cite{AI9129108} study an antenna selection problem in user-centric cell-free networks. This study converts a sum-rate problem into an MDP, then uses reinforcement learning to solve the problem. In this context, an algorithm called multi-threaded asynchronous advantage actor-critic is used to assess the quality of the actions in the MDP framework and perform the reinforcement learning.

The studies in~\cite{AIismath2020deep, AIismath2021deep} test the use of AI in user-centric millimeter wave networks. The scheme used is denoted as deep contextual bandit, and it corresponds to the case when a deep neural network is used to predict the reward for the actions of the agent in an action-reward environment. These studies analyze topics such as beam sweeping and initial access for access point with sleep-state mode. We note that beam sweeping means choosing the beamformer used to serve the user from a predefined quantized beam codebook. The intuition from using AI is that it can provide fast and efficient response to unexpected loss of connection in millimeter wave communication.

Finally, machine learning for wireless communication is a huge topic that is usually addressed in dedicated survey papers, e.g.,~\cite{surveyMLwirelessNet8743390}. Readers interested in this topic are advised to check such survey papers.

\subsection{Lessons Learned}\label{sec:Misc_LL}
Scalability is defined as the ability of the system to accommodate a growing amount of work gracefully~\cite{bondi2000Scalabilitycharacteristics}. It is one of the milestones for the solutions developed for cell-free communications. Distributed resource allocation schemes is one way of achieving scalability.

As a concept, IRS is an interesting technology because it targets the channel quality instead of targeting the transmitter or the receiver as most technologies do. It can be used to further boost EE in cell-free networks. In this regard, it can support a dense deployment of the network at a lower operating cost compared to a DU-only network. In future, a mature IRS technology could help network operators to decrease their electricity bills in a densely deployed cell-free network.

Machine learning is a mighty tool to boost the performance of cell-free communications. Studies to quantify the performance gains in cell-free networks are needed. These studies can include beamforming, resource allocation, user-association, DU-CU association management, spectrum management, mobility management. It is expected that more applications for machine learning in cell-free communications will appear.

\begin{figure*}[t]
	\centering
	\includegraphics[width=1\linewidth]{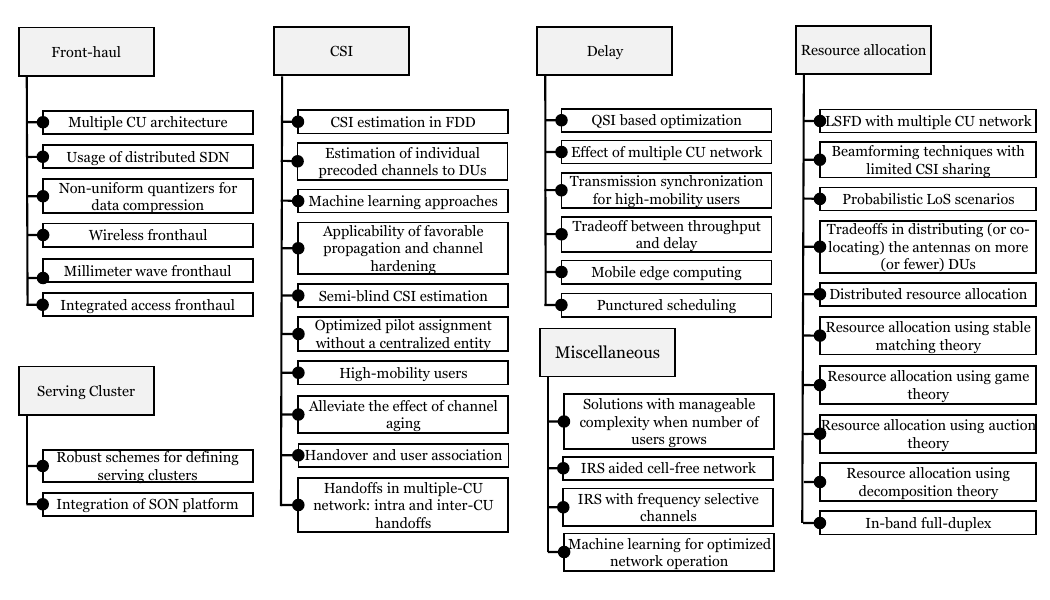}
	\vspace{-2em}
	\caption{List for some open issues that need to be studied for user-centric cell-free MIMO networks.}
	\label{fig:ReviewOpenIssues}
	\vspace{-1em}
\end{figure*}

\section{Possible Future Research Directions}\label{sec:openResearchProb}
In this section, we summarize some open technical issues for user-centric cell-free MIMO scheme. Despite of the importance of these topics, they are much less than fully addressed in the relevant literature.

\subsection{Fronthaul}
The multiple-CU architecture, e.g.,~\cite{8761828, bjornson2019scalable9064545}, for the user-centric cell-free network has not been as well investigated as the single-CU architecture. Thus, studying the effect of the former architecture on the fronthaul is critical. An interesting topic would be to compare the two architectures under the same network assumptions. Furthermore, some network solutions, like the LSFD for example, may need some modifications to be useful in a network with multiple CUs.

Another research area involves integrating the distributed SDN framework, discussed on page~\pageref{page:SDNdiscussion}, to dynamically assign the DUs to the CU. What is interesting in SDN is the flexible and programmable features offered to the network, which allows network operators to implement features like automatic migration of DUs between neighboring CUs based on fronthaul capacity. Furthermore, for a dense deployment of DUs, integrating platforms like distributed SDN will be crucial, because the network will be hard to manage.

Compression for data and CSI on the fronthaul, under the different scenarios shown in Fig.~\ref{fig:ReviewCompressScenarios}, is indispensable. The studies in the literature are based on uniform-quantizers, however, uniform quantizers do not perform well when the signal levels are skewed toward one region. Indeed, optimizing the quantization levels can lower the quantization error and enhance signal quality~\cite{6226311}. Thus, non-uniform quantizers should be able to provide smaller distortion for the signal compared to uniform quantizers. A famous algorithm to perform non-uniform signal compression is the Lloyd-Max quantizer~\cite{Compression1456239}, which minimizes the mean-square quantization error under a fixed number of quantization levels. Another design is to use a compander (compressor-expander) that is composed of a uniform quantizer preceded by a non-liner transformation function. Rate-distortion theory~\cite{cover1999elements} is also useful to analyze this topic.

Problem formulations that employ wireless fronthaul are different from those that use a fixed-capacity fronthaul. In the former problems, the capacity of the fronthaul will change with the problem parameters, e.g., the locations of the DUs~\cite{ammarWirelessbackahulJournal}. Hence, studying the performance under wireless fronthaul could produce some novel analysis and conclusions for cell-free communications. We note that in practice, wireless fronthaul could be used only for part of the network where wired fronthaul is hard to deploy.

Millimeter wave with cell-free communication is gaining the interest of researchers. As discussed previously, this area is very promising because the two areas, i.e., millimeter wave and cell-free communications can complement each other. However, currently, millimeter wave communication is mostly being used on the access channel, e.g., \cite{EnergyEfficiencyMmwave8676377, EECellFreeUCMMwave8292302, EECellFreeUCMMwave8516938}, so, its effect on the system performance when used on the fronthaul still needs to be studied.

Another interesting topic is integrated access and backhaul (IAB)~\cite{3GPPTR38.874}, which can provide an additional spectrum resource for the access channel or fronthaul in scenarios where dedicated spectrum is not possible. It can serve as a cost-effective alternative for wired fronthaul. System level simulations are needed to study the benefits of IAB in cell-free communications. Designing a high-performance IAB for conventional networks is still an open research challenge~\cite{IAB9040265} and it rarely has been studied for cell-free communications. One idea would be to optimize the classification of network nodes as IAB-donors (with the wired connections) or IAB nodes to enhance the performance and minimize deployment costs. Another topic would be to perform end-to-end performance evaluations in a cell-free network architecture using simulators such as~ns-3~\cite{ns3}.

\subsection{Channel State Information}
As discussed previously, the main disadvantage of the FDD mode compared to the TDD mode is that the channel is not reciprocal in the UL and DL, and unless advanced signal processing techniques are applied, we need to use both UL and DL pilot training phases. The FDD mode in cell-free communication requires further enhancements to decrease the number of pilots needed for both UL and DL CSI estimation, such enhancements may focus on exploiting the reciprocity of angle of arrival. In this context, the studies in~\cite{PowerClusteringCellFreeFDD8968400} can serve as a starting point. In-depth studies on the applicability of favorable propagation and channel hardening under different network scenarios are advised. The study can use a PPP, as done in~\cite{cellFreeStochasticGeometry8379438}, or any other suitable point process models. 

For downlink CSI estimation schemes, the user usually estimates the effective SISO precoded channel from all the serving DUs. However, some applications, such as SIC and mobility related measurements, could require from the user to estimate the individual precoded channels from each serving DU. In this case, the pilot sequence used in the downlink will be dependent on the number of DUs, and this is a problem when the number of DUs is very large. A novel research idea could be to solve or mitigate the impact of this problem by proposing alternate options. Machine learning approaches, such as FFDNet and DnCNN, could be a good direction~\cite{FFDNet8365806, DeepLearning8815888}.

Semi-blind CSI estimation techniques are very promising, especially for massive MIMO systems. However, it seems that there are no studies that target cell-free communications, making this an open research topic. The investigation in~\cite{9016270} reports that the space-alternating generalized expectation-maximization (SAGE) based semi-blind estimator provides a considerable improvement over existing pilot-aided schemes, where gains of $6$~dB for ${\rm SNR} \ge 10$~dB for massive MIMO systems are reported. Similarly, the study in~\cite{JavadMirzaei8490886} reports a $10$~dB reduction in normalized MSE (NMSE) for semi-blind time-domain CSI estimation compared to frequency domain discrete Fourier Transform based interpolation for frequency-selective massive MIMO networks. This encourages the use of semi-blind channel estimation for cell-free communications.

Most studies that optimize PA policies assume that the pilots are optimized using a centralized entity, e.g., the CU in a single-CU network. However, centralized optimization for PA may not be feasible in an actual deployment. Thus, distributed optimization for PA is needed, which is a good research topic that needs more investigation. 

The performance of a user-centric cell-free MIMO scheme under high-mobility profiles for users is not well investigated, and it is a wide open research area. Some of the topics can include studying pilot designs, channel aging, users association, enhancements for handoff, and synchronization issues. Due to channel aging, robust pilot designs under channel aging are needed. In this regard, time-multiplexed pilots may not perform well when the delay of the received signals is taken into account in the study. However, even conventional pilot book designs that are based on the discrete Fourier Transform, Walsh-Hadamard matrix, or Zadoff-Chu sequences~\cite{sesia2011lte} cannot perform well because the orthogonality of the pilots will be destroyed due to channel aging. Thus, intelligent designs will be useful for such scenarios. As shown in~\cite{channelAgingMassiveMIMO8122014}, optimally selecting the frame duration can alleviate channel aging, so this can be a starting point to launch further studies.

Moreover, high-mobility will lead to complications in frequent redefinition of serving clusters for users, which will affect network performance. Hence, predictive and stable handoffs that do not degrade the performance are needed~\cite{504082, machineLearningChannelAging8979256, Synchronization4287203}. Additionally, metrics like handoff rate~\cite{Ravi7006787}, sojourn time~\cite{6477064}, and handoff probability~\cite{6477064} must be redefined for the user-centric cell-free networks, which will help in studying the effect of handoffs on network performance. For example, inside a multiple-CU network, two types of handoff can be defined. The first type can be denoted as intra-CU handoff, where the DUs involved belong to the same CU. While the second type is an inter-CU handoff, where the DUs involved belong to different CUs. Definitely, the inter-CU handoff will have more impact on the performance, because it requires more signaling.

\subsection{Resource Allocation}
As shown in~\eqref{eq:JointDetection}, the LSFD approach to perform joint multiuser data detection relies on gathering the data signals multiplied by the conjugate of the users' channels at the CU. However, in a multiple-CU network, this may not possible because the serving DUs may be under the control of different CUs. This means that a modification for the LSFD needs to be introduced, else it will not be directly applicable. 

Sharing the CSI between the serving DUs and CUs is beneficial for some beamforming techniques, e.g., ZF. However, this sharing imposes large load on the fronthaul and centralized beamforming. On the other hand, some beamforming techniques like local MMSE combining~\cite{differentCooperationLevels8845768}, the weighted MMSE~\cite{MSE5756489, ammarC_RA_UC}, and PPZF~\cite{LocaPartialZFBF9069486} can be a better candidate because they limit the CSI sharing. Thus, future studies should focus on the ability of limiting CSI sharing for a scalable beamforming scheme.

The high density of the deployed DUs in cell-free massive MIMO networks raises the possibility of LoS communication. Based on this insight, a probabilistic LoS channel model~\cite{ITU:P.1410_5, 8859647} can be optionally used in the studies. This can increase the reliability of the conclusions  being derived in the studies, however, it may increase the complexity of the study. All in all, using probabilistic LoS channel models can be considered an option.

A useful topic is to study the effect of distributing (or co-locating) the antennas on more (or less) DUs. This topic can be studied under different scenarios like limited fronthaul capacity or wireless fronthaul for example. In~\cite{ammarWirelessbackahulJournal}, this topic is studied under wireless fronthaul, where the locations of the DUs is optimized, however, the adopted network scheme is a cell-centric scheme.

One important direction for future research for cell-free communications is to focus on implementing distributed resource allocation schemes. The distributed implementation is very attractive as it allows an easy deployment and lower computational complexity on DUs and CUs. It also allows for lower load on the fronthaul, because the information needed to perform resource allocation, e.g., the CSI, does not need to be transmitted to a single node in the network.

In the literature, mathematical tools like game theory and stable matching theory do not seem to be used in a cell-free communications framework. Thus, there is a novelty in using such tools and they can definitely help in deriving important conclusions about the operation and performance of cell-free networks. Another less visited topic is in-band full-duplex communications for cell-free networks. It is important to investigate under which scenario and system parameters, a full-duplex scheme can be beneficial for cell-free communications.

\subsection{Delay, Serving Cluster, Other Areas}
Studies focusing on delay related metrics for cell-free communications are rare. The QSI is a main metric that should be taken into account when performing resource allocation. However, it is rarely studied in the literature. One novel topic is to study the effect of using multiple CUs on the delay, for example, the density of the CUs can be optimized. Another possible topic is to target the synchronization of transmissions for high mobility users.

Cell-free communication can be a key to boost the reliability constraint for uRLLC application. A critical topic to study is the tradeoffs between the data rate and the delay. In this regard, MEC can be a good platform to manage delay-sensitive applications. Thus, the use of multiple CUs in a network to implement MEC needs to be investigated. Punctured scheduling is another novel topic that is not studied for user-centric cell-free networks. 

The integration of management platforms such as the distributed SDN and self-organizing networks (SON) platform to mange cell-free networks is critical. One of the important related topics is managing the construction of the serving clusters to allow a seamless service for the users. This is very important for network operators, especially with dense deployments of DUs. Moreover, the solutions proposed in the future should take into account scalability, so when the number of users grows the proposed solutions should be able to keep computational complexity manageable.

Usage of IRS in cell-free networks is still a novel topic. The current expected benefit for cell-free networks is an enhanced EE~\cite{IRS9308937, IRS9363171, IRS9298843, IRS9352948}. One of the future directions that can be tackled is to consider IRS-aided cell-free networks with frequency-selective channels.

\label{page:ML_future}So far, most optimization frameworks suffers from a main challenge, which is the computational complexity that could prevent the CUs and the DUs from taking real-time decisions. Machine learning come to the rescue in this matter because technically we can train our network offline to perform optimal decisions then deploy our trained neural network and perform real-time optimized decisions. As this survey paper is not specialized in machine learning, we refer the readers to dedicated survey papers on machine learning for wireless communications~\cite{surveyMLwirelessNet8743390}.


\section{Conclusion}\label{sec:conclusionFutureDirections}
We have presented a detailed survey of the main challenges, solutions and opportunities for user-centric cell-free massive MIMO networks. This architecture is very promising as it can revolutionize future mobile networks by alleviating the problem of cell-edge users and non-uniform coverage of the current cellular networks. Further, it can boost the network performance by providing enhanced connectivity, signal power, interference management, and macro diversity. This review paper provides a comprehensive starting point for readers interested in the topic, where we provide a wide range of subjects being discussed in the literature.

Based on the literature, the TDD system is more appropriate than FDD for the user-centric cell-free MIMO network because it allows for exploitation of channel reciprocity with hardware calibration. Currently, FDD systems may not be desirable, because the downlink pilots will depend on the number of DUs~\cite{8768014, interdonato2019ubiquitous}. It also may not be possible for communication scenarios with short coherence times. It is worth emphasizing that even TDD systems suffer when coherence times are short, because pilot training, uplink and downlink data transmission need to be executed in one coherence time. Generally, deploying an FDD system may require advanced signal processing techniques and exploiting of methods such as angle of arrival reciprocity. Furthermore, exchanging CSI between a wide range of DUs will not be convenient~\cite{channelAcquisition5462883}, thus schemes that use local CSI are very attractive.

There are some promising techniques to intelligently use the fronthaul capacity. New fronthaul solutions that can integrate both the DUs and their wired connections are good candidates to deploy the fronthaul. However, such solutions require low deployment and maintenance costs. A candidate technology can be the radio stripes system~\cite{frenger2019antenna, radioStripsSystem}. Wireless fronthaul solutions can be used to backup the wired fronthaul and can be used in places where a wired fronthaul is not possible.

User-centric cell-free MIMO and millimeter wave systems can be a very interesting combination~\cite{DeepLearning8815888, mmWaveConnectivity8579590, mmwave6732923, EnergyEfficiencyMmwave8676377, EECellFreeUCMMwave8292302, EECellFreeUCMMwave8516938, 9149862} to provide a wireless Gbit/s experience. Cell-free communication can enhance the connectivity of the millimeter wave communication, and the small range and large bandwidth of millimeter wave can help in deploying a very dense cell-free network. In this regard, the two technologies complement each other. 

SDN seems a promising platform to deploy a user-centric cell-free MIMO network. The benefit of such platform lies in allowing the deployment of multiple-CU cell-free networks with dynamic association between the CUs and the DUs. Also, management platforms such as SON can allow the deployment of optimization algorithms for resource allocation. However, complete network models and solutions still need to be built to study the capabilities of such platforms.

Developing distributed solutions for resource allocation is a crucial research direction. Densifying the network will push the DUs to not share the CSI or limit this sharing to wired connections. Moreover, optimization schemes that can solely depend on long-term channel statistics without a large loss in performance are very important. The performance can be dependent on the degree that the channel hardening and favorable propagation apply.

The performance of high-mobility users is still an open research area that is rarely visited in the literature. Studying the effect of high-mobility still needs to be investigated, and the effect of topics such as channel aging, pilot design and channel estimation for high speed users, frequent formation of serving clusters and synchronization issues. Studies targeting latency under user-centric cell-free massive MIMO networks are rarely investigated. Hence, future works studying this imporatnt metric are needed, especially for uRLLC applications.

In summary, we have developed a comprehensive view of the motivation, state-of-the-art and future of user-centric, cell-free, MIMO networks.

\ifCLASSOPTIONcaptionsoff
\newpage
\fi

\footnotesize
\bibliography{Review_HA_References}

\begin{thebibliography}{100}

\bibitem{cellFreeUserCentricPower8901451}
S.~{Buzzi}, C.~{D’Andrea}, A.~{Zappone}, and C.~{D’Elia}, ``User-centric
  {5G} cellular networks: Resource allocation and comparison with the cell-free
  massive {MIMO} approach,'' {\em IEEE Transactions on Wireless
  Communications}, vol.~19, pp.~1250--1264, Feb 2020.

\bibitem{cellFreeVersusSmallCells7827017}
H.~Q. {Ngo}, A.~{Ashikhmin}, H.~{Yang}, E.~G. {Larsson}, and T.~L. {Marzetta},
  ``Cell-free massive {MIMO} versus small cells,'' {\em IEEE Transactions on
  Wireless Communications}, vol.~16, pp.~1834--1850, March 2017.

\bibitem{differentCooperationLevels8845768}
E.~{Björnson} and L.~{Sanguinetti}, ``Making cell-free massive {MIMO}
  competitive with {MMSE} processing and centralized implementation,'' {\em
  IEEE Transactions on Wireless Communications}, vol.~19, pp.~77--90, Jan 2020.

\bibitem{8385475}
G.~{Interdonato}, P.~{Frenger}, and E.~G. {Larsson}, ``Utility-based downlink
  pilot assignment in cell-free massive {MIMO},'' in {\em WSA 2018; 22nd
  International ITG Workshop on Smart Antennas}, pp.~1--8, March 2018.

\bibitem{han2019sparse}
D.~Han, J.~Park, and N.~Lee, ``Sparse joint transmission for cell-free massive
  {MIMO}: A sparse {PCA} approach,'' {\em arXiv preprint arXiv:1912.05231},
  2019.

\bibitem{8901196}
Z.~{Chen}, E.~{Björnson}, and E.~G. {Larsson}, ``Dynamic resource allocation
  in co-located and cell-free massive {MIMO},'' {\em IEEE Transactions on Green
  Communications and Networking}, pp.~1--1, 2019.

\bibitem{powerControlCellFree7917284}
E.~{Nayebi}, A.~{Ashikhmin}, T.~L. {Marzetta}, H.~{Yang}, and B.~D. {Rao},
  ``Precoding and power optimization in cell-free massive {MIMO} systems,''
  {\em IEEE Transactions on Wireless Communications}, vol.~16, pp.~4445--4459,
  July 2017.

\bibitem{HardwareImpairements8891922}
H.~{Masoumi} and M.~J. {Emadi}, ``Performance analysis of cell-free massive
  {MIMO} system with limited fronthaul capacity and hardware impairments,''
  {\em IEEE Transactions on Wireless Communications}, vol.~19, no.~2,
  pp.~1038--1053, 2020.

\bibitem{LocaPartialZFBF9069486}
G.~{Interdonato}, M.~{Karlsson}, E.~{Björnson}, and E.~G. {Larsson}, ``Local
  partial zero-forcing precoding for cell-free massive {MIMO},'' {\em IEEE
  Transactions on Wireless Communications}, pp.~1--1, 2020.

\bibitem{9113273}
J.~{Zhang}, E.~{Björnson}, M.~{Matthaiou}, D.~W.~K. {Ng}, H.~{Yang}, and D.~J.
  {Love}, ``Prospective multiple antenna technologies for beyond 5g,'' {\em
  IEEE Journal on Selected Areas in Communications}, vol.~38, no.~8,
  pp.~1637--1660, 2020.

\bibitem{heath2018foundations}
R.~W. Heath~Jr and A.~Lozano, {\em Foundations of {MIMO} communication}.
\newblock Cambridge University Press, 2018.

\bibitem{marzetta2016fundamentals}
T.~L. {Marzetta}, E.~G. {Larsson}, H.~{Yang}, and H.~Q. {Ngo}, {\em
  Fundamentals of massive {MIMO}}.
\newblock Cambridge University Press, 2016.

\bibitem{massiveMIMO5595728}
T.~L. {Marzetta}, ``Noncooperative cellular wireless with unlimited numbers of
  base station antennas,'' {\em IEEE Transactions on Wireless Communications},
  vol.~9, no.~11, pp.~3590--3600, 2010.

\bibitem{massiveMIMO6798744}
L.~{Lu}, G.~Y. {Li}, A.~L. {Swindlehurst}, A.~{Ashikhmin}, and R.~{Zhang}, ``An
  overview of massive mimo: Benefits and challenges,'' {\em IEEE Journal of
  Selected Topics in Signal Processing}, vol.~8, no.~5, pp.~742--758, 2014.

\bibitem{massiveMIMODistributed6810508}
K.~T. {Truong} and R.~W. {Heath}, ``The viability of distributed antennas for
  massive {MIMO} systems,'' in {\em 2013 Asilomar Conference on Signals,
  Systems and Computers}, pp.~1318--1323, 2013.

\bibitem{massiveMIMO7080890}
E.~{Björnson}, M.~{Matthaiou}, and M.~{Debbah}, ``Massive {MIMO} with
  non-ideal arbitrary arrays: Hardware scaling laws and circuit-aware design,''
  {\em IEEE Transactions on Wireless Communications}, vol.~14, no.~8,
  pp.~4353--4368, 2015.

\bibitem{massiveMIMO6736761}
E.~G. {Larsson}, O.~{Edfors}, F.~{Tufvesson}, and T.~L. {Marzetta}, ``Massive
  {MIMO} for next generation wireless systems,'' {\em IEEE Communications
  Magazine}, vol.~52, no.~2, pp.~186--195, 2014.

\bibitem{PDPUsercentricVsDisjoint8969384}
H.~A. {Ammar} and R.~{Adve}, ``Power delay profile in coordinated distributed
  networks: User-centric v/s disjoint clustering,'' in {\em 2019 IEEE Global
  Conference on Signal and Information Processing (GlobalSIP)}, pp.~1--5, Nov
  2019.

\bibitem{clustersbasedonNetperf7105966}
D.~{Liu}, S.~{Han}, C.~{Yang}, and Q.~{Zhang}, ``Semi-dynamic user-specific
  clustering for downlink cloud radio access network,'' {\em IEEE Transactions
  on Vehicular Technology}, vol.~65, no.~4, pp.~2063--2077, 2016.

\bibitem{ammarC_RA_UC}
H.~A. Ammar, R.~Adve, S.~Shahbazpanahi, G.~Boudreau, and K.~V. Srinivas,
  ``Downlink resource allocation in multiuser cell-free {MIMO} networks with
  user-centric clustering,'' {\em IEEE Transactions on Wireless
  Communications}, pp.~1--1, 2021.

\bibitem{ammarDistributed_RA_UC}
H.~A. Ammar, R.~Adve, S.~Shahbazpanahi, G.~Boudreau, and K.~V. Srinivas,
  ``Distributed resource allocation optimization for user-centric cell-free
  {MIMO} networks,'' {\em IEEE Transactions on Wireless Communications},
  pp.~1--1, 2021.

\bibitem{8768014}
J.~{Zhang}, S.~{Chen}, Y.~{Lin}, J.~{Zheng}, B.~{Ai}, and L.~{Hanzo},
  ``Cell-free massive {MIMO}: A new next-generation paradigm,'' {\em IEEE
  Access}, vol.~7, pp.~99878--99888, 2019.

\bibitem{EnergyE2020towards}
A.~Papazafeiropoulos, H.~Q. Ngo, P.~Kourtessis, S.~Chatzinotas, and J.~M.
  Senior, ``Towards optimal energy efficiency in cell-free massive {MIMO}
  systems,'' {\em arXiv preprint arXiv:2005.07459}, 2020.

\bibitem{DelayAware6180015}
Y.~{Cui}, V.~K.~N. {Lau}, and Y.~{Wu}, ``Delay-aware {BS} discontinuous
  transmission control and user scheduling for energy harvesting downlink
  coordinated {MIMO} systems,'' {\em IEEE Transactions on Signal Processing},
  vol.~60, pp.~3786--3795, July 2012.

\bibitem{8886730}
P.~{Liu}, K.~{Luo}, D.~{Chen}, and T.~{Jiang}, ``Spectral efficiency analysis
  of cell-free massive {MIMO} systems with zero-forcing detector,'' {\em IEEE
  Transactions on Wireless Communications}, vol.~19, no.~2, pp.~795--807, 2020.

\bibitem{cellFreeStochasticGeom8972478}
A.~{Papazafeiropoulos}, P.~{Kourtessis}, M.~D. {Renzo}, S.~{Chatzinotas}, and
  J.~M. {Senior}, ``Performance analysis of cell-free massive {MIMO} systems: A
  stochastic geometry approach,'' {\em IEEE Transactions on Vehicular
  Technology}, vol.~69, no.~4, pp.~3523--3537, 2020.

\bibitem{UserCentricvsCellFreeBackhaul8000355}
S.~{Buzzi} and C.~{D’Andrea}, ``Cell-free massive {MIMO}: User-centric
  approach,'' {\em IEEE Wireless Communications Letters}, vol.~6, no.~6,
  pp.~706--709, 2017.

\bibitem{EnergyEfficiencyMmwave8676377}
M.~{Alonzo}, S.~{Buzzi}, A.~{Zappone}, and C.~{D’Elia}, ``Energy-efficient
  power control in cell-free and user-centric massive {MIMO} at millimeter
  wave,'' {\em IEEE Transactions on Green Communications and Networking},
  vol.~3, no.~3, pp.~651--663, 2019.

\bibitem{EECellFreeUCMMwave8292302}
M.~{Alonzo} and S.~{Buzzi}, ``Cell-free and user-centric massive {MIMO} at
  millimeter wave frequencies,'' in {\em 2017 IEEE 28th Annual International
  Symposium on Personal, Indoor, and Mobile Radio Communications (PIMRC)},
  pp.~1--5, 2017.

\bibitem{EECellFreeUCMMwave8516938}
M.~{Alonzo}, S.~{Buzzi}, and A.~{Zappone}, ``Energy-efficient downlink power
  control in mmwave cell-free and user-centric massive mimo,'' in {\em 2018
  IEEE 5G World Forum (5GWF)}, pp.~493--496, 2018.

\bibitem{NetworkMIMO6095627}
H.~{Huh}, A.~M. {Tulino}, and G.~{Caire}, ``Network {MIMO} with linear
  zero-forcing beamforming: Large system analysis, impact of channel
  estimation, and reduced-complexity scheduling,'' {\em IEEE Transactions on
  Information Theory}, vol.~58, pp.~2911--2934, May 2012.

\bibitem{3GPP:TS36.819}
3GPP, {\em Coordinated multi-point operation for {LTE} physical layer aspects},
  2013.
\newblock
  \url{http://www.3gpp.org/ftp/Specs/archive/36_series/36.819/36819-b20.zip}.

\bibitem{CRAN6897914}
A.~{Checko}, H.~L. {Christiansen}, Y.~{Yan}, L.~{Scolari}, G.~{Kardaras}, M.~S.
  {Berger}, and L.~{Dittmann}, ``Cloud {RAN} for mobile networks—{A}
  technology overview,'' {\em IEEE Communications Surveys and Tutorials},
  vol.~17, pp.~405--426, Firstquarter 2015.

\bibitem{MultiCellMIMO5594708}
D.~{Gesbert}, S.~{Hanly}, H.~{Huang}, S.~{Shamai Shitz}, O.~{Simeone}, and
  W.~{Yu}, ``Multi-cell {MIMO} cooperative networks: {A} new look at
  interference,'' {\em IEEE Journal on Selected Areas in Communications},
  vol.~28, no.~9, pp.~1380--1408, 2010.

\bibitem{VirtualMIMO6601776}
X.~{Hong}, Y.~{Jie}, C.~{Wang}, J.~{Shi}, and X.~{Ge}, ``Energy-spectral
  efficiency trade-off in virtual {MIMO} cellular systems,'' {\em IEEE Journal
  on Selected Areas in Communications}, vol.~31, no.~10, pp.~2128--2140, 2013.

\bibitem{smallCell6525591}
I.~{Hwang}, B.~{Song}, and S.~S. {Soliman}, ``A holistic view on hyper-dense
  heterogeneous and small cell networks,'' {\em IEEE Communications Magazine},
  vol.~51, no.~6, pp.~20--27, 2013.

\bibitem{CRANfronthaul8113473}
I.~A. {Alimi}, A.~L. {Teixeira}, and P.~P. {Monteiro}, ``Toward an efficient
  {C-RAN} optical fronthaul for the future networks: A tutorial on
  technologies, requirements, challenges, and solutions,'' {\em IEEE
  Communications Surveys and Tutorials}, vol.~20, no.~1, pp.~708--769, 2018.

\bibitem{perlman2015introduction}
S.~Perlman and A.~Forenza, ``An introduction to {pCell},'' {\em Artemis white
  paper}, 2015.

\bibitem{burr2018ultra}
A.~Burr, M.~Bashar, and D.~Maryopi, ``Ultra-dense radio access networks for
  smart cities: Cloud-{RAN}, fog-{RAN} and" cell-free" massive {MIMO},'' {\em
  arXiv preprint arXiv:1811.11077}, 2018.

\bibitem{3GPPTR21.915}
3GPP TR 21.915, {\em Release 15 Description, Technical Specification Group
  Services and System Aspects}, 2019.
\newblock \url{www.3gpp.org/ftp/Specs/archive/21_series/21.915/}.

\bibitem{StochasticUserCentric8449213}
C.~{Zhu} and W.~{Yu}, ``Stochastic modeling and analysis of user-centric
  network {MIMO} systems,'' {\em IEEE Transactions on Communications}, vol.~66,
  no.~12, pp.~6176--6189, 2018.

\bibitem{9120231}
F.~{Marzouk}, J.~P. {Barraca}, and A.~{Radwan}, ``On energy efficient resource
  allocation in shared {RANs}: Survey and qualitative analysis,'' {\em IEEE
  Communications Surveys and Tutorials}, vol.~22, no.~3, pp.~1515--1538, 2020.

\bibitem{7839266}
S.~{Bassoy}, H.~{Farooq}, M.~A. {Imran}, and A.~{Imran}, ``Coordinated
  multi-point clustering schemes: A survey,'' {\em IEEE Communications Surveys
  Tutorials}, vol.~19, pp.~743--764, Secondquarter 2017.

\bibitem{CRAN7018201}
J.~{Wu}, Z.~{Zhang}, Y.~{Hong}, and Y.~{Wen}, ``Cloud radio access network
  {(C-RAN)}: a primer,'' {\em IEEE Network}, vol.~29, no.~1, pp.~35--41, 2015.

\bibitem{RA_CRAN7143328}
H.~{Dahrouj}, A.~{Douik}, O.~{Dhifallah}, T.~Y. {Al-Naffouri}, and
  M.~{Alouini}, ``Resource allocation in heterogeneous cloud radio access
  networks: advances and challenges,'' {\em IEEE Wireless Communications},
  vol.~22, no.~3, pp.~66--73, 2015.

\bibitem{backhaulSmallCells7306536}
N.~{Wang}, E.~{Hossain}, and V.~K. {Bhargava}, ``Backhauling {5G} small cells:
  A radio resource management perspective,'' {\em IEEE Wireless
  Communications}, vol.~22, no.~5, pp.~41--49, 2015.

\bibitem{wirelessBackhaulSmallCells7306534}
U.~{Siddique}, H.~{Tabassum}, E.~{Hossain}, and D.~I. {Kim}, ``Wireless
  backhauling of {5G} small cells: challenges and solution approaches,'' {\em
  IEEE Wireless Communications}, vol.~22, no.~5, pp.~22--31, 2015.

\bibitem{LimitsOfCooperation6482234}
A.~{Lozano}, R.~W. {Heath}, and J.~G. {Andrews}, ``Fundamental limits of
  cooperation,'' {\em IEEE Transactions on Information Theory}, vol.~59, no.~9,
  pp.~5213--5226, 2013.

\bibitem{interdonato2019ubiquitous}
G.~Interdonato, E.~Bj{o}rnson, H.~Q. Ngo, P.~Frenger, and E.~G. Larsson,
  ``Ubiquitous cell-free massive {MIMO} communications,'' {\em EURASIP Journal
  on Wireless Communications and Networking}, vol.~2019, no.~1, p.~197, 2019.

\bibitem{8761828}
G.~{Interdonato}, P.~{Frenger}, and E.~G. {Larsson}, ``Scalability aspects of
  cell-free massive {MIMO},'' in {\em ICC 2019 - 2019 IEEE International
  Conference on Communications (ICC)}, pp.~1--6, 2019.

\bibitem{bjornson2019scalable9064545}
E.~{Björnson} and L.~{Sanguinetti}, ``Scalable cell-free massive {MIMO}
  systems,'' {\em IEEE Transactions on Communications}, pp.~1--1, 2020.

\bibitem{mmWaveHierarchicalBackhaul7904705}
Y.~{Chiang} and W.~{Liao}, ``{mw-HierBack}: A cost-effective and robust
  millimeter wave hierarchical backhaul solution for {HetNets},'' {\em IEEE
  Transactions on Mobile Computing}, vol.~16, no.~12, pp.~3445--3458, 2017.

\bibitem{distributedSDN8187644}
F.~{Bannour}, S.~{Souihi}, and A.~{Mellouk}, ``Distributed {SDN} control:
  Survey, taxonomy, and challenges,'' {\em IEEE Communications Surveys and
  Tutorials}, vol.~20, no.~1, pp.~333--354, 2018.

\bibitem{distributedSDN6838330}
K.~{Phemius}, M.~{Bouet}, and J.~{Leguay}, ``{DISCO}: Distributed multi-domain
  {SDN} controllers,'' in {\em 2014 IEEE Network Operations and Management
  Symposium (NOMS)}, pp.~1--4, 2014.

\bibitem{DynamicMapping8108113}
H.~A. {Ammar}, Y.~{Nasser}, and A.~{Kayssi}, ``Dynamic {SDN}
  controllers-switches mapping for load balancing and controller failure
  handling,'' in {\em 2017 International Symposium on Wireless Communication
  Systems (ISWCS)}, pp.~216--221, 2017.

\bibitem{frenger2019antenna}
P.~Frenger, J.~Hederen, M.~Hessler, and G.~Interdonato, ``Antenna arrangement
  for distributed massive {MIMO},'' Nov.~28 2019.
\newblock US Patent App. 16/435,054.

\bibitem{radioStripsSystem}
E.~Ernfors, ``Radio stripes: re-thinking mobile networks,'' Feb 25, 2019.

\bibitem{EnergyEfficiency8097026}
H.~Q. {Ngo}, L.~{Tran}, T.~Q. {Duong}, M.~{Matthaiou}, and E.~G. {Larsson},
  ``On the total energy efficiency of cell-free massive {MIMO},'' {\em IEEE
  Transactions on Green Communications and Networking}, vol.~2, no.~1,
  pp.~25--39, 2018.

\bibitem{UC_CellFreOpticalFronthaule020}
P.~Agheli, M.~Emadi, and H.~Beyranvand, ``Performance analysis of cell-free and
  user-centric {MIMO} networks with optical fronthaul and backhaul links,''
  {\em arXiv preprint arXiv:2011.06680}, 2020.

\bibitem{9130689}
X.~{Zhang}, J.~{Wang}, and H.~V. {Poor}, ``Statistical delay and error-rate
  bounded qos provisioning over {mmWave} cell-free {M-MIMO} and {FBC-HARQ-IR}
  based {6G} wireless networks,'' {\em IEEE Journal on Selected Areas in
  Communications}, vol.~38, no.~8, pp.~1661--1677, 2020.

\bibitem{FullDuplex9110914}
H.~V. {Nguyen}, V.~D. {Nguyen}, O.~A. {Dobre}, S.~K. {Sharma},
  S.~{Chatzinotas}, B.~{Ottersten}, and O.~S. {Shin}, ``On the spectral and
  energy efficiencies of full-duplex cell-free massive {MIMO},'' {\em IEEE
  Journal on Selected Areas in Communications}, vol.~38, no.~8, pp.~1698--1718,
  2020.

\bibitem{HardwareImpairements9004558}
J.~{Zheng}, J.~{Zhang}, L.~{Zhang}, X.~{Zhang}, and B.~{Ai}, ``Efficient
  receiver design for uplink cell-free massive {MIMO} with hardware
  impairments,'' {\em IEEE Transactions on Vehicular Technology}, vol.~69,
  no.~4, pp.~4537--4541, 2020.

\bibitem{clusters6415394}
M.~{Hong}, R.~{Sun}, H.~{Baligh}, and Z.~{Luo}, ``Joint base station clustering
  and beamformer design for partial coordinated transmission in heterogeneous
  networks,'' {\em IEEE Journal on Selected Areas in Communications}, vol.~31,
  pp.~226--240, February 2013.

\bibitem{NoncoherentCRAN8482453}
C.~{Pan}, H.~{Ren}, M.~{Elkashlan}, A.~{Nallanathan}, and L.~{Hanzo}, ``The
  non-coherent ultra-dense {C-RAN} is capable of outperforming its coherent
  counterpart at a limited fronthaul capacity,'' {\em IEEE Journal on Selected
  Areas in Communications}, vol.~36, no.~11, pp.~2549--2560, 2018.

\bibitem{ammarC_RA_UC_conf}
H.~A. Ammar, R.~Adve, S.~Shahbazpanahi, G.~Boudreau, and K.~Srinivas,
  ``Resource allocation and scheduling in non-coherent user-centric cell-free
  {MIMO},'' in {\em ICC 2021 - IEEE International Conference on
  Communications}, pp.~1--6, 2021.

\bibitem{ammarWirelessbackahulJournal}
H.~A. Ammar, R.~Adve, S.~Shahbazpanahi, and G.~Boudreau, ``Analysis and design
  of distributed {MIMO} networks with a wireless fronthaul,'' {\em IEEE
  Transactions on Communications}, pp.~1--1, 2021.

\bibitem{mckeown2008openflow}
N.~M. \textit{et al.}, ``Openflow: enabling innovation in campus networks,''
  {\em ACM SIGCOMM computer communication review}, vol.~38, no.~2, pp.~69--74,
  2008.

\bibitem{floodlightController}
Project Floodlight, {\em Floodlight Controller}, 2021.
\newblock \url{https://floodlight.atlassian.net/wiki}.

\bibitem{maxMinRate8756286}
M.~{Bashar}, K.~{Cumanan}, A.~G. {Burr}, H.~Q. {Ngo}, M.~{Debbah}, and
  P.~{Xiao}, ``Max–min rate of cell-free massive {MIMO} uplink with optimal
  uniform quantization,'' {\em IEEE Transactions on Communications}, vol.~67,
  no.~10, pp.~6796--6815, 2019.

\bibitem{DeepLearningFronthaul9110901}
M.~{Bashar}, A.~{Akbari}, K.~{Cumanan}, H.~Q. {Ngo}, A.~G. {Burr}, P.~{Xiao},
  M.~{Debbah}, and J.~{Kittler}, ``Exploiting deep learning in
  limited-fronthaul cell-free massive {MIMO} uplink,'' {\em IEEE Journal on
  Selected Areas in Communications}, vol.~38, no.~8, pp.~1678--1697, 2020.

\bibitem{8693830}
S.~{Das} and M.~{Ruffini}, ``A variable rate fronthaul scheme for cloud radio
  access networks,'' {\em Journal of Lightwave Technology}, vol.~37,
  pp.~3153--3165, July 2019.

\bibitem{7120102}
T.~A. {Khan}, P.~{Orlik}, K.~J. {Kim}, and R.~W. {Heath}, ``Performance
  analysis of cooperative wireless networks with unreliable backhaul links,''
  {\em IEEE Communications Letters}, vol.~19, pp.~1386--1389, Aug 2015.

\bibitem{7981320}
H.~T. {Nguyen}, J.~{Zhang}, N.~{Yang}, T.~Q. {Duong}, and W.~{Hwang}, ``Secure
  cooperative single carrier systems under unreliable backhaul and dense
  networks impact,'' {\em IEEE Access}, vol.~5, pp.~18310--18324, 2017.

\bibitem{limitedFronthaulMmwave8678745}
G.~{Femenias} and F.~{Riera-Palou}, ``Cell-free millimeter-wave massive {MIMO}
  systems with limited fronthaul capacity,'' {\em IEEE Access}, vol.~7,
  pp.~44596--44612, 2019.

\bibitem{5560739}
P.~{Zillmann}, ``Relationship between two distortion measures for memoryless
  nonlinear systems,'' {\em IEEE Signal Processing Letters}, vol.~17, no.~11,
  pp.~917--920, 2010.

\bibitem{bussgang1952crosscorrelation}
J.~J. Bussgang, ``Crosscorrelation functions of amplitude-distorted gaussian
  signals,'' 1952.

\bibitem{8730536}
D.~{Maryopi}, M.~{Bashar}, and A.~{Burr}, ``On the uplink throughput of zero
  forcing in cell-free massive {MIMO} with coarse quantization,'' {\em IEEE
  Transactions on Vehicular Technology}, vol.~68, no.~7, pp.~7220--7224, 2019.

\bibitem{EnergyEfficiency8781848}
M.~{Bashar}, K.~{Cumanan}, A.~G. {Burr}, H.~Q. {Ngo}, E.~G. {Larsson}, and
  P.~{Xiao}, ``Energy efficiency of the cell-free massive {MIMO} uplink with
  optimal uniform quantization,'' {\em IEEE Transactions on Green
  Communications and Networking}, vol.~3, no.~4, pp.~971--987, 2019.

\bibitem{EE9212395}
M.~{Bashar}, H.~Q. {Ngo}, K.~{Cumanan}, A.~G. {Burr}, P.~{Xiao},
  E.~{Björnson}, and E.~G. {Larsson}, ``Uplink spectral and energy efficiency
  of cell-free massive {MIMO} with optimal uniform quantization,'' {\em IEEE
  Transactions on Communications}, pp.~1--1, 2020.

\bibitem{6226311}
D.~{Samardzija}, J.~{Pastalan}, M.~{MacDonald}, S.~{Walker}, and
  R.~{Valenzuela}, ``Compressed transport of baseband signals in radio access
  networks,'' {\em IEEE Transactions on Wireless Communications}, vol.~11,
  no.~9, pp.~3216--3225, 2012.

\bibitem{Compression1456239}
A.~K. {Jain}, ``Image data compression: A review,'' {\em Proceedings of the
  IEEE}, vol.~69, no.~3, pp.~349--389, 1981.

\bibitem{ComputerAndForward7962724}
Q.~{Huang} and A.~{Burr}, ``Compute-and-forward in cell-free massive {MIMO}:
  Great performance with low backhaul load,'' in {\em 2017 IEEE International
  Conference on Communications Workshops (ICC Workshops)}, pp.~601--606, 2017.

\bibitem{4787140}
Y.~H. {Gan}, C.~{Ling}, and W.~H. {Mow}, ``Complex lattice reduction algorithm
  for low-complexity full-diversity {MIMO} detection,'' {\em IEEE Transactions
  on Signal Processing}, vol.~57, no.~7, pp.~2701--2710, 2009.

\bibitem{cover1999elements}
T.~M. Cover, {\em Elements of information theory}.
\newblock John Wiley \& Sons, 1999.

\bibitem{6920005}
B.~{Dai} and W.~{Yu}, ``Sparse beamforming and user-centric clustering for
  downlink cloud radio access network,'' {\em IEEE Access}, vol.~2,
  pp.~1326--1339, 2014.

\bibitem{7581201}
C.~{Hua}, Y.~{Luo}, and H.~{Liu}, ``Wireless backhaul resource allocation and
  user-centric clustering in ultra-dense wireless networks,'' {\em IET
  Communications}, vol.~10, no.~15, pp.~1858--1864, 2016.

\bibitem{CompressedSensing1614066}
D.~L. {Donoho}, ``Compressed sensing,'' {\em IEEE Transactions on Information
  Theory}, vol.~52, pp.~1289--1306, April 2006.

\bibitem{7287780}
V.~N. {Ha}, L.~B. {Le}, and N.~{Dao}, ``Coordinated multipoint transmission
  design for cloud-{RANs} with limited fronthaul capacity constraints,'' {\em
  IEEE Transactions on Vehicular Technology}, vol.~65, pp.~7432--7447, Sep.
  2016.

\bibitem{channelAcquisition5462883}
E.~{Björnson}, R.~{Zakhour}, D.~{Gesbert}, and B.~{Ottersten}, ``Cooperative
  multicell precoding: Rate region characterization and distributed strategies
  with instantaneous and statistical {CSI},'' {\em IEEE Transactions on Signal
  Processing}, vol.~58, no.~8, pp.~4298--4310, 2010.

\bibitem{7996634}
B.~{Huang}, Y.~{Chiang}, and W.~{Liao}, ``Remote radio head ({RRH}) deployment
  in flexible {C-RAN} under limited fronthaul capacity,'' in {\em 2017 IEEE
  International Conference on Communications (ICC)}, pp.~1--6, May 2017.

\bibitem{7837706}
A.~{Adhikary}, A.~{Ashikhmin}, and T.~L. {Marzetta}, ``Uplink interference
  reduction in large-scale antenna systems,'' {\em IEEE Transactions on
  Communications}, vol.~65, no.~5, pp.~2194--2206, 2017.

\bibitem{Scalable9174860}
S.~{Chen}, J.~{Zhang}, E.~{Björnson}, J.~{Zhang}, and B.~{Ai}, ``Structured
  massive access for scalable cell-free massive {MIMO} systems,'' {\em IEEE
  Journal on Selected Areas in Communications}, pp.~1--1, 2020.

\bibitem{ammarWirelessbackahulConf}
H.~A. Ammar, R.~Adve, S.~Shahbazpanahi, and G.~Boudreau, ``Optimizing {RRH}
  placement under a noise-limited point-to-point wireless backhaul,'' in {\em
  ICC 2021 - IEEE International Conference on Communications}, pp.~1--6, 2021.

\bibitem{Arin8529184}
A.~{Minasian}, R.~S. {Adve}, S.~{Shahbazpanahi}, and G.~{Boudreau}, ``On {RRH}
  placement for multi-user distributed massive {MIMO} systems,'' {\em IEEE
  Access}, vol.~6, pp.~70597--70614, 2018.

\bibitem{5GBackhaulNetworks6963798}
X.~{Ge}, H.~{Cheng}, M.~{Guizani}, and T.~{Han}, ``{5G} wireless backhaul
  networks: challenges and research advances,'' {\em IEEE Network}, vol.~28,
  no.~6, pp.~6--11, 2014.

\bibitem{DeepLearning8815888}
Y.~{Jin}, J.~{Zhang}, S.~{Jin}, and B.~{Ai}, ``Channel estimation for cell-free
  {mmWave} massive {MIMO} through deep learning,'' {\em IEEE Transactions on
  Vehicular Technology}, vol.~68, no.~10, pp.~10325--10329, 2019.

\bibitem{3GPPTR38.874}
3GPP, {\em Study on Integrated Access and Backhaul}, 2018.
\newblock \url{3gpp.org/ftp/Specs/archive/38_series/38.874/}.

\bibitem{IAB9040265}
M.~{Polese}, M.~{Giordani}, T.~{Zugno}, A.~{Roy}, S.~{Goyal}, D.~{Castor}, and
  M.~{Zorzi}, ``Integrated access and backhaul in {5G} mmwave networks:
  Potential and challenges,'' {\em IEEE Communications Magazine}, vol.~58,
  no.~3, pp.~62--68, 2020.

\bibitem{WSRjointAccessBack8786917}
E.~{Chen}, M.~{Tao}, and N.~{Zhang}, ``User-centric joint access-backhaul
  design for full-duplex self-backhauled wireless networks,'' {\em IEEE
  Transactions on Communications}, vol.~67, pp.~7980--7993, Nov 2019.

\bibitem{6844864}
M.~A. {Khalighi} and M.~{Uysal}, ``Survey on free space optical communication:
  A communication theory perspective,'' {\em IEEE Communications Surveys
  Tutorials}, vol.~16, no.~4, pp.~2231--2258, 2014.

\bibitem{powerAllocEnergyEfficiency9136914}
T.~{Van Chien}, E.~{Björnson}, and E.~G. {Larsson}, ``Joint power allocation
  and load balancing optimization for energy-efficient cell-free massive {MIMO}
  networks,'' {\em IEEE Transactions on Wireless Communications}, vol.~19,
  no.~10, pp.~6798--6812, 2020.

\bibitem{FeedbackOverhead6449246}
X.~{Hou} and C.~{Yang}, ``Feedback overhead analysis for base station
  cooperative transmission,'' {\em IEEE Transactions on Wireless
  Communications}, vol.~15, pp.~4491--4504, July 2016.

\bibitem{DLTrainingCSI8799031}
G.~{Interdonato}, H.~Q. {Ngo}, P.~{Frenger}, and E.~G. {Larsson}, ``Downlink
  training in cell-free massive {MIMO}: A blessing in disguise,'' {\em IEEE
  Transactions on Wireless Communications}, vol.~18, no.~11, pp.~5153--5169,
  2019.

\bibitem{Rician9099874}
S.~{Jin}, D.~{Yue}, and H.~H. {Nguyen}, ``Spectral and energy efficiency in
  cell-free massive {MIMO} systems over correlated rician fading,'' {\em IEEE
  Systems Journal}, pp.~1--12, 2020.

\bibitem{kay1993fundamentals}
S.~M. Kay, {\em Fundamentals of statistical signal processing}.
\newblock Prentice Hall PTR, 1993.

\bibitem{ChannelNonReciprocity8910383}
J.~{Morte Palacios}, O.~{Raeesi}, A.~{Gokceoglu}, and M.~{Valkama}, ``Impact of
  channel non-reciprocity in cell-free massive {MIMO},'' {\em IEEE Wireless
  Communications Letters}, vol.~9, no.~3, pp.~344--348, 2020.

\bibitem{SLNRnonIdealReciprocity6589033}
S.~{Han}, C.~{Yang}, G.~{Wang}, D.~{Zhu}, and M.~{Lei}, ``Coordinated
  multi-point transmission strategies for {TDD} systems with non-ideal channel
  reciprocity,'' {\em IEEE Transactions on Communications}, vol.~61,
  pp.~4256--4270, October 2013.

\bibitem{o2017mimo}
C.~O'keeffe and M.~O'brien, ``{MIMO} antenna calibration device, integrated
  circuit and method for compensating phase mismatch,'' Apr.~18 2017.
\newblock US Patent 9,628,256.

\bibitem{stayton2011systems}
G.~T. Stayton, ``Systems and methods for antenna calibration,'' Nov.~1 2011.
\newblock US Patent 8,049,662.

\bibitem{PowerClusteringCellFreeFDD8968400}
A.~{Abdallah} and M.~M. {Mansour}, ``Efficient angle-domain processing for
  {FDD}-based cell-free massive {MIMO} systems,'' {\em IEEE Transactions on
  Communications}, pp.~1--1, 2020.

\bibitem{32276}
R.~{Roy} and T.~{Kailath}, ``{ESPRIT}-estimation of signal parameters via
  rotational invariance techniques,'' {\em IEEE Transactions on Acoustics,
  Speech, and Signal Processing}, vol.~37, no.~7, pp.~984--995, 1989.

\bibitem{8446023}
S.~{Kim} and B.~{Shim}, ``{FDD}-based cell-free massive {MIMO} systems,'' in
  {\em 2018 IEEE 19th International Workshop on Signal Processing Advances in
  Wireless Communications (SPAWC)}, pp.~1--5, 2018.

\bibitem{7524027}
H.~{Xie}, F.~{Gao}, S.~{Zhang}, and S.~{Jin}, ``A unified transmission strategy
  for {TDD/FDD} massive {MIMO} systems with spatial basis expansion model,''
  {\em IEEE Transactions on Vehicular Technology}, vol.~66, no.~4,
  pp.~3170--3184, 2017.

\bibitem{cellFreeFDD9014542}
S.~{Kim}, J.~W. {Choi}, and B.~{Shim}, ``Downlink pilot precoding and
  compressed channel feedback for {FDD}-based cell-free systems,'' {\em IEEE
  Transactions on Wireless Communications}, vol.~19, no.~6, pp.~3658--3672,
  2020.

\bibitem{CoherentCRAN8606433}
C.~{Pan}, H.~{Ren}, M.~{Elkashlan}, A.~{Nallanathan}, and L.~{Hanzo},
  ``Weighted sum-rate maximization for the ultra-dense user-centric {TDD}
  {C-RAN} downlink relying on imperfect {CSI},'' {\em IEEE Tran. on Wireless
  Comm.}, vol.~18, no.~2, pp.~1182--1198, 2019.

\bibitem{FFDNet8365806}
K.~{Zhang}, W.~{Zuo}, and L.~{Zhang}, ``{FFDNet}: Toward a fast and flexible
  solution for {CNN}-based image denoising,'' {\em IEEE Transactions on Image
  Processing}, vol.~27, no.~9, pp.~4608--4622, 2018.

\bibitem{EAsymptotic6457363}
H.~Q. {Ngo}, E.~G. {Larsson}, and T.~L. {Marzetta}, ``Energy and spectral
  efficiency of very large multiuser {MIMO} systems,'' {\em IEEE Transactions
  on Communications}, vol.~61, no.~4, pp.~1436--1449, 2013.

\bibitem{FavProp7903703}
X.~{Wu}, N.~C. {Beaulieu}, and D.~{Liu}, ``On favorable propagation in massive
  {MIMO} systems and different antenna configurations,'' {\em IEEE Access},
  vol.~5, pp.~5578--5593, 2017.

\bibitem{ChannelHardening1327795}
B.~M. {Hochwald}, T.~L. {Marzetta}, and V.~{Tarokh}, ``Multiple-antenna channel
  hardening and its implications for rate feedback and scheduling,'' {\em IEEE
  Transactions on Information Theory}, vol.~50, no.~9, pp.~1893--1909, 2004.

\bibitem{ChannelHardeningDegree7880691}
H.~Q. {Ngo} and E.~G. {Larsson}, ``No downlink pilots are needed in {TDD}
  massive {MIMO},'' {\em IEEE Transactions on Wireless Communications},
  vol.~16, no.~5, pp.~2921--2935, 2017.

\bibitem{cellFreeStochasticGeometry8379438}
Z.~{Chen} and E.~{Björnson}, ``Channel hardening and favorable propagation in
  cell-free massive {MIMO} with stochastic geometry,'' {\em IEEE Transactions
  on Communications}, vol.~66, no.~11, pp.~5205--5219, 2018.

\bibitem{CorrelatedShadowing4357088}
Z.~{Wang}, E.~K. {Tameh}, and A.~R. {Nix}, ``Joint shadowing process in urban
  peer-to-peer radio channels,'' {\em IEEE Transactions on Vehicular
  Technology}, vol.~57, no.~1, pp.~52--64, 2008.

\bibitem{CorrelatedShadowing104090}
M.~{Gudmundson}, ``Correlation model for shadow fading in mobile radio
  systems,'' {\em Electronics Letters}, vol.~27, no.~23, pp.~2145--2146, 1991.

\bibitem{randomVsStructuredPilots8403508}
M.~{Attarifar}, A.~{Abbasfar}, and A.~{Lozano}, ``Random vs structured pilot
  assignment in cell-free massive {MIMO} wireless networks,'' in {\em 2018 IEEE
  International Conference on Communications Workshops (ICC Workshops)},
  pp.~1--6, 2018.

\bibitem{pilotAssign9178782}
S.~{Buzzi}, C.~{D’Andrea}, M.~{Fresia}, Y.~{Zhang}, and S.~{Feng}, ``Pilot
  assignment in cell-free massive {MIMO} based on the hungarian algorithm,''
  {\em IEEE Wireless Communications Letters}, pp.~1--1, 2020.

\bibitem{kuhn1955hungarian}
H.~W. Kuhn, ``The hungarian method for the assignment problem,'' {\em Naval
  research logistics quarterly}, vol.~2, no.~1-2, pp.~83--97, 1955.

\bibitem{bien2011hierarchical}
J.~{Bien} and R.~{Tibshirani}, ``Hierarchical clustering with prototypes via
  minimax linkage,'' {\em Journal of the American Statistical Association},
  vol.~106, no.~495, pp.~1075--1084, 2011.

\bibitem{karypis2000comparison}
M.~G. Karypis, V.~Kumar, and M.~Steinbach, ``A comparison of document
  clustering techniques,'' in {\em TextMining Workshop at KDD2000 (May 2000)},
  2000.

\bibitem{GraphCodePilotContamin8487005}
J.~H. {Sorensen}, E.~{de Carvalho}, C.~{Stefanovic}, and P.~{Popovski}, ``Coded
  pilot random access for massive {MIMO} systems,'' {\em IEEE Transactions on
  Wireless Communications}, vol.~17, no.~12, pp.~8035--8046, 2018.

\bibitem{masoumi2020cell}
H.~Masoumi, M.~J. Emadi, and S.~Buzzi, ``Cell-free massive {MIMO} with
  underlaid {D2D} communications and low resolution {ADCs},'' {\em arXiv
  preprint arXiv:2005.10068}, 2020.

\bibitem{GraphColoringPilot9110802}
H.~{Liu}, J.~{Zhang}, S.~{Jin}, and B.~{Ai}, ``Graph coloring based pilot
  assignment for cell-free massive {MIMO} systems,'' {\em IEEE Transactions on
  Vehicular Technology}, vol.~69, no.~8, pp.~9180--9184, 2020.

\bibitem{PiltAssigGraphColoring7217795}
X.~{Zhu}, L.~{Dai}, and Z.~{Wang}, ``Graph coloring based pilot allocation to
  mitigate pilot contamination for multi-cell massive {MIMO} systems,'' {\em
  IEEE Communications Letters}, vol.~19, no.~10, pp.~1842--1845, 2015.

\bibitem{graphColoringUsercentric7080877}
Z.~{Chen}, X.~{Hou}, and C.~{Yang}, ``Training resource allocation for
  user-centric base station cooperation networks,'' {\em IEEE Transactions on
  Vehicular Technology}, vol.~65, no.~4, pp.~2729--2735, 2016.

\bibitem{ashikhmin2017pilot}
A.~Ashikhmin, H.~Q. Ngo, T.~L. Marzetta, and H.~Yang, ``Pilot assignment in
  cell free massive {MIMO} wireless systems,'' Apr.~4 2017.
\newblock US Patent 9,615,384.

\bibitem{pilotPowerControl8450041}
T.~C. {Mai}, H.~Q. {Ngo}, M.~{Egan}, and T.~Q. {Duong}, ``Pilot power control
  for cell-free massive {MIMO},'' {\em IEEE Transactions on Vehicular
  Technology}, vol.~67, no.~11, pp.~11264--11268, 2018.

\bibitem{8914726}
H.~{Liu}, J.~{Zhang}, X.~{Zhang}, A.~{Kurniawan}, T.~{Juhana}, and B.~{Ai},
  ``Tabu-search-based pilot assignment for cell-free massive {MIMO} systems,''
  {\em IEEE Transactions on Vehicular Technology}, vol.~69, no.~2,
  pp.~2286--2290, 2020.

\bibitem{TabuSearch5957382}
N.~{Srinidhi}, T.~{Datta}, A.~{Chockalingam}, and B.~S. {Rajan}, ``Layered tabu
  search algorithm for large-{MIMO} detection and a lower bound on {ML}
  performance,'' {\em IEEE Transactions on Communications}, vol.~59, no.~11,
  pp.~2955--2963, 2011.

\bibitem{CSI6415397}
H.~{Yin}, D.~{Gesbert}, M.~{Filippou}, and Y.~{Liu}, ``A coordinated approach
  to channel estimation in large-scale multiple-antenna systems,'' {\em IEEE
  Journal on Selected Areas in Communications}, vol.~31, no.~2, pp.~264--273,
  2013.

\bibitem{uplinkPilotcellFreeMassiveMIMO2020}
J.~Gao, Y.~Wu, Y.~Wang, W.~Zhang, and F.~Wei, ``Uplink transmission design for
  crowded correlated cell-free massive {MIMO-OFDM} systems,'' {\em arXiv
  preprint arXiv:2011.00203}, 2020.

\bibitem{mMTC8053903}
E.~{de Carvalho}, E.~{Björnson}, J.~H. {Sorensen}, E.~G. {Larsson}, and
  P.~{Popovski}, ``Random pilot and data access in massive {MIMO} for
  machine-type communications,'' {\em IEEE Transactions on Wireless
  Communications}, vol.~16, no.~12, pp.~7703--7717, 2017.

\bibitem{PowerControlConventional7801046}
P.~{Liu}, S.~{Jin}, T.~{Jiang}, Q.~{Zhang}, and M.~{Matthaiou}, ``Pilot power
  allocation through user grouping in multi-cell massive {MIMO} systems,'' {\em
  IEEE Transactions on Communications}, vol.~65, no.~4, pp.~1561--1574, 2017.

\bibitem{RicianFadingPhaseShifts8809413}
O.~{Özdogan}, E.~{Björnson}, and J.~{Zhang}, ``Performance of cell-free
  massive {MIMO} with rician fading and phase shifts,'' {\em IEEE Transactions
  on Wireless Communications}, vol.~18, no.~11, pp.~5299--5315, 2019.

\bibitem{nonReciprocityCellFreeDNN9098852}
N.~{Athreya}, V.~{Raj}, and S.~{Kalyani}, ``Beyond {5G}: Leveraging cell free
  {TDD} massive {MIMO} using cascaded deep learning,'' {\em IEEE Wireless
  Communications Letters}, vol.~9, no.~9, pp.~1533--1537, 2020.

\bibitem{ArinNonReciprocity8421240}
A.~{Minasian}, S.~{Shahbazpanahi}, and R.~S. {Adve}, ``Distributed massive
  {MIMO} systems with non-reciprocal channels: Impacts and robust
  beamforming,'' {\em IEEE Transactions on Communications}, vol.~66, no.~11,
  pp.~5261--5277, 2018.

\bibitem{CSIAging6608213}
K.~T. {Truong} and R.~W. {Heath}, ``Effects of channel aging in massive {MIMO}
  systems,'' {\em Journal of Communications and Networks}, vol.~15, no.~4,
  pp.~338--351, 2013.

\bibitem{1512123}
K.~E. {Baddour} and N.~C. {Beaulieu}, ``Autoregressive modeling for fading
  channel simulation,'' {\em IEEE Transactions on Wireless Communications},
  vol.~4, no.~4, pp.~1650--1662, 2005.

\bibitem{jakes1994microwave}
W.~C. Jakes and D.~C. Cox, {\em Microwave mobile communications}.
\newblock Wiley-IEEE Press, 1994.

\bibitem{channelAgingMassiveMIMO7307172}
C.~{Kong}, C.~{Zhong}, A.~K. {Papazafeiropoulos}, M.~{Matthaiou}, and
  Z.~{Zhang}, ``Sum-rate and power scaling of massive {MIMO} systems with
  channel aging,'' {\em IEEE Transactions on Communications}, vol.~63, no.~12,
  pp.~4879--4893, 2015.

\bibitem{channelAgingMassiveMIMO7473866}
A.~K. {Papazafeiropoulos}, ``Impact of general channel aging conditions on the
  downlink performance of massive {MIMO},'' {\em IEEE Transactions on Vehicular
  Technology}, vol.~66, no.~2, pp.~1428--1442, 2017.

\bibitem{channelAgingMassiveMIMO8122014}
R.~{Chopra}, C.~R. {Murthy}, H.~A. {Suraweera}, and E.~G. {Larsson},
  ``Performance analysis of {FDD} massive {MIMO} systems under channel aging,''
  {\em IEEE Transactions on Wireless Communications}, vol.~17, no.~2,
  pp.~1094--1108, 2018.

\bibitem{3GPP:TR36.881}
3GPP, {\em Study on latency reduction techniques for LTE}, 2016.
\newblock \url{https://www.3gpp.org/ftp/Specs/archive/36_series/36.881/}.

\bibitem{zhang2020seamless}
Y.~Zhang, M.~H. Fong, P.~Zong, and A.~Davydov, ``Seamless mobility for 5g and
  lte systems and devices,'' Apr.~23 2020.
\newblock US Patent App. 16/584,651.

\bibitem{504082}
X.~{Lagrange} and P.~{Godlewski}, ``Performance of a hierarchical cellular
  network with mobility-dependent hand-over strategies,'' in {\em Proceedings
  of Vehicular Technology Conference - VTC}, vol.~3, pp.~1868--1872 vol.3,
  1996.

\bibitem{johnson1996dynamic}
D.~B. Johnson and D.~A. Maltz, ``Dynamic source routing in ad hoc wireless
  networks,'' in {\em Mobile computing}, pp.~153--181, Springer, 1996.

\bibitem{roy2011handbook}
R.~R. Roy, {\em Handbook of mobile ad hoc networks for mobility models},
  vol.~170.
\newblock Springer, 2011.

\bibitem{5343061}
J.~{Harri}, F.~{Filali}, and C.~{Bonnet}, ``Mobility models for vehicular ad
  hoc networks: a survey and taxonomy,'' {\em IEEE Communications Surveys
  Tutorials}, vol.~11, no.~4, pp.~19--41, 2009.

\bibitem{Ravi7006787}
S.~{Sadr} and R.~S. {Adve}, ``Handoff rate and coverage analysis in multi-tier
  heterogeneous networks,'' {\em IEEE Transactions on Wireless Communications},
  vol.~14, no.~5, pp.~2626--2638, 2015.

\bibitem{6477064}
X.~{Lin}, R.~K. {Ganti}, P.~J. {Fleming}, and J.~G. {Andrews}, ``Towards
  understanding the fundamentals of mobility in cellular networks,'' {\em IEEE
  Transactions on Wireless Communications}, vol.~12, no.~4, pp.~1686--1698,
  2013.

\bibitem{CSIAgingzheng2020cell}
J.~Zheng, J.~Zhang, E.~Bj{\"o}rnson, and B.~Ai, ``Cell-free massive {MIMO} with
  channel aging and pilot contamination,'' {\em arXiv preprint
  arXiv:2008.10827}, 2020.

\bibitem{machineLearningChannelAging8979256}
J.~{Yuan}, H.~Q. {Ngo}, and M.~{Matthaiou}, ``Machine learning-based channel
  prediction in massive {MIMO} with channel aging,'' {\em IEEE Transactions on
  Wireless Communications}, vol.~19, no.~5, pp.~2960--2973, 2020.

\bibitem{Synchronization4287203}
M.~{Morelli}, C.~.~J. {Kuo}, and M.~{Pun}, ``Synchronization techniques for
  orthogonal frequency division multiple access ({OFDMA}): A tutorial review,''
  {\em Proceedings of the IEEE}, vol.~95, pp.~1394--1427, July 2007.

\bibitem{NetworkSlicing8320765}
I.~{Afolabi}, T.~{Taleb}, K.~{Samdanis}, A.~{Ksentini}, and H.~{Flinck},
  ``Network slicing and softwarization: A survey on principles, enabling
  technologies, and solutions,'' {\em IEEE Communications Surveys Tutorials},
  vol.~20, no.~3, pp.~2429--2453, 2018.

\bibitem{8491240}
T.~M. {Shami}, D.~{Grace}, A.~{Burr}, and M.~D. {Zakaria}, ``Radio resource
  management for user-centric {JT-CoMP},'' in {\em 2018 15th International
  Symposium on Wireless Communication Systems (ISWCS)}, pp.~1--5, Aug 2018.

\bibitem{8281464}
T.~H. {Nguyen}, T.~K. {Nguyen}, H.~D. {Han}, and V.~D. {Nguyen}, ``Optimal
  power control and load balancing for uplink cell-free multi-user massive
  {MIMO},'' {\em IEEE Access}, vol.~6, pp.~14462--14473, 2018.

\bibitem{6328484}
P.~{Baracca}, F.~{Boccardi}, and V.~{Braun}, ``A dynamic joint clustering
  scheduling algorithm for downlink {CoMP} systems with limited csi,'' in {\em
  2012 International Symposium on Wireless Communication Systems (ISWCS)},
  pp.~830--834, Aug 2012.

\bibitem{Ahmad9084256}
A.~A. {Khan}, R.~{Adve}, and W.~{Yu}, ``Optimizing downlink resource allocation
  in multiuser {MIMO} networks via fractional programming and the hungarian
  algorithm,'' {\em IEEE Transactions on Wireless Communications}, pp.~1--1,
  2020.

\bibitem{PrecodingDistrib2020Atzeni}
I.~Atzeni, B.~Gouda, and A.~Tölli, ``Distributed precoding design via
  over-the-air signaling for cell-free massive {MIMO},'' {\em arXiv preprint
  arXiv:2004.00299}, 2020.

\bibitem{6151868}
S.~{Kaviani}, O.~{Simeone}, W.~A. {Krzymien}, and S.~{Shamai}, ``Linear
  precoding and equalization for network {MIMO} with partial cooperation,''
  {\em IEEE Transactions on Vehicular Technology}, vol.~61, pp.~2083--2096, Jun
  2012.

\bibitem{WSRMSE8307115}
J.~{Kaleva}, A.~{Tölli}, M.~{Juntti}, R.~A. {Berry}, and M.~L. {Honig},
  ``Decentralized joint precoding with pilot-aided beamformer estimation,''
  {\em IEEE Transactions on Signal Processing}, vol.~66, no.~9, pp.~2330--2341,
  2018.

\bibitem{yu2011adaptive}
W.~Yu, T.~Kwon, and C.~Shin, {\em Adaptive resource allocation in cooperative
  cellular networks}, ch.~9, pp.~233--256.
\newblock Cambridge Univ. Press, 2011.

\bibitem{Cell-freeWSR2005.12331}
Q.~{Vu}, L.~{Tran}, and M.~{Juntti}, ``Noncoherent joint transmission
  beamforming for dense small cell networks: Global optimality, efficient
  solution and distributed implementation,'' {\em arXiv preprint
  arXiv:2005.12331}, 2020.

\bibitem{marks1978general}
B.~R. Marks and G.~P. Wright, ``A general inner approximation algorithm for
  nonconvex mathematical programs,'' {\em Operations research}, vol.~26, no.~4,
  pp.~681--683, 1978.

\bibitem{tuy2005monotonic}
H.~Tuy, F.~Al-Khayyal, and P.~T. Thach, ``Monotonic optimization: Branch and
  cut methods,'' in {\em Essays and Surveys in Global Optimization},
  pp.~39--78, Springer, 2005.

\bibitem{grant2014cvx}
M.~Grant and S.~Boyd, ``Cvx: Matlab software for disciplined convex
  programming, version 2.1,'' 2014.
\newblock [Online]. Available: \url{http://cvxr.com/cvx/}.

\bibitem{bjornson2013optimal}
E.~Bj{\"o}rnson and E.~Jorswieck, {\em Optimal resource allocation in
  coordinated multi-cell systems}.
\newblock Now Publishers Inc, 2013.

\bibitem{IABStochastic2019}
C.~Saha and H.~S. Dhillon, ``Load balancing in {5G} hetnets with millimeter
  wave integrated access and backhaul,'' {\em arXiv preprint arXiv:1902.06300},
  2019.

\bibitem{HardwareImpairements8476516}
J.~{Zhang}, Y.~{Wei}, E.~{Björnson}, Y.~{Han}, and S.~{Jin}, ``Performance
  analysis and power control of cell-free massive {MIMO} systems with hardware
  impairments,'' {\em IEEE Access}, vol.~6, pp.~55302--55314, 2018.

\bibitem{8360138}
T.~M. {Hoang}, H.~Q. {Ngo}, T.~Q. {Duong}, H.~D. {Tuan}, and A.~{Marshall},
  ``Cell-free massive {MIMO} networks: Optimal power control against active
  eavesdropping,'' {\em IEEE Transactions on Communications}, vol.~66, no.~10,
  pp.~4724--4737, 2018.

\bibitem{IEEE1588_7949184}
``{IEEE} standard for a precision clock synchronization protocol for networked
  measurement and control systems - redline,'' {\em IEEE Std 1588-2008
  (Revision of IEEE Std 1588-2002) - Redline}, pp.~1--300, July 2008.

\bibitem{synchronizationCalibration6760595}
R.~{Rogalin}, O.~Y. {Bursalioglu}, H.~{Papadopoulos}, G.~{Caire}, A.~F.
  {Molisch}, A.~{Michaloliakos}, V.~{Balan}, and K.~{Psounis}, ``Scalable
  synchronization and reciprocity calibration for distributed multiuser
  {MIMO},'' {\em IEEE Transactions on Wireless Communications}, vol.~13, no.~4,
  pp.~1815--1831, 2014.

\bibitem{8974591}
S.~{Mukherjee} and J.~{Lee}, ``Edge computing-enabled cell-free massive {MIMO}
  systems,'' {\em IEEE Transactions on Wireless Communications}, vol.~19,
  no.~4, pp.~2884--2899, 2020.

\bibitem{mobileEdgeComputing7901477}
T.~X. {Tran}, A.~{Hajisami}, P.~{Pandey}, and D.~{Pompili}, ``Collaborative
  mobile edge computing in {5G} networks: New paradigms, scenarios, and
  challenges,'' {\em IEEE Communications Magazine}, vol.~55, no.~4, pp.~54--61,
  2017.

\bibitem{7498684}
M.~{Chiang} and T.~{Zhang}, ``Fog and {IoT}: An overview of research
  opportunities,'' {\em IEEE Internet of Things Journal}, vol.~3, no.~6,
  pp.~854--864, 2016.

\bibitem{puncturedScheduling9011578}
A.~{Anand}, G.~{de Veciana}, and S.~{Shakkottai}, ``Joint scheduling of {URLLC}
  and {eMBB} traffic in {5G} wireless networks,'' {\em IEEE/ACM Transactions on
  Networking}, vol.~28, no.~2, pp.~477--490, 2020.

\bibitem{PowerAlloc8630677}
M.~{Bashar}, K.~{Cumanan}, A.~G. {Burr}, M.~{Debbah}, and H.~Q. {Ngo}, ``On the
  uplink max–min {SINR} of cell-free massive {MIMO} systems,'' {\em IEEE
  Transactions on Wireless Communications}, vol.~18, no.~4, pp.~2021--2036,
  2019.

\bibitem{DistribResourceAllo7676375}
A.~{Zappone}, E.~{Jorswieck}, and A.~{Leshem}, ``Distributed resource
  allocation for energy efficiency in {MIMO} {OFDMA} wireless networks,'' {\em
  IEEE Journal on Selected Areas in Communications}, vol.~34, no.~12,
  pp.~3451--3465, 2016.

\bibitem{EEdistributedAntenna7482747}
H.~{Ren}, N.~{Liu}, C.~{Pan}, and C.~{He}, ``Energy efficiency optimization for
  {MIMO} distributed antenna systems,'' {\em IEEE Transactions on Vehicular
  Technology}, vol.~66, no.~3, pp.~2276--2288, 2017.

\bibitem{9236635}
N.~{Akbar}, E.~{Björnson}, N.~{Yang}, and E.~G. {Larsson}, ``Max-min power
  control in downlink massive {MIMO} with distributed antenna arrays,'' {\em
  IEEE Transactions on Communications}, pp.~1--1, 2020.

\bibitem{8094949}
E.~{Björnson}, J.~{Hoydis}, and L.~{Sanguinetti}, ``Massive {MIMO} has
  unlimited capacity,'' {\em IEEE Transactions on Wireless Communications},
  vol.~17, no.~1, pp.~574--590, 2018.

\bibitem{clusteringOverlappingCoordinationClusters5594709}
C.~T.~K. {Ng} and H.~{Huang}, ``Linear precoding in cooperative {MIMO} cellular
  networks with limited coordination clusters,'' {\em IEEE Journal on Selected
  Areas in Communications}, vol.~28, pp.~1446--1454, December 2010.

\bibitem{NOMA6736749}
G.~W. \textit{et al.}, ``{5GNOW}: non-orthogonal, asynchronous waveforms for
  future mobile applications,'' {\em IEEE Communications Magazine}, vol.~52,
  no.~2, pp.~97--105, 2014.

\bibitem{NOMACellFree8368267}
Y.~{Li} and G.~A. {Aruma Baduge}, ``{NOMA}-aided cell-free massive {MIMO}
  systems,'' {\em IEEE Wireless Communications Letters}, vol.~7, no.~6,
  pp.~950--953, 2018.

\bibitem{NOMA8895763}
M.~{Bashar}, K.~{Cumanan}, A.~G. {Burr}, H.~Q. {Ngo}, L.~{Hanzo}, and
  P.~{Xiao}, ``On the performance of cell-free massive {MIMO} relying on
  adaptive {NOMA/OMA} mode-switching,'' {\em IEEE Transactions on
  Communications}, vol.~68, no.~2, pp.~792--810, 2020.

\bibitem{CellFreeNOMA9024101}
F.~{Rezaei}, C.~{Tellambura}, A.~A. {Tadaion}, and A.~R. {Heidarpour}, ``Rate
  analysis of cell-free massive {MIMO-NOMA} with three linear precoders,'' {\em
  IEEE Transactions on Communications}, vol.~68, no.~6, pp.~3480--3494, 2020.

\bibitem{CellFreeNOMA9130101}
T.~K. {Nguyen}, H.~H. {Nguyen}, and H.~D. {Tuan}, ``Max-min {QoS} power control
  in generalized cell-free massive {MIMO-NOMA} with optimal backhaul
  combining,'' {\em IEEE Transactions on Vehicular Technology}, vol.~69,
  no.~10, pp.~10949--10964, 2020.

\bibitem{CellFreeNOMA8957510}
F.~{Rezaei}, A.~R. {Heidarpour}, C.~{Tellambura}, and A.~{Tadaion}, ``Underlaid
  spectrum sharing for cell-free massive {MIMO-NOMA},'' {\em IEEE
  Communications Letters}, vol.~24, no.~4, pp.~907--911, 2020.

\bibitem{pochet2006production}
Y.~Pochet and L.~A. Wolsey, {\em Production planning by mixed integer
  programming}.
\newblock Springer Science \& Business Media, 2006.

\bibitem{rossi2006handbook}
F.~Rossi, P.~Van~Beek, and T.~Walsh, {\em Handbook of constraint programming}.
\newblock Elsevier, 2006.

\bibitem{hooker2007integrated}
J.~N. Hooker, {\em Integrated methods for optimization}, vol.~100.
\newblock Springer Science \& Business Media, 2007.

\bibitem{6415388}
J.~{Hoydis}, S.~{ten Brink}, and M.~{Debbah}, ``Massive {MIMO} in the {UL/DL}
  of cellular networks: How many antennas do we need?,'' {\em IEEE Journal on
  Selected Areas in Communications}, vol.~31, no.~2, pp.~160--171, 2013.

\bibitem{MSE5756489}
Q.~{Shi}, M.~{Razaviyayn}, Z.~{Luo}, and C.~{He}, ``An iteratively weighted
  {MMSE} approach to distributed sum-utility maximization for a {MIMO}
  interfering broadcast channel,'' {\em IEEE Transactions on Signal
  Processing}, vol.~59, no.~9, pp.~4331--4340, 2011.

\bibitem{8599043}
M.~{Attarifar}, A.~{Abbasfar}, and A.~{Lozano}, ``Modified conjugate
  beamforming for cell-free massive {MIMO},'' {\em IEEE Wireless Communications
  Letters}, vol.~8, no.~2, pp.~616--619, 2019.

\bibitem{201114573}
S.~Mukherjee and R.~Chopra, ``Performance analysis of cell free massive {MIMO}
  systems in {LoS/ NLoS} channels,'' {\em arXiv preprint arXiv:2011.14573},
  2020.

\bibitem{ITU:P.1410_5}
ITU, {\em Propagation data and prediction methods required for the design of
  terrestrial broadband radio access systems operating in a frequnecy range
  from 3 to 60 {GHz}}, Feb. 2012.
\newblock \url{https://www.itu.int/rec/R-REC-P.1410-5-201202-I}.

\bibitem{8859647}
D.~{Kim}, J.~{Lee}, and T.~Q.~S. {Quek}, ``Multi-layer unmanned aerial vehicle
  networks: Modeling and performance analysis,'' {\em IEEE Transactions on
  Wireless Communications}, vol.~19, no.~1, pp.~325--339, 2020.

\bibitem{multicastCellFree8017421}
T.~X. {Doan}, H.~Q. {Ngo}, T.~Q. {Duong}, and K.~{Tourki}, ``On the performance
  of multigroup multicast cell-free massive {MIMO},'' {\em IEEE Communications
  Letters}, vol.~21, no.~12, pp.~2642--2645, 2017.

\bibitem{FullDuplex8943119}
D.~{Wang}, M.~{Wang}, P.~{Zhu}, J.~{Li}, J.~{Wang}, and X.~{You}, ``Performance
  of network-assisted full-duplex for cell-free massive {MIMO},'' {\em IEEE
  Transactions on Communications}, vol.~68, no.~3, pp.~1464--1478, 2020.

\bibitem{EE8777141}
X.~{Yu}, W.~{Xu}, S.~{Leung}, Q.~{Shi}, and J.~{Chu}, ``Power allocation for
  energy efficient optimization of distributed {MIMO} system with
  beamforming,'' {\em IEEE Transactions on Vehicular Technology}, vol.~68,
  no.~9, pp.~8966--8981, 2019.

\bibitem{cellfreeIofTPilots2020}
S.~Rao, A.~Ashikhmin, and H.~Yang, ``Cell-free massive {MIMO} with
  nonorthogonal pilots for internet of things,'' {\em arXiv preprint}, 2020.

\bibitem{9079911}
T.~C. {Mai}, H.~Q. {Ngo}, and T.~Q. {Duong}, ``Downlink spectral efficiency of
  cell-free massive {MIMO} systems with multi-antenna users,'' {\em IEEE
  Transactions on Communications}, vol.~68, no.~8, pp.~4803--4815, 2020.

\bibitem{SuccessiveLowerBoundRazaviyayn2013unified}
M.~Razaviyayn, M.~Hong, and Z.-Q. Luo, ``A unified convergence analysis of
  block successive minimization methods for nonsmooth optimization,'' {\em SIAM
  Journal on Optimization}, vol.~23, no.~2, pp.~1126--1153, 2013.

\bibitem{EE8695055}
G.~{Dong}, H.~{Zhang}, S.~{Jin}, and D.~{Yuan}, ``Energy-efficiency-oriented
  joint user association and power allocation in distributed massive {MIMO}
  systems,'' {\em IEEE Transactions on Vehicular Technology}, vol.~68, no.~6,
  pp.~5794--5808, 2019.

\bibitem{EnergyEfficiency8689095}
J.~{Shi}, H.~{Xu}, Z.~{Yang}, and M.~{Chen}, ``Energy efficient beamforming for
  user-centric virtual cell networks,'' {\em IEEE Transactions on Green
  Communications and Networking}, vol.~3, pp.~575--590, Sep. 2019.

\bibitem{EE7900388}
L.~D. {Nguyen}, T.~Q. {Duong}, H.~Q. {Ngo}, and K.~{Tourki}, ``Energy
  efficiency in cell-free massive {MIMO} with zero-forcing precoding design,''
  {\em IEEE Communications Letters}, vol.~21, no.~8, pp.~1871--1874, 2017.

\bibitem{zappone2015energy}
A.~Zappone and E.~Jorswieck, ``Energy efficiency in wireless networks via
  fractional programming theory,'' {\em Foundations and Trends in
  Communications and Information Theory}, vol.~11, no.~3-4, pp.~185--396, 2015.

\bibitem{dinkelbach1967nonlinear}
W.~Dinkelbach, ``On nonlinear fractional programming,'' {\em Management
  science}, vol.~13, no.~7, pp.~492--498, 1967.

\bibitem{LambertFuncthoorfar2008inequalities}
A.~Hoorfar and M.~Hassani, ``Inequalities on the lambert w function and
  hyperpower function,'' {\em J. Inequal. Pure and Appl. Math}, vol.~9, no.~2,
  pp.~5--9, 2008.

\bibitem{IRS9352948}
Y.~Zhang, B.~Di, H.~Zhang, J.~Lin, C.~Xu, D.~Zhang, Y.~Li, and L.~Song,
  ``Beyond cell-free {MIMO}: Energy efficient reconfigurable intelligent
  surface aided cell-free {MIMO} communications,'' {\em IEEE Transactions on
  Cognitive Communications and Networking}, pp.~1--1, 2021.

\bibitem{IRS9363171}
Q.~N. Le, V.-D. Nguyen, O.~A. Dobre, and R.~Zhao, ``Energy efficiency
  maximization in {RIS}-aided cell-free network with limited backhaul,'' {\em
  IEEE Communications Letters}, pp.~1--1, 2021.

\bibitem{IRS8910627}
Q.~Wu and R.~Zhang, ``Towards smart and reconfigurable environment: Intelligent
  reflecting surface aided wireless network,'' {\em IEEE Communications
  Magazine}, vol.~58, no.~1, pp.~106--112, 2020.

\bibitem{IRS9308937}
T.~Zhou, K.~Xu, X.~Xia, W.~Xie, and J.~Xu, ``Achievable rate optimization for
  aerial intelligent reflecting surface-aided cell-free massive {MIMO}
  system,'' {\em IEEE Access}, vol.~9, pp.~3828--3837, 2021.

\bibitem{IRS9298843}
S.~Huang, Y.~Ye, M.~Xiao, H.~V. Poor, and M.~Skoglund, ``Decentralized
  beamforming design for intelligent reflecting surface-enhanced cell-free
  networks,'' {\em IEEE Wireless Communications Letters}, vol.~10, no.~3,
  pp.~673--677, 2021.

\bibitem{3GPP:TR36.902}
3GPP, {\em Self-configuring and self-optimizing network (SON) use cases and
  solutions}, 2011.
\newblock \url{3gpp.org/ftp/Specs/archive/38_series/38.874/}.

\bibitem{alsharif2019energy}
M.~H. Alsharif, A.~H. Kelechi, J.~Kim, and J.~H. Kim, ``Energy efficiency and
  coverage trade-off in {5G} for eco-friendly and sustainable cellular
  networks,'' {\em Symmetry}, vol.~11, no.~3, p.~408, 2019.

\bibitem{EricssonMobilityReport}
Ericsson, {\em Ericsson Mobility Report}, November 2020.
\newblock
  \url{https://www.ericsson.com/4adc87/assets/local/mobility-report/documents/2020/november-2020-ericsson-mobility-report.pdf}.

\bibitem{footprintComm5978416}
A.~{Fehske}, G.~{Fettweis}, J.~{Malmodin}, and G.~{Biczok}, ``The global
  footprint of mobile communications: The ecological and economic
  perspective,'' {\em IEEE Communications Magazine}, vol.~49, no.~8,
  pp.~55--62, 2011.

\bibitem{6815733}
L.~{Venturino}, A.~{Zappone}, C.~{Risi}, and S.~{Buzzi}, ``Energy-efficient
  scheduling and power allocation in downlink {OFDMA} networks with base
  station coordination,'' {\em IEEE Transactions on Wireless Communications},
  vol.~14, pp.~1--14, Jan 2015.

\bibitem{FR8314727}
K.~{Shen} and W.~{Yu}, ``Fractional programming for communication
  systems—part {I}: Power control and beamforming,'' 2018.

\bibitem{FR8310563}
K.~{Shen} and W.~{Yu}, ``Fractional programming for communication
  systems—part {II}: Uplink scheduling via matching,'' {\em IEEE Transactions
  on Signal Processing}, vol.~66, pp.~2631--2644, May 2018.

\bibitem{7437385}
B.~{Dai} and W.~{Yu}, ``Energy efficiency of downlink transmission strategies
  for cloud radio access networks,'' {\em IEEE Journal on Selected Areas in
  Communications}, vol.~34, no.~4, pp.~1037--1050, 2016.

\bibitem{6056688}
S.~{Tombaz}, A.~{Vastberg}, and J.~{Zander}, ``Energy- and cost-efficient
  ultra-high-capacity wireless access,'' {\em IEEE Wireless Communications},
  vol.~18, no.~5, pp.~18--24, 2011.

\bibitem{7456325}
J.~{Zuo}, J.~{Zhang}, C.~{Yuen}, W.~{Jiang}, and W.~{Luo}, ``Energy-efficient
  downlink transmission for multicell massive {DAS} with pilot contamination,''
  {\em IEEE Transactions on Vehicular Technology}, vol.~66, no.~2,
  pp.~1209--1221, 2017.

\bibitem{7031971}
E.~{Björnson}, L.~{Sanguinetti}, J.~{Hoydis}, and M.~{Debbah}, ``Optimal
  design of energy-efficient multi-user {MIMO} systems: Is massive {MIMO} the
  answer?,'' {\em IEEE Transactions on Wireless Communications}, vol.~14,
  no.~6, pp.~3059--3075, 2015.

\bibitem{8970501}
G.~{Femenias}, N.~{Lassoued}, and F.~{Riera-Palou}, ``Access point switch
  {ON/OFF} strategies for green cell-free massive {MIMO} networking,'' {\em
  IEEE Access}, vol.~8, pp.~21788--21803, 2020.

\bibitem{EnergyHarvesting9201540}
S.~{Kusaladharma}, W.~P. {Zhu}, W.~{Ajib}, and G.~A.~A. {Baduge}, ``Stochastic
  geometry based performance characterization of {SWIPT} in cell-free massive
  {MIMO},'' {\em IEEE Transactions on Vehicular Technology}, vol.~69, no.~11,
  pp.~13357--13370, 2020.

\bibitem{9149862}
J.~{García-Morales}, G.~{Femenias}, and F.~{Riera-Palou}, ``Energy-efficient
  access-point sleep-mode techniques for cell-free mmwave massive {MIMO}
  networks with non-uniform spatial traffic density,'' {\em IEEE Access},
  vol.~8, pp.~137587--137605, 2020.

\bibitem{mmWaveConnectivity8579590}
M.~{Gapeyenko}, V.~{Petrov}, D.~{Moltchanov}, M.~R. {Akdeniz}, S.~{Andreev},
  N.~{Himayat}, and Y.~{Koucheryavy}, ``On the degree of multi-connectivity in
  {5G} millimeter-wave cellular urban deployments,'' {\em IEEE Transactions on
  Vehicular Technology}, vol.~68, no.~2, pp.~1973--1978, 2019.

\bibitem{mmwave6732923}
S.~{Rangan}, T.~S. {Rappaport}, and E.~{Erkip}, ``Millimeter-wave cellular
  wireless networks: Potentials and challenges,'' {\em Proceedings of the
  IEEE}, vol.~102, no.~3, pp.~366--385, 2014.

\bibitem{7446253}
S.~{Buzzi}, C.~{I}, T.~E. {Klein}, H.~V. {Poor}, C.~{Yang}, and A.~{Zappone},
  ``A survey of energy-efficient techniques for 5g networks and challenges
  ahead,'' {\em IEEE Journal on Selected Areas in Communications}, vol.~34,
  no.~4, pp.~697--709, 2016.

\bibitem{EnergyHarvestignEnvironmental7010878}
S.~{Ulukus}, A.~{Yener}, E.~{Erkip}, O.~{Simeone}, M.~{Zorzi}, P.~{Grover}, and
  K.~{Huang}, ``Energy harvesting wireless communications: A review of recent
  advances,'' {\em IEEE Journal on Selected Areas in Communications}, vol.~33,
  no.~3, pp.~360--381, 2015.

\bibitem{EnergyHarvestignRF6951347}
X.~{Lu}, P.~{Wang}, D.~{Niyato}, D.~I. {Kim}, and Z.~{Han}, ``Wireless networks
  with rf energy harvesting: A contemporary survey,'' {\em IEEE Communications
  Surveys and Tutorials}, vol.~17, no.~2, pp.~757--789, 2015.

\bibitem{localCaching7150324}
M.~{Ji}, G.~{Caire}, and A.~F. {Molisch}, ``Wireless device-to-device caching
  networks: Basic principles and system performance,'' {\em IEEE Journal on
  Selected Areas in Communications}, vol.~34, no.~1, pp.~176--189, 2016.

\bibitem{6151778}
X.~{Zhou}, B.~{Maham}, and A.~{Hjorungnes}, ``Pilot contamination for active
  eavesdropping,'' {\em IEEE Transactions on Wireless Communications}, vol.~11,
  no.~3, pp.~903--907, 2012.

\bibitem{she2020tutorial}
C.~She, C.~Sun, Z.~Gu, Y.~Li, C.~Yang, H.~Poor, and B.~Vucetic, ``A tutorial of
  ultra-reliable and low-latency communications in {6G}: Integrating
  theoretical knowledge into deep learning,'' {\em arXiv preprint
  arXiv:2009.06010}, 2020.

\bibitem{interferencePricing1626432}
{Jianwei Huang}, R.~A. {Berry}, and M.~L. {Honig}, ``Distributed interference
  compensation for wireless networks,'' {\em IEEE Journal on Selected Areas in
  Communications}, vol.~24, no.~5, pp.~1074--1084, 2006.

\bibitem{levin2003supermodular}
J.~Levin, ``Supermodular games,'' {\em Lectures Notes, Department of Economics,
  Stanford University}, 2003.

\bibitem{bertsekas1979distributed}
D.~P. Bertsekas, ``A distributed algorithm for the assignment problem,'' {\em
  Lab. for Information and Decision Systems Working Paper, MIT}, 1979.

\bibitem{OptPowerGP4275017}
M.~{Chiang}, C.~W. {Tan}, D.~P. {Palomar}, D.~{O'neill}, and D.~{Julian},
  ``Power control by geometric programming,'' {\em IEEE Transactions on
  Wireless Communications}, vol.~6, no.~7, pp.~2640--2651, 2007.

\bibitem{FederatedLearning9124715}
T.~T. {Vu}, D.~T. {Ngo}, N.~H. {Tran}, H.~Q. {Ngo}, M.~N. {Dao}, and R.~H.
  {Middleton}, ``Cell-free massive {MIMO} for wireless federated learning,''
  {\em IEEE Transactions on Wireless Communications}, vol.~19, no.~10,
  pp.~6377--6392, 2020.

\bibitem{federatedLearning9084352}
T.~{Li}, A.~K. {Sahu}, A.~{Talwalkar}, and V.~{Smith}, ``Federated learning:
  Challenges, methods, and future directions,'' {\em IEEE Signal Processing
  Magazine}, vol.~37, no.~3, pp.~50--60, 2020.

\bibitem{mcmahan2017communication}
B.~M. \textit{et al.}, ``Communication-efficient learning of deep networks from
  decentralized data,'' in {\em Artificial Intelligence and Statistics},
  pp.~1273--1282, PMLR, 2017.

\bibitem{edgeAI9134426}
Y.~{Shi}, K.~{Yang}, T.~{Jiang}, J.~{Zhang}, and K.~B. {Letaief},
  ``Communication-efficient edge {AI}: Algorithms and systems,'' {\em IEEE
  Communications Surveys \& Tutorials}, vol.~22, no.~4, pp.~2167--2191, 2020.

\bibitem{edgeIntelligence8736011}
Z.~{Zhou}, X.~{Chen}, E.~{Li}, L.~{Zeng}, K.~{Luo}, and J.~{Zhang}, ``Edge
  intelligence: Paving the last mile of artificial intelligence with edge
  computing,'' {\em Proceedings of the IEEE}, vol.~107, no.~8, pp.~1738--1762,
  2019.

\bibitem{WMMSE5756489}
Q.~{Shi}, M.~{Razaviyayn}, Z.~{Luo}, and C.~{He}, ``An iteratively weighted
  {MMSE} approach to distributed sum-utility maximization for a {MIMO}
  interfering broadcast channel,'' {\em IEEE Transactions on Signal
  Processing}, vol.~59, no.~9, pp.~4331--4340, 2011.

\bibitem{mMTC7565189}
C.~{Bockelmann}, N.~{Pratas}, H.~{Nikopour}, K.~{Au}, T.~{Svensson},
  C.~{Stefanovic}, P.~{Popovski}, and A.~{Dekorsy}, ``Massive machine-type
  communications in {5G}: physical and {MAC}-layer solutions,'' {\em IEEE
  Communications Magazine}, vol.~54, no.~9, pp.~59--65, 2016.

\bibitem{bertsekas1995dynamic}
D.~P. Bertsekas, {\em Dynamic programming and optimal control}, vol.~1.
\newblock Athena scientific Belmont, MA, 1995.

\bibitem{4455486}
M.~J. {Neely}, E.~{Modiano}, and C.~{Li}, ``Fairness and optimal stochastic
  control for heterogeneous networks,'' {\em IEEE/ACM Transactions on
  Networking}, vol.~16, no.~2, pp.~396--409, 2008.

\bibitem{tactileInternet7470948}
M.~{Maier}, M.~{Chowdhury}, B.~P. {Rimal}, and D.~P. {Van}, ``The tactile
  internet: vision, recent progress, and open challenges,'' {\em IEEE
  Communications Magazine}, vol.~54, no.~5, pp.~138--145, 2016.

\bibitem{tactileInternet8605315}
O.~H. \textit{et al.}, ``The {IEEE} 1918.1 “tactile internet” standards
  working group and its standards,'' {\em Proceedings of the IEEE}, vol.~107,
  no.~2, pp.~256--279, 2019.

\bibitem{Telehealthdorsey2016state}
E.~R. Dorsey and E.~J. Topol, ``State of telehealth,'' {\em New England Journal
  of Medicine}, vol.~375, no.~2, pp.~154--161, 2016.

\bibitem{7891004}
S.~{Huang}, B.~{Liang}, and J.~{Li}, ``Distributed interference and delay aware
  design for {D2D} communication in large wireless networks with adaptive
  interference estimation,'' {\em IEEE Transactions on Wireless
  Communications}, vol.~16, no.~6, pp.~3924--3939, 2017.

\bibitem{7010886}
F.~{Shan}, J.~{Luo}, W.~{Wu}, M.~{Li}, and X.~{Shen}, ``Discrete rate
  scheduling for packets with individual deadlines in energy harvesting
  systems,'' {\em IEEE Journal on Selected Areas in Communications}, vol.~33,
  no.~3, pp.~438--451, 2015.

\bibitem{uRLLC8403963}
H.~{Ji}, S.~{Park}, J.~{Yeo}, Y.~{Kim}, J.~{Lee}, and B.~{Shim},
  ``Ultra-reliable and low-latency communications in {5G} downlink: Physical
  layer aspects,'' {\em IEEE Wireless Communications}, vol.~25, no.~3,
  pp.~124--130, 2018.

\bibitem{HARQIR6954538}
D.~{To}, H.~X. {Nguyen}, Q.~{Vien}, and L.~{Huang}, ``Power allocation for
  {HARQ-IR} systems under {QoS} constraints and limited feedback,'' {\em IEEE
  Transactions on Wireless Communications}, vol.~14, no.~3, pp.~1581--1594,
  2015.

\bibitem{FBC7134725}
B.~{Makki}, T.~{Svensson}, and M.~{Zorzi}, ``Finite block-length analysis of
  spectrum sharing networks using rate adaptation,'' {\em IEEE Transactions on
  Communications}, vol.~63, no.~8, pp.~2823--2835, 2015.

\bibitem{8322772}
Z.~{Wu}, F.~{Zhao}, and X.~{Liu}, ``Signal space diversity aided dynamic
  multiplexing for {eMBB} and {URLLC} traffics,'' in {\em 2017 3rd IEEE
  International Conference on Computer and Communications (ICCC)},
  pp.~1396--1400, 2017.

\bibitem{8538471}
A.~A. {Esswie} and K.~I. {Pedersen}, ``Multi-user preemptive scheduling for
  critical low latency communications in {5G} networks,'' in {\em 2018 IEEE
  Symposium on Computers and Communications (ISCC)}, pp.~00136--00141, 2018.

\bibitem{8408793}
A.~A. {Esswie} and K.~I. {Pedersen}, ``Opportunistic spatial preemptive
  scheduling for {URLLC} and {eMBB} coexistence in multi-user {5G} networks,''
  {\em IEEE Access}, vol.~6, pp.~38451--38463, 2018.

\bibitem{5910719}
V.~{Kotzsch} and G.~{Fettweis}, ``On synchronization requirements and
  performance limitations for {CoMP} systems in large cells,'' in {\em 2011 8th
  International Workshop on Multi-Carrier Systems Solutions}, pp.~1--5, May
  2011.

\bibitem{davydov2016timing}
A.~Davydov, G.~Morozov, A.~Maltsev, V.~Sergeyev, and I.~Bolotin, ``Timing
  synchronization for downlink ({DL}) transmissions in coordinated multipoint
  ({CoMP}) systems,'' May~24 2016.
\newblock US Patent 9,351,277.

\bibitem{schenk2008rf}
T.~Schenk, {\em {RF} imperfections in high-rate wireless systems: impact and
  digital compensation}.
\newblock Springer Science \& Business Media, 2008.

\bibitem{valkama201115}
M.~Valkama, ``15 rf impairment compensation for future radio systems,'' {\em
  Multi-Mode/Multi-Band RF Transceivers for Wireless Communications: Advanced
  Techniques, Architectures, and Trends}, p.~453, 2011.

\bibitem{residualImpairments7106472}
X.~{Zhang}, M.~{Matthaiou}, M.~{Coldrey}, and E.~{Björnson}, ``Impact of
  residual transmit {RF} impairments on training-based {MIMO} systems,'' {\em
  IEEE Transactions on Communications}, vol.~63, no.~8, pp.~2899--2911, 2015.

\bibitem{POMDP7895211}
L.~Ferrari, Q.~Zhao, and A.~Scaglione, ``Utility maximizing sequential sensing
  over a finite horizon,'' {\em IEEE Transactions on Signal Processing},
  vol.~65, no.~13, pp.~3430--3445, 2017.

\bibitem{bondi2000Scalabilitycharacteristics}
A.~Bondi, ``Characteristics of scalability and their impact on performance,''
  in {\em Proceedings of the 2nd international workshop on Software and
  performance}, pp.~195--203, 2000.

\bibitem{duboc2010framework}
A.~L. d. C.~L. Duboc, {\em A framework for the Characterization and Analysis of
  Software Systems Scalability}.
\newblock PhD thesis, Citeseer, 2010.

\bibitem{hill1990scalability}
M.~D. Hill, ``What is scalability?,'' {\em ACM SIGARCH Computer Architecture
  News}, vol.~18, no.~4, pp.~18--21, 1990.

\bibitem{weinstock2006system}
C.~B. Weinstock and J.~B. Goodenough, ``On system scalability,'' tech. rep.,
  Carnegie-Mellon Univ. Pittsburgh PA Software Engineering Inst., 2006.

\bibitem{AI9129108}
X.~Chai, H.~Gao, J.~Sun, X.~Su, T.~Lv, and J.~Zeng, ``Reinforcement learning
  based antenna selection in user-centric massive {MIMO},'' in {\em 2020 IEEE
  91st Vehicular Technology Conference (VTC2020-Spring)}, pp.~1--6, 2020.

\bibitem{AIismath2020deep}
I.~Ismath, S.~Manosha, S.~Ali, N.~Rajatheva, and M.~Latva-aho, ``Deep
  reinforcement learning based fast initial access for {mmWave} based
  user-centric systems,'' {\em arXiv preprint arXiv:2009.06974}, 2020.

\bibitem{AIismath2021deep}
I.~Ismath, S.~Ali, N.~Rajatheva, and M.~Latva-aho, ``Deep contextual bandits
  for fast neighbor-aided initial access in {mmWave} cell-free networks,'' {\em
  arXiv preprint arXiv:2103.09694}, 2021.

\bibitem{surveyMLwirelessNet8743390}
Y.~Sun, M.~Peng, Y.~Zhou, Y.~Huang, and S.~Mao, ``Application of machine
  learning in wireless networks: Key techniques and open issues,'' {\em IEEE
  Communications Surveys \& Tutorials}, vol.~21, no.~4, pp.~3072--3108, 2019.

\bibitem{ns3}
{\em Network Simulator ns-3}, 2021. [Online].
\newblock \url{https://www.nsnam.org/}.

\bibitem{9016270}
K.~{Mawatwal}, D.~{Sen}, and R.~{Roy}, ``Performance analysis of a {SAGE}-based
  semi-blind channel estimator for pilot contaminated {MU} massive {MIMO}
  systems,'' {\em IEEE Access}, vol.~8, pp.~46682--46700, 2020.

\bibitem{JavadMirzaei8490886}
J.~{Mirzaei}, R.~S. {Adve}, and S.~{Shahbazpanahi}, ``Semi-blind time-domain
  channel estimation for frequency-selective multiuser massive {MIMO}
  systems,'' {\em IEEE Transactions on Communications}, vol.~67, no.~2,
  pp.~1045--1058, 2019.

\bibitem{sesia2011lte}
S.~Sesia, I.~Toufik, and M.~Baker, {\em {LTE}-the {UMTS} long term evolution:
  from theory to practice}.
\newblock John Wiley \& Sons, 2011.

\end{thebibliography}
\bibliographystyle{ieeetr}

\begin{IEEEbiography}[{\includegraphics[width=25mm,height=32mm,clip,keepaspectratio]{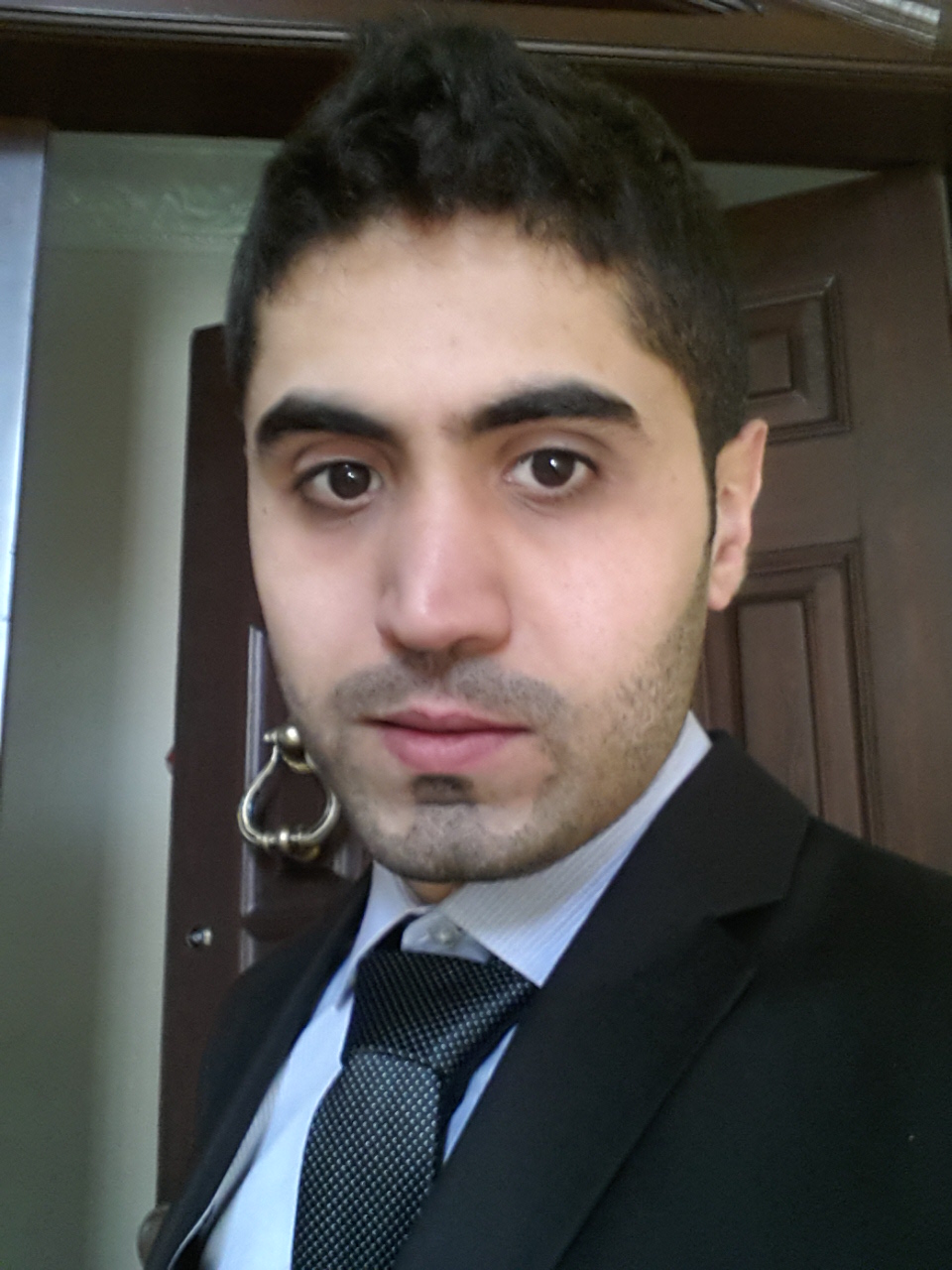}}]
	{Hussein~A.~Ammar}
	(S'19) received his Master's degree in electrical and computer engineering from the American University of Beirut (AUB) in 2017. From February 2018 to July 2018, he worked as a research assistant at the Mobile and Distributed Computing Laboratory at AUB and as a R\&D engineer in the information and communications technology industry. He is currently pursuing the Ph.D. degree at the University of Toronto, Toronto, ON, Canada, where he is working on the optimization of coordinated distributed MIMO systems. His research interests include wireless communications, coordinated distributed MIMO, statistical signal processing, mathematical optimization, and point process theory. He is a recipient of the University of Toronto Fellowship, and the Edward S. Rogers Sr. Graduate Scholarship.
\end{IEEEbiography}
\begin{IEEEbiography}[{\includegraphics[width=25mm,height=32mm,clip,keepaspectratio]{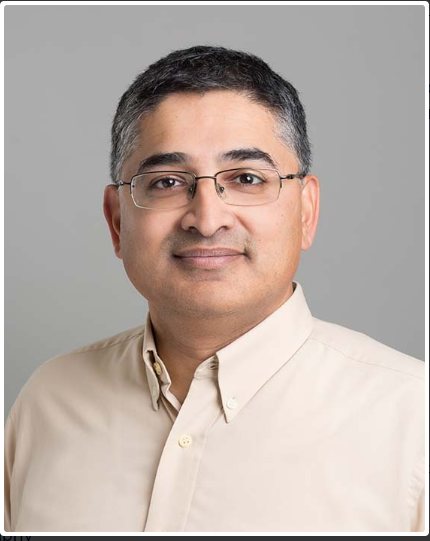}}]
	{Raviraj~Adve}
	(S'88, M'97, SM'06, F'17) was born in Bombay, India. He received his B. Tech. in Electrical Engineering from IIT, Bombay, in 1990 and his Ph.D. from Syracuse University in 1996, His thesis received the Syracuse University	Outstanding Dissertation Award. Between 1997 and August 2000, he worked for Research Associates for Defense Conversion Inc. on contract with the Air Force Research Laboratory at Rome, NY. He joined the faculty at the University of Toronto in August 2000 where he is currently a Professor. Dr. Adve's research interests include analysis and design techniques for cooperative and heterogeneous networks, energy harvesting networks and in signal processing techniques for radar and sonar systems. He received the 2009 Fred Nathanson Young Radar Engineer of the Year award. Dr. Adve is a Fellow of the IEEE.
\end{IEEEbiography}
\begin{IEEEbiography}[\vspace{-1em}{\includegraphics[width=25mm,height=32mm,clip,keepaspectratio]{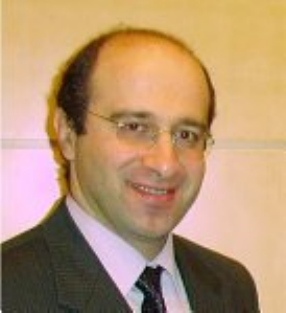}}]{Shahram~Shahbazpanahi}
	(M'02, SM'10) was born in Sanandaj, Kurdistan, Iran. He	received the B.Sc., M.Sc., and Ph.D. degrees in electrical engineering from Sharif University of Technology, Tehran, Iran, in 1992, 1994, and 2001, respectively. From September 1994 to September 1996, he was an instructor with the Department of Electrical Engineering, Razi University, Kermanshah, Iran. From July 2001 to March 2003, he was a Postdoctoral Fellow with the Department of Electrical and Computer Engineering, McMaster University, Hamilton, ON, Canada. From April 2003 to September 2004, he was a Visiting Researcher with the Department of Communication Systems, University of Duisburg-Essen, Duisburg, Germany. From September 2004 to April 2005,	he was a Lecturer and Adjunct Professor with the Department of Electrical and Computer Engineering, McMaster University. In July 2005, he joined the Faculty of Engineering and Applied Science, University of Ontario Institute of Technology, Oshawa, ON, Canada, where he currently holds a Professor position. His research interests include statistical and array signal processing; space-time adaptive processing; detection and estimation; multi-antenna, multi-user, and cooperative communications; spread spectrum techniques; DSP programming; and hardware/real-time software design for telecommunication systems. Dr. Shahbazpanahi has served as an Associate Editor for the IEEE TRANSACTIONS ON SIGNAL PROCESSING and the IEEE SIGNAL	PROCESSING LETTERS. He has also served as a Senior Area Editor for the IEEE SIGNAL PROCESSING LETTERS. He was an elected  member of the Sensor Array and Multichannel (SAM) Technical Committee of the IEEE Signal Processing Society. He has received several awards, including the Early Researcher Award from Ontario's Ministry of Research and Innovation, the NSERC Discovery Grant (three awards), the Research Excellence Award from the Faculty of Engineering and Applied Science, the University of Ontario Institute of Technology, and the Research Excellence Award, Early Stage, from the University of Ontario Institute of Technology.
\end{IEEEbiography}
\begin{IEEEbiography}[{\includegraphics[width=25mm,height=32mm,clip,keepaspectratio]{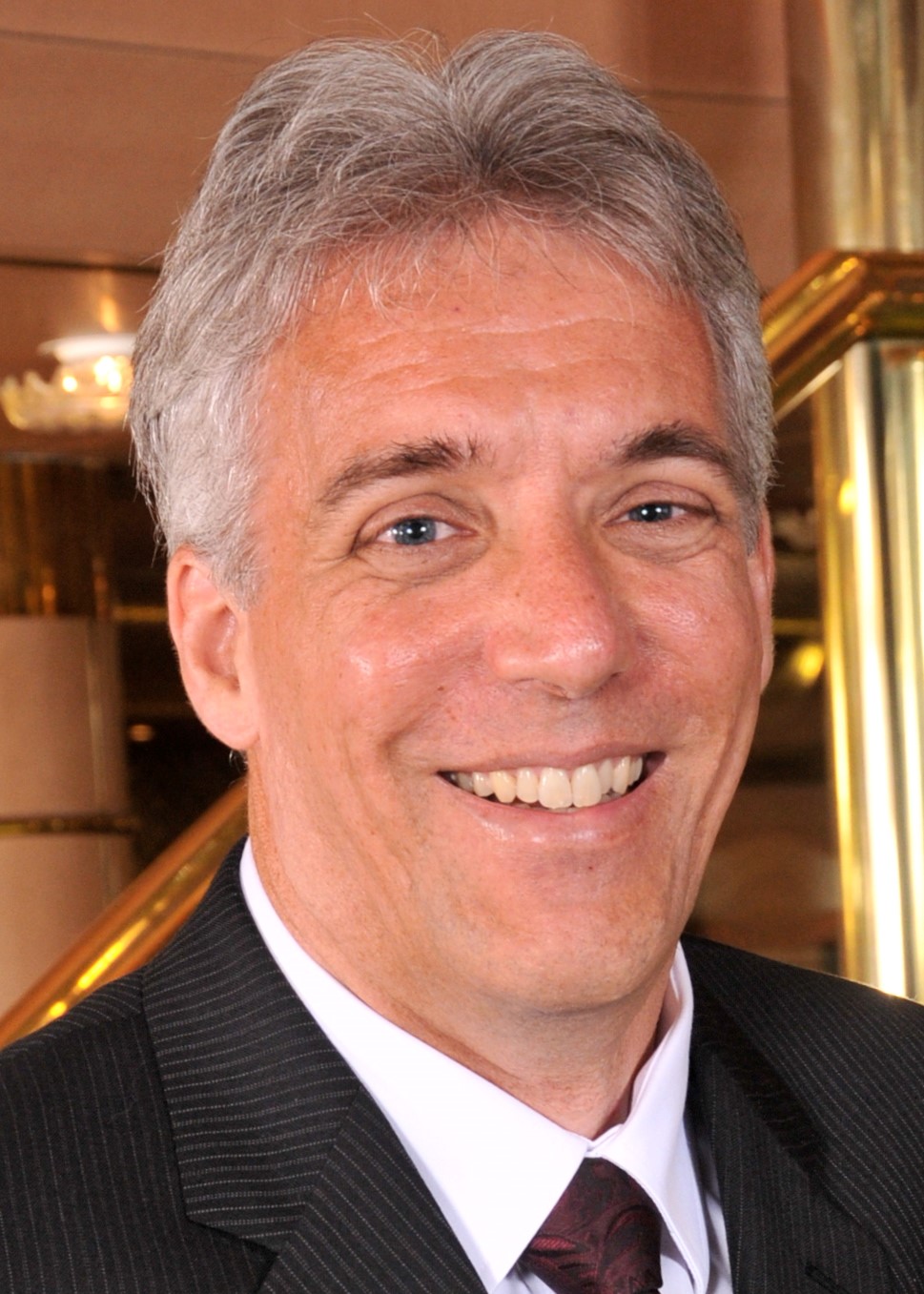}}]{Gary~Boudreau}
	(M'84-SM'11) received a B.A.Sc. in Electrical Engineering from the University of Ottawa in 1983, an M.A.Sc. in Electrical Engineering from Queens University in 1984 and a Ph.D. in electrical engineering from Carleton University in 1989. From 1984 to 1989 he was employed as a communications systems engineer with Canadian Astronautics Limited and from 1990 to 1993 he worked as a satellite systems engineer for MPR Teltech Ltd. For the period spanning 1993 to 2009 he was employed by Nortel Networks in a variety of wireless systems and management roles within the CDMA and LTE basestation product groups. In 2010 he joined Ericsson Canada where he is currently employed in the 5G systems architecture group. His interests include digital and wireless communications as well as digital signal processing.
\end{IEEEbiography}
\begin{IEEEbiography}[{\includegraphics[width=25mm,height=32mm,clip,keepaspectratio]{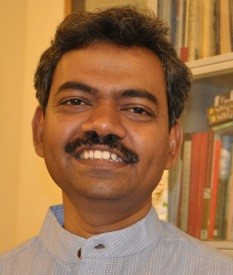}}]{Kothapalli~Venkata~Srinivas}
	was born in Vijayawada, India. He received the B.E. degree in electronics and communications engineering from the Andhra University College of Engineering, Vishakhapatnam, India, in June 1996, the M.Tech. degree from the Indian Institute of Technology, Kanpur, India, in 1998, and the Ph.D degree from the Indian Institute of Technology Madras, Chennai, India, in 2009, both in electrical engineering.\\ 
	He was a Postdoctoral Fellow at the Department of Electrical and Computer Engineering, University of Toronto, from March 2009 to October 2011. From January 2015 to December 2018, he was a Faculty member at the Indian Institute of Technology (BHU), Varanasi, India. His past industry experience includes working at the Indian Space Research Organisation, Samsung Electronics and Nokia Networks. In 2019 he joined Ericsson Canada where he is currently employed in the 5G systems group. His research interests include wireless communications, with emphasis on physical and MAC layer algorithms, and machine learning.
\end{IEEEbiography}

\end{document}